\documentclass[a4paper,12pt]{elsarticle}
\usepackage{fullpage}
\usepackage{subcaption}
\usepackage{hyperref}
\usepackage{amssymb}
\usepackage{amsmath}
\usepackage{mathrsfs}
\usepackage{graphicx}
\usepackage{epstopdf}
\usepackage{xcolor}
\usepackage{multirow}
\usepackage{array}
\usepackage{tikz} 
\usepackage{soul}
\usepackage[normalem]{ulem}
\usetikzlibrary{shapes.geometric,arrows} 
\usetikzlibrary{shapes,positioning}
\usepackage{ upgreek }
\newcolumntype{M}[1]{>{\centering\arraybackslash}m{#1}}
 
\makeatletter
\newcommand*\bigcdot{\mathpalette\bigcdot@{.5}}
\newcommand*\bigcdot@[2]{\mathbin{\vcenter{\hbox{\scalebox{#2}{$\m@th#1\bullet$}}}}}
\makeatother

\begin{document}

\begin{frontmatter}

\title{An adaptive approach to remove tensile instability in SPH for weakly compressible fluids}

\author[a,b]{Kanishka Bhattacharya}
\author[a]{Tapan Jana}
\author[a]{Amit~Shaw\corref{cor1}}
\ead{abshaw@civil.iitkgp.ac.in}
\cortext[cor1]{Corresponding author}

\author[a]{{L. S. Ramachandra}}
\author[c]{{Vishal Mehera}}

\address[a]{Civil Engineering Department, Indian Institute of Technology Kharagpur, West Bengal, India}
\address[b]{CSIR-Structural Engineering Research Centre, Chennai, India}
\address[c]{Bhabha Atomic Research Centre, Visakhapatnam, India}

\begin{abstract}
Smoothed Particle Hydrodynamics (SPH) is plagued by the phenomenon of \textit{tensile instability}, which is the occurrence of short wavelength zero energy modes resulting in unphysical clustering of particles. The root cause of the instability is the shape of derivative of the compactly supported kernel function which may yield negative stiffness in the particle interaction under certain circumstances. In this work, an adaptive algorithm is developed to remove \textit{tensile instability} in SPH for weakly compressible fluids. Herein, a B-spline function is used as the SPH kernel and the knots of the B-spline are adapted to change the shape of the kernel, thereby satisfying the condition associated with stability. The knot-shifting criterion is based on the particle movement within the influence domain. This enables the prevention of instability in fluid problems where excessive rearrangement of particle positions occurs. A 1D dispersion analysis of an Oldroyd B fluid material model is performed to show how the algorithm prevents instabilities for short wavelengths but ensures accuracy at large wavelengths. The efficacy of the approach is demonstrated through a few benchmark fluid dynamics simulations where a visco-elastic Oldroyd B material model and a non-viscous Eulerian fluid material model are considered. 

\end{abstract}


\end{frontmatter}

\section{Introduction}
\label{s1}
Smoothed Particle Hydrodynamics (SPH) is a particle-based method that has picked up much attention in the past few decades as an alternative to the traditional mesh-based methods. SPH was first developed by \citet{lucy1977numerical} and \citet{gingold1977smoothed} to simulate astrodynamical problems. Since then, SPH has been widely used in fluid dynamics problems. A lot of work has been done in the areas of incompressible flows (\citep{cummins1999sph},\citep{ellero2007incompressible},\citep{lind2012incompressible},\citep{pozorski2002sph},\citep{BOCKMANN2012138}), multiphase fluid flows (\citep{colagrossi2003numerical},\citep{wang2016overview},\citep{monaghan1995sph},\citep{CAO20187},\citep{YANG201998}), viscoelastic flows (\citep{fang2006numerical},\citep{rafiee2007incompressible},\citep{jiang2010sph},\citep{xu2013sph},\citep{xu2016improved}) and fluid-structure interaction (\citep{antoci2007numerical},\citep{khayyer2018enhanced},\citep{rafiee2009sph},\citep{REBOUILLAT2010739},\citep{KHANPOUR201667},\citep{NASAR2019563}). In the last few years, SPH has also been used in solid mechanics problems \citep{shaw2015beyond}. Few of the studies include fracture modeling (\citep{chakraborty2013pseudo},\citep{benz1995simulations}, \citep{zhao2023simulation}), high velocity impact and blast modeling (\citep{stellingwerf1993impact},\citep{rabczuk2003simulation},\citep{chakraborty2015prognosis},\citep{islam2017computational},\citep{FENG201377}, \citep{karmakar2021response}) and geotechnical simulations (\citep{bui2008lagrangian},\citep{bui2011slope},\citep{chen2012numerical},\citep{peng2015sph}). Despite its potential and exploration in several areas of computational mechanics, one major drawback of SPH is the \textit{tensile instability}, which, if unattended, may ruin the simulation.

\textit{Tensile instability} is the occurrence of small wavelength zero energy modes which pollute the solution and sometimes even change the entire dynamics of the problem. The root of the instability has been studied by many researchers and is now well documented (\citep{schuessler1981comments},\cite{phillips1985numerical},\cite{swegle1995smoothed},\cite{morris1996analysis},\citep{morris1996study}). As two SPH particles move away from each other due to negative pressure (tension), the magnitude of the gradient of the SPH kernel first increases, reaches a maximum and then decreases. The force between two SPH particles is proportional to the gradient of the kernel; consequently, the force also initially increases, reaches a maximum and then decreases. However, a decreasing force with increasing distance between two SPH particles results in negative stiffness, which ultimately causes an unphysical separation of the particles. This is the genesis of the \textit{tensile instability}. The same argument can be made for positive pressure. As two SPH particles approach each other, the repulsive force first increases, but after a point starts decreasing, which results in particle clumping. \citet{swegle1995smoothed} performed a detailed study of these instabilities. Via a 1D linear perturbation analysis, he arrived at an instability criterion which depends on the sign of the product of the stress and the second derivative of the SPH kernel function at the nearest neighbour.

A few remedies are available in the literature to tackle the problem of \textit{tensile instability}. \citet{schuessler1981comments} proposed a kernel whose $1$-st derivative monotonically increases as particles approach each other, thereby preventing the clumping of particles in compression. However, the $1$-st derivative of the kernel is discontinuous, and also, the kernel will not be able to prevent the instability in tension. Some other researchers (\citep{wen1994stabilizing}, \citep{guenther1994conservative}, \citep{hicks1997conservative}, \citep{randles1996smoothed}) used conservative smoothing on SPH variables, which effectively introduced a diffusive term in the conservation equations to attenuate the short wavelengths associated with the instability. \citet{guenther1994conservative} also showed how the conservative smoothing could be used as a more accurate dissipative mechanism than the standard artificial viscosity. \citet{dyka1995approach} and \citet{dyka1997stress}, in a 1D setting, introduced dual sets of particles: the standard SPH particles carried velocity, while `stress particles' were introduced between SPH particles, where stresses were calculated. Though this eliminated the tensile instability, carrying this forward to 2D becomes computationally intensive due to the tracking of the two different sets of particles and the mapping of properties from one set to the other (\citet{randles2000normalized}). \citet{monaghan2000sph} and \citet{gray2001sph} developed the artificial stress method. To prevent the clumping of particles due to the \textit{tensile instability}, they suggested the introduction of a small repulsive force between the particles. Using a dispersion analysis, they showed how the parameters associated with the repulsive force could be estimated to prohibit \textit{tensile instability} as well as ensure accuracy. Because the instability was noticeable only in tension, they provided the repulsive force only to particles in tension. For the modelling of fluid flows at low and moderate Reynold's numbers, a background compressive pressure was added to ensure that the entire domain is in compression (\citet{morris1997modeling}, \citet{marrone2013accurate}). This approach was successful in preventing the instabilities from arising in regions of negative pressure. The drawback with this approach is the setting of the background pressure, as too large a value results in numerical noise. \citet{yang2014smoothed} proposed a hyperbolic-shaped kernel to remove the instability in viscous fluids under compression. Similar to \citep{schuessler1981comments}, the value of the $1$-st derivative of the kernel increases as particles approach each other. Though it has been shown that the kernel is able to remove the instability in compression, it will not be able to prevent the instability in tension. Another method to tackle \textit{tensile instability} is the particle shifting method. When the equations of motion are solved, the SPH particles follow the streamlines of motion, which makes the particle distribution anisotropic, resulting in a breakdown of the solution at later stages. To tackle this, the particle shifting method was introduced in an Incompressible SPH setting (\citep{xu2009accuracy},\citep{lind2012incompressible}). The same particle shifting technique can be utilised to tackle the instability in Weakly Compressible SPH. Fick's law of diffusion is used to shift particles from regions of high concentration to regions of low concentration (\citep{sun2017deltaplus},\citep{xu2018technique}), thereby effectively preventing the clumping of particles.

The corrective measures mentioned above are either computationally intensive or require some parameters which need to be judiciously chosen \textit{a-priori}.  Recently, we proposed an adaptive approach \cite{lahiri2020stable} where the shape of the kernel at a particle is modified, on the basis of the state of stress. Using this approach, we were able to show how the issue of \textit{tensile instability} can be resolved in elastic dynamics problems. Based on a similar concept, a stable SPH computational framework for the simulation of Weakly Compressible fluids is developed in this paper. A B-spline basis function constructed over a variable knot vector is taken as the kernel, and its shape is adapted by changing the location of the intermediate knots to satisfy the Swegle's condition of preventing instability \citep{swegle1995smoothed}. Most of the studies (\citep{morris1997modeling},\citep{marrone2013accurate},\citep{monaghan2000sph},\citep{gray2001sph}) have shown that compressive stresses do not show any visible signs of instability; hence the remedies aim to remove the instability in tension. In the simulations performed in this paper, too, it was the instability in tension that affected the results. Hence, in this work, the shape of the kernel is modified in a bid to satisfy Swegle's condition for tension for the \emph{farthest immediate neighbour}, which automatically ensures the stability of all the other nearest neighbour points in tension. \citet{yang2014smoothed} had used a hyperbolic kernel to eliminate instability in problems involving positive pressure. Although the problem explored by \citet{yang2014smoothed} is not investigated in this paper, it is shown how the kernel used in this study can be adapted to mimic the properties of the hyperbolic kernel, thereby satisfying Swegle's condition for compression.

In this work, two benchmark problems viz. an impacting visco-elastic fluid drop and the rotation of an inviscid Eulerian fluid patch are considered. The governing equations for the visco-elastic fluid are presented in Section \ref{s2}, and the SPH discretisation of the same equations is given in Section \ref{s3}. A 1D perturbation analysis of the exact equations and the SPH discretised equations are performed in Section \ref{s4}. The proposed algorithm to tackle the instability is presented in Section \ref{s5}. The efficacy of the algorithm is demonstrated in Section \ref{s6}. Finally, the concluding remarks are highlighted in Section \ref{s7}.   

\section{Governing equations for a visco-elastic fluid}
\label{s2}
The conservation equations for a fluid in indicial notation are;
\begin{subequations}
\label{conserv_eq}
\begin{align}
&\frac{d\rho}{dt}=-\rho\frac{\partial v^\beta}{\partial x^\beta}, \label{conserv_eq_1} \\
&\frac{dv^\alpha}{dt}=\frac{1}{\rho}\frac{\partial \sigma^{\alpha\beta}}{\partial x^\beta}+g^\alpha,\label{conserv_eq_2}
\end{align}
\end{subequations}where $\rho$ is the density, $t$ is the time, $x^\beta$ and $v^\beta$ are the $\beta^{th}$ components of the position and velocity vector respectively, $\sigma^{\alpha\beta}$ is the $(\alpha,\beta)^{th}$ component of the stress tensor and $g^\alpha$ is the $\alpha^{th}$ component of the vector corresponding to the acceleration due to gravity. Einstein summation convention is followed, i.e. summation is taken over repeated indices.  

The stress tensor is expressed as the sum of the hydrostatic pressure ($P$) and a deviatoric stress. For an Oldroyd B fluid, which may be considered as a polymer solution, the deviatoric stress can be composed as the sum of a Newtonian solvent contribution ($\uptau_s^{\alpha\beta}$) and a polymeric contribution ($\uptau_p^{\alpha\beta}$). This gives,
\begin{equation}
\label{stress_dist}
\begin{split}
\sigma^{\alpha\beta}=-P\delta^{\alpha\beta}+\uptau_s^{\alpha\beta}+\theta\uptau_p^{\alpha\beta},
\end{split}
\end{equation}
where $\delta^{\alpha\beta}$ is the Kronecker Delta. A standard procedure in SPH is to consider a Weakly Compressible fluid with an equation of state for the calculation of the pressure as, 
\begin{equation}
\label{Pressure_EOS}
\begin{split}
P=\frac{\rho_0 c_0^2}{\gamma}\Big(\Big(\frac{\rho}{\rho_0}\Big)^{\gamma}-1\Big),
\end{split}
\end{equation}
where $c_0$ denotes the speed of sound, $\rho_0$ is the initial density, and $\gamma$ is taken to be 7 to make the equation stiff. The value of the speed of sound is set at least ten times the maximum fluid velocity. This keeps the Mach number ($M$) below $0.1$, and because $\frac{\delta \rho}{\rho} \sim M^2$, this ensures that the variation in density is less than $1\%$, and thus, the behaviour of the fluid is close to that of an incompressible fluid.

The solvent contribution of the deviatoric stress is linearly related to the rate of deformation tensor $d^{\alpha\beta}=\frac{1}{2}(\frac{\partial v^\alpha}{\partial x^\beta}+\frac{\partial v^\beta}{\partial x^\alpha})$ as
\begin{equation}
\label{tau_s}
\begin{split}
\uptau_s^{\alpha\beta}=2\eta_s d^{\alpha\beta},
\end{split}
\end{equation}
where $\eta_s$ is the solvent viscosity.
The polymer contribution can be obtained from the following differential equation:
\begin{equation}
\label{tau_p_1}
\begin{split}
\uptau_p^{\alpha\beta}+\lambda_1 \overset{\nabla}{\uptau_p^{\alpha\beta}}=2\eta_p d^{\alpha\beta},
\end{split}
\end{equation}
where $\lambda_1$ is the relaxation time of the fluid, $\eta_p$ is the polymer contribution to the viscosity, and $\overset{\nabla}{\uptau_p^{\alpha\beta}}$ is the upper convected derivative of $\uptau_p^{\alpha\beta}$ which is defined as
\begin{equation}
\label{tau_p_2}
\begin{split}
\overset{\nabla}{\uptau_p^{\alpha\beta}}=\frac{d\uptau_p^{\alpha\beta}}{dt}-\frac{\partial v^\alpha}{\partial x^\gamma}\uptau_p^{\gamma\beta}-\frac{\partial v^\beta}{\partial x^\gamma}\uptau_p^{\alpha\gamma}.
\end{split}
\end{equation} 
Substituting Equation \eqref{tau_p_2} in Equation \eqref{tau_p_1} we get
\begin{equation}
\label{tau_p_3}
\begin{split}
\frac{d\uptau_p^{\alpha\beta}}{dt}=\frac{\partial v^\alpha}{\partial x^\gamma}\uptau_p^{\gamma\beta	}+\frac{\partial v^\beta}{\partial x^\gamma}\uptau_p^{\alpha\gamma}-\frac{1}{\lambda_1}\uptau_p^{\alpha\beta}+\frac{2\eta_p}{\lambda_1}d^{\alpha\beta}.
\end{split}
\end{equation}   
In Equation \eqref{stress_dist}, $\theta=1$ gives an Oldroyd B model while $\theta=0$ gives a Newtonian model. An inviscid Eulerian fluid may be obtained by taking $\theta = 0$ and setting the viscosities ($\eta_s$ and $\eta_p$) to 0. 

\section{SPH equations}
\label{s3}
In SPH, the domain is discretised into particles, and at a given particle, a local continuous field over its neighbouring particles is created through a kernel function. Following Fang et. al., \citep{fang2006numerical}, the SPH discretised form of Equations \eqref{conserv_eq}, \eqref{tau_s} and \eqref{tau_p_3} may be written as;
\begin{subequations}
\label{SPH_discret}
\begin{align}
&\frac{d\rho_i}{dt}=\sum_j m_j (v^\beta_{i} - v^\beta_{j})\frac{\partial W_{ij}}{\partial x^\beta_i},\label{SPH_discret_1} \\
&\frac{dv_i^\alpha}{dt} = \sum_j m_j(\frac{\sigma_i^{\alpha\beta}}{\rho_i^2}+\frac{\sigma_j^{\alpha\beta}}{\rho_j^2}-\Pi_{ij}\delta^{\alpha\beta})\frac{\partial W_{ij}}{\partial x_i^\beta} + g^\alpha,\label{SPH_discret_2} \\
&\uptau_{s,i}^{\alpha\beta} = \eta_s(k_i^{\alpha\beta}+k_i^{\beta\alpha}),\label{SPH_discret_3} \\
&\frac{d\uptau_{p,i}^{\alpha\beta}}{dt}=k_i^{\alpha\gamma}\uptau_{p,i}^{\gamma\beta}+k_i^{\beta\gamma}\uptau_{p,i}^{\gamma\alpha}-\frac{1}{\lambda_1}\uptau_{p,i}^{\alpha\beta}+\frac{\eta_p}{\lambda_1}(k_i^{\alpha\beta}+k_i^{\beta\alpha}),\label{SPH_discret_4}
\end{align}
\end{subequations}    
where 
\begin{equation}
\label{velocity_grad}
\begin{split}
k_i^{\alpha\beta}=\frac{\partial v_i^\alpha}{\partial x^\beta}=\sum_j \frac{m_j}{\rho_j}(v_j^\alpha-v_i^\alpha)\frac{\partial W_{ij}}{\partial x_i^\beta}.
\end{split}
\end{equation}  
In Equation \eqref{SPH_discret_2}, $\Pi_{ij}$ is the artificial viscosity which is required to stabilise the computation in the presence of a shock or a sharp gradient. The following form of the artificial viscosity is used in the present study;
\begin{equation}
\Pi_{ij}=
\begin{cases}
		\frac{-\gamma_1\overline{c}_{ij}\mu_{ij} + \gamma_2\mu^2_{ij}}{\overline{\rho}_{ij}} & \text{for} \, \boldsymbol{x}_{ij}.\boldsymbol{v}_{ij} < 0,\\
		0 & \text{otherwise}, \\
\end{cases}
\label{artvisc}
\end{equation}
where, $\mu_{ij}= \frac{h\left(\boldsymbol{v}_{ij}.\boldsymbol{x}_{ij}\right)}{|\boldsymbol{x}_{ij}|^2 + \epsilon h^2}$; $\overline{c}_{ij} = \frac{c_i + c_j}{2}$; $\overline{\rho}_{ij} = \frac{\rho_i + \rho_j}{2}$; $\gamma_1$ and $\gamma_2$ are parameters which control the intensity of the artificial viscosity; $\epsilon$ is a small number to avoid singularity when two interacting particles ($i$ and $j$) are  close to each other; $c_i$ and $c_j$ are the wave propagation speeds evaluated at the $i$-th and $j$-th particles respectively; and $\boldsymbol{v}_{ij} = \boldsymbol{v}_i- \boldsymbol{v}_j$ and $\boldsymbol{x}_{ij} = \boldsymbol{x}_i- \boldsymbol{x}_j$ indicate the relative velocity and position of the $i-j$ particle pair. 

\section{Dispersion Analysis}
\label{s4} 
From the dispersion relation, one can obtain the wavelengths, which are Zero Energy Modes and due to which the instabilities in the system arise. The exact and the SPH dispersion relations for an Oldroyd B fluid are derived in this section. These relations are later on used in Section \ref{s5.4} to show how the approach outlined in this paper can prevent \textit{tensile instability}. 
\subsection{The Exact Dispersion Analysis}
\label{s4.1} 
First, the exact dispersion relation is derived for an Oldroyd B fluid. A 1D infinite expanse of fluid is considered, which is initially at rest. It is assumed that this 1D continuum has initial uniform stress $\overline{\sigma}=-\overline{P}+\overline{\uptau}_p$. From Equation \eqref{tau_s} and Equation \eqref{tau_p_3}, it can be understood that theoretically, a 1D continuum at rest cannot have non-zero values of ${\uptau}_s$, but can have non-zero values of ${\uptau}_p$. A perturbation is given to the initial state, and the resulting variables are
\begin{equation}
\label{exact_perturb}
\begin{split}
&v=Ve^{i(k\overline{x}-\omega t)}, \\
&\rho = \overline{\rho}+ \delta \rho, \\
&\delta \rho = De^{i(k\overline{x}-\omega t)}, \\
&P = \overline{P}+M\delta \rho, \\
&M = {c_0}^2(\overline{\rho}/\rho_0)^{\gamma -1}, \\
&\uptau_s = T_se^{i(k\overline{x}-\omega t)}, \\
&\uptau_p = \overline{\uptau}_p + T_pe^{i(k\overline{x}-\omega t)}, 
\end{split}
\end{equation}
where the initial state variables are denoted by a bar on the top. $\overline{x}$ is the spatial coordinate at the initial state. $V$, $D$, $M$, $T_s$ and $T_p$ are the amplitudes of the perturbations to $v$, $\rho$, $P$, $\uptau_s$ and $\uptau_p$ respectively. Substituting these perturbed variables in the continuity equation (Equation \eqref{conserv_eq_1}) yields
\begin{equation}
\label{exact_pert_continuity}
D = \frac{\overline{\rho}}{\omega}kV.
\end{equation}
The linear momentum conservation equation (Equation \eqref{conserv_eq_2}) upon perturbation becomes
\begin{equation}
\label{exact_pert_momentum}
\overline{\rho}\omega V = k(MD-T_s-\theta T_p).
\end{equation}
Upon substituting the perturbed variables from Equation \eqref{exact_perturb} in the equation for the solvent contribution, ${\uptau}_s$ (Equation \eqref{tau_s}) and the polymer contribution ${\uptau}_p$ (Equation \eqref{tau_p_3}) of the deviatoric stress, we obtain;
\begin{subequations}
\label{exact_pert_taus_taup}
\begin{align}
&T_s = 2i\eta_s kV,\label{exact_pert_taus_taup_1} \\
&T_p = \frac{2(\overline{\uptau_p}+\eta_p/\lambda_1)}{(\frac{1}{\lambda_1}-i\omega)}ikV.\label{exact_pert_taus_taup_2}
\end{align}
\end{subequations}
Upon using $T_p$ from Equation \eqref{exact_pert_taus_taup_2} in Equation \eqref{exact_pert_momentum}, an analytical expression for the dispersion relation cannot be obtained. Now, the exact dispersion relation is going to be used to validate the accuracy of the SPH dispersion relation for long wavelengths, i.e. $k \rightarrow 0$. From $\omega = c k$, we see that if $k \rightarrow 0$, then $\omega \rightarrow 0$. Now, $\lambda_1 = 0.02$ for the impact drop problem in Section \ref{s6.2}, hence we can say, $|i\omega| << |\frac{1}{\lambda_1}|$ for large wavelengths, and obtain a simplified equation for $T_p$;
\begin{equation}
\begin{split}
\label{exact_pert_taup_simpl}
&T_p = 2ikV(\overline{\uptau_p}+\eta_p/\lambda_1)\lambda_1.
\end{split}
\end{equation}
Finally, upon substitution of Equations \eqref{exact_pert_continuity}, \eqref{exact_pert_taus_taup_1} and \eqref{exact_pert_taup_simpl} in Equation \eqref{exact_pert_momentum} we obtain a quadratic equation in $\omega$ as,
\begin{equation}
\label{exact_dipersion}
\overline{\rho}\omega^2+2i k^2 Z \omega -M\overline{\rho}k^2=0,
\end{equation}
where $Z = (\eta_s + \theta (\overline{\uptau_p}+\eta_p/\lambda_1)\lambda_1)$. Solving for $\omega$ we get
\begin{equation}
\label{exact_omega}
\omega = -\frac{k^2Z}{\overline{\rho}}i \pm \sqrt{Mk^2-\frac{k^4Z^2}{\overline{\rho}^2}}.
\end{equation}
So, we obtain $\omega$ in the form $\omega=Re(\omega)+iIm(\omega)$. Now, in the perturbation of the velocity, we get $v=Ve^{i(k\overline{x}-Re(\omega)t)}e^{Im(\omega)t}$. From the harmonic component of the perturbation, we obtain the wave speed as 
\begin{equation}
\label{exact_wavespd}
c=Re(\omega)/k=\sqrt{M-\frac{k^2Z^2}{\overline{\rho}^2}} ,
\end{equation}
which is the exact dispersion relation for a 1D Oldroyd B continuum.

\subsection{The SPH Dispersion Analysis}
\label{s4.2} 
In this section, the SPH Dispersion relation is derived. A 1D infinite expanse of SPH particles with uniform spacing $\Delta p$, at rest, is considered. Similar to the exact dispersion analysis, it is assumed that this 1D continuum has initial uniform stress $\overline{\sigma}=-P+\overline{\uptau}_p$. Now, a harmonic perturbation is given to these SPH particles. The perturbation in position and velocity of particle $a$ is
\begin{equation}
\label{SPH_perturb}
\begin{split}
&x_a = \overline{x}_a + \delta x_a, \\
&\delta x_a = Xe^{i(k\overline{x}_a-\omega t)}, \\
&\delta v_a = Ve^{i(k\overline{x}_a-\omega t)}.
\end{split}
\end{equation}
The perturbation in density, pressure and stresses are the same as in Equation \eqref{exact_perturb} with a subscript $a$, denoting the variable value at particle $a$. Here $\overline{x}_a$ denotes the initial position of particle $a$. The continuity equation (Equation \eqref{SPH_discret_1}) upon perturbation is 
\begin{equation}
\label{SPH_pert_continuity}
\begin{split}
\frac{d(\delta \rho_a)}{dt}=-\sum_b \overline{\rho}\Delta p(\delta v_b - \delta v_a)(\frac{\partial W_{ab}}{\partial \overline{x}_a}+\frac{\partial^2 W_{ab}}{\partial \overline{x}^2_a}(\delta x_a - \delta x_b)),
\end{split}
\end{equation}
where the summation is over particles $b$ within the domain of $a$.
It is assumed that the 1D bar has a unit cross-sectional area, i.e. $m=\overline{\rho}\Delta p$. Considering only the first-order terms and substituting for the perturbed variables, we obtain
\begin{equation}
\label{SPH_perturb_D}
\begin{split}
D = \frac{\overline{\rho}\Delta p V}{\omega}\sum_b \sin k\xi \frac{\partial W_{ab}}{\partial \overline{x}_a}
\end{split}
\end{equation}
where $\xi = \overline{x}_a - \overline{x}_b$. The linear momentum conservation equation (Equation \eqref{SPH_discret_2}) reads
\begin{equation}
\label{SPH_pert_momentum}
\begin{split}
\frac{(d\delta v_a)}{dt}=\sum_b \overline{\rho}\Delta p[\frac{\overline{\sigma}+\delta \sigma_a}{(\overline{\rho}	+\delta \rho_a)^2}+\frac{\overline{\sigma}+\delta \sigma_b}{(\overline{\rho}	+\delta \rho_b)^2}][\frac{\partial W_{ab}}{\partial \overline{x}_a}+\frac{\partial^2 W_{ab}}{\partial \overline{x}^2_a}(\delta x_a - \delta x_b)].
\end{split}
\end{equation}
Only keeping the first-order terms in Equation \eqref{SPH_pert_momentum} gives
\begin{equation}
\label{SPH_pert_momentum_2}
\begin{split}
(-i\omega \delta v_a)&=\frac{2\overline{\sigma}\Delta p}{\overline{\rho}}\sum_b (\delta x_a - \delta x_b)\frac{\partial^2 W_{ab}}{\partial \overline{x}^2_a}+\frac{\Delta p}{\overline{\rho}}\sum_b \delta \sigma_b\frac{\partial W_{ab}}{\partial \overline{x}_a} - \frac{2\overline{\sigma}\Delta p}{\overline{\rho}^2}\sum_b \delta \rho_b\frac{\partial W_{ab}}{\partial \overline{x}_a} \\
&=[\frac{2\overline{\sigma}\Delta p}{\overline{\rho}}X\sum_b (1-\cos k\xi)\frac{\partial^2 W_{ab}}{\partial \overline{x}^2_a}+\frac{\Delta p}{\overline{\rho}}(-MD+T_s+\theta	T_p)\sum_b i\sin k\xi\frac{\partial W_{ab}}{\partial \overline{x}_a} \\
&-2\overline{\sigma}\frac{\Delta pD}{\overline{\rho}^2}\sum_b i\sin k\xi\frac{\partial W_{ab}}{\partial \overline{x}_a}]e^{i(k\overline{x}_a-\omega t)}. 
\end{split}
\end{equation}
In the above equation, $\delta\sigma_b=-\delta P_b + \delta\uptau_{s,b} +\theta \delta\uptau_{p,b}$ is used. By substituting the perturbed variables in Equation \eqref{SPH_discret_3} and \eqref{SPH_discret_4} we arrive at
\begin{subequations}
\label{SPH_perturb_Ts_Tp}
\begin{align}
&T_s = 2\eta_s\Delta pVi\sum_b \sin k\xi\frac{\partial W_{ab}}{\partial \overline{x}_a},\label{SPH_perturb_Ts_Tp_1} \\
&T_p=2\Delta p\frac{(\overline{\uptau}_p+\eta_P/\lambda_1)}{(\frac{1}{\lambda_1}-i\omega)}Vi\sum_b \sin k\xi\frac{\partial W_{ab}}{\partial \overline{x}_a}.\label{SPH_perturb_Ts_Tp_2}
\end{align}
\end{subequations}
Now, similar to the discussion in the previous section, if $T_p$ from Equation \eqref{SPH_perturb_Ts_Tp_2} is used, an analytical expression of the dispersion relation may not be possible. The SPH dispersion relation is being derived to compare its accuracy with the exact dispersion relation for long wavelength modes and also to investigate the \textit{tensile instability} for short wavelength modes. As already discussed in Section \ref{s4.1}, the approximation $|i\omega| << |\frac{1}{\lambda_1}|$ can be used for long wavelength modes. 
Now, the shortest wavelength is $\lambda=2\Delta p$, which gives us $\sum_b \sin k\xi\frac{\partial W_{ab}}{\partial \overline{x}_a}=0$.  Hence from Equation (\ref{SPH_pert_momentum_2}) we arrive at $\omega = \sqrt{\frac{-2\overline{\sigma}\Delta p B}{\overline{\rho}}}$, where $B=\sum_b (1-\cos k\xi)\frac{\partial^2 W_{ab}}{\partial \overline{x}^2_a}$. Without any loss of generalisation, $\uptau_p$ is ignored, i.e. $\overline{\sigma}=-\overline{P}$. Now, for a range of density ratios $(\frac{\rho}{\rho_0})$ from $0.95 - 1.05$, and for a range of $h/\Delta p$ from $0.8 - 2$, from $\omega = \sqrt{\frac{-2\overline{\sigma}\Delta p B}{\overline{\rho}}}$ we calculate that the magnitude of $\omega$ is atleast one order of magnitude less than $1/\lambda_1$. Hence for short wavelengths as well, the approximation $|i\omega| << |\frac{1}{\lambda_1}|$ can be used. Hence we obtain the modified equation of $T_p$ as;
\begin{equation}
\label{SPH_perturb_Tp_simpl}
\begin{split}
T_p=2\Delta p (\overline{\uptau}_p+\eta_P/\lambda_1) \lambda_1 Vi\sum_b \sin k\xi\frac{\partial W_{ab}}{\partial \overline{x}_a}. 
\end{split}
\end{equation}
Substituting Equations \eqref{SPH_perturb_D}, \eqref{SPH_perturb_Ts_Tp_1} and \eqref{SPH_perturb_Tp_simpl} in Equation \eqref{SPH_pert_momentum_2} we obtain a quadratic equation in $\omega$;
\begin{equation}
\label{SPH_dispersion}
\begin{split}
\overline{\rho}\omega^2+2iA^2(\Delta p)^2Z\omega-M\overline{\rho}(\Delta p)^2A^2 - 2\overline{\sigma}(\Delta p)^2A^2 + 2\overline{\sigma}\Delta pB=0,
\end{split}
\end{equation}
where $A=\sum_b \sin k\xi\frac{\partial W_{ab}}{\partial \overline{x}_a}$. Solving the quadratic equation gives 
\begin{equation}
\label{SPH_omega}
\begin{split}
\omega=\frac{-Z\Delta p^2A^2i}{\overline{\rho}}\pm\sqrt{\frac{-Z^2\Delta p^4A^4}{\overline{\rho}^2}-\frac{2\overline{\sigma}\Delta pB}{\overline{\rho}}+A^2\Delta p^2(M+\frac{2\overline{\sigma}}{\overline{\rho}})}.
\end{split}
\end{equation}
Hence, the wave speed is obtained as 
\begin{equation}
\label{SPH_wavespd}
\begin{split}
c=\frac{1}{k}\sqrt{\frac{-Z^2\Delta p^4A^4}{\overline{\rho}^2}-\frac{2\overline{\sigma}\Delta pB}{\overline{\rho}}+A^2\Delta p^2(M+\frac{2\overline{\sigma}}{\overline{\rho}})}.
\end{split}
\end{equation}

\section{Adaptive Algorithm for Stable SPH computation}
\label{s5}
As mentioned in the \nameref{s1}, Swegle's stability analysis \cite{swegle1995smoothed} constitutes the premise of the adaptive algorithm developed in this work. Herein, the shape of the kernel at a given particle location is continuously modified, such that the condition which may cause instability does not arise. However, while doing so, it is also important to ensure that the adaptive exercise does not become computationally intensive. To this end, a B-spline basis function defined over a set of variable knots is considered as the kernel. The advantage of a B-Spline basis function is that the shape of the kernel can be modified by changing the position of the knots. The algorithm and its implementation steps are discussed in this section. 

First, the B-Spline basis function for a variable knot vector is presented in Section \ref{s5.1}. Using this basis function as the kernel, it is shown how the adaptive algorithm works in Section \ref{s5.2}. In Section \ref{s5.3}, it is shown how the \emph{farthest immediate neighbour} is estimated. Finally, in Section \ref{s5.4}, the 1D dispersion relation for the Oldroyd B material is plotted to show how the zero energy modes can be eliminated.
\subsection{B-Spline Basis Function as Kernel} 
\label{s5.1} 
We use the deBoor, Cox and Mansfield recurrence formula (\citep{piegl1996nurbs}) to define the B-Spline basis functions. Let $\Xi=\lbrace\zeta_1,\zeta_2, \zeta_3,...,\zeta_{m} | \zeta_{I} \in \mathbb{R} \rbrace$ be a non-decreasing sequence of real numbers called as the knot vector with $\zeta_{I}$ being the position of the $I$-th knot. The $I$-th B-Spline basis function of $P$-th degree denoted by $N_{I,P}(\zeta)$ is defined as;
\begin{equation}
\label{B-spline basis}
\begin{split}
  &N_{I,0}=\begin{cases}
    1, & \text{if $\zeta_{I}\leq \zeta<\zeta_{I+1}$}.\\
    0, & \text{otherwise}.
  \end{cases} \\
N_{I,P}(\zeta)&=\frac{\zeta-\zeta_{I}}{\zeta_{I+P}-\zeta_I}N_{I,P-1}(\zeta)+\frac{\zeta_{I+P+1}-\zeta}{\zeta_{I+P+1}-\zeta_{I+1}}N_{I+1,P-1}(\zeta).    
\end{split}
\end{equation}

\begin{figure}
\begin{subfigure}[b]{0.5\textwidth}
\includegraphics[trim={2cm 0 1.5cm 0}, width=\textwidth]{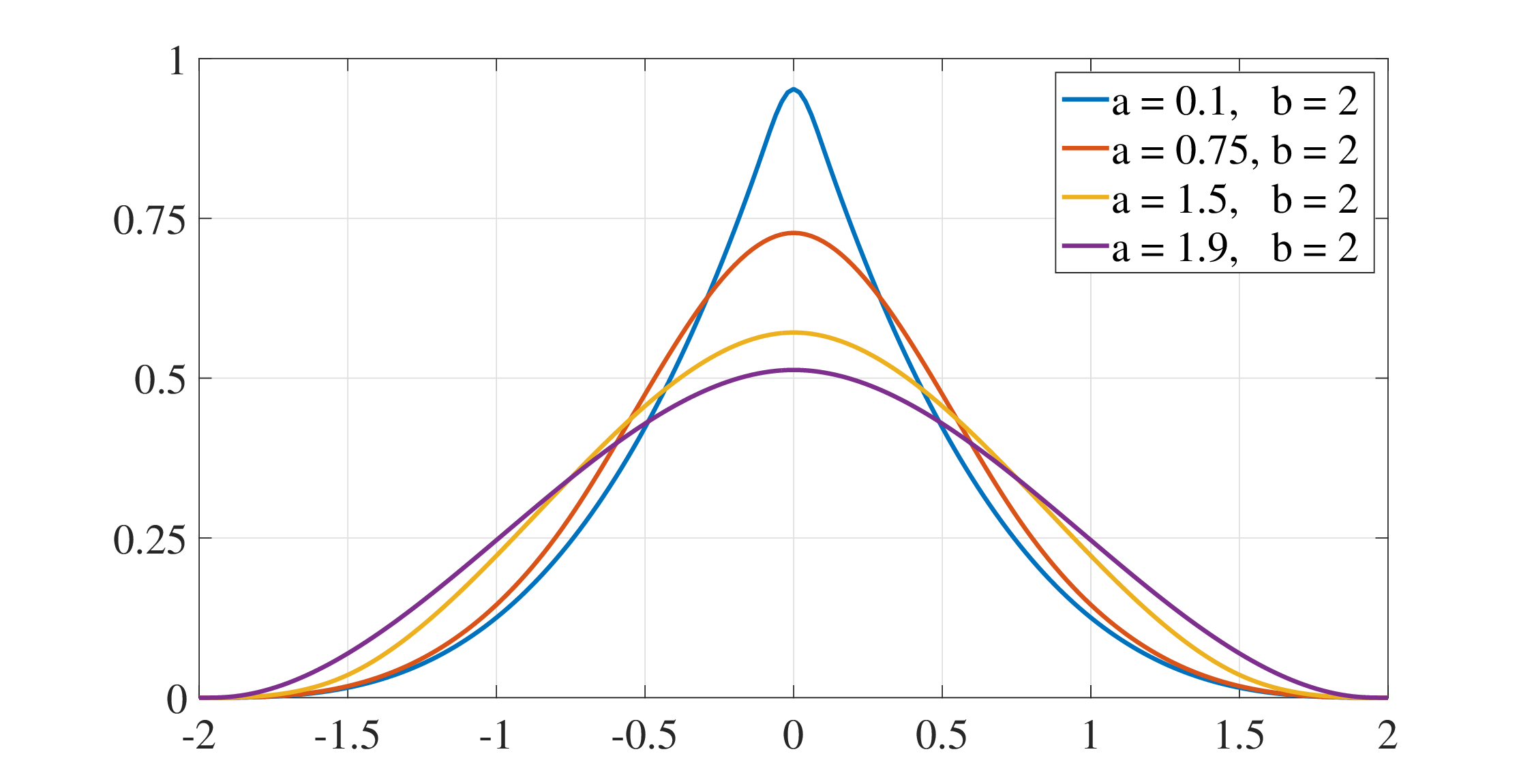}
\caption{}
\end{subfigure}
\begin{subfigure}[b]{0.5\textwidth}
\includegraphics[trim={1.5cm 0 2cm 0}, width=\textwidth]{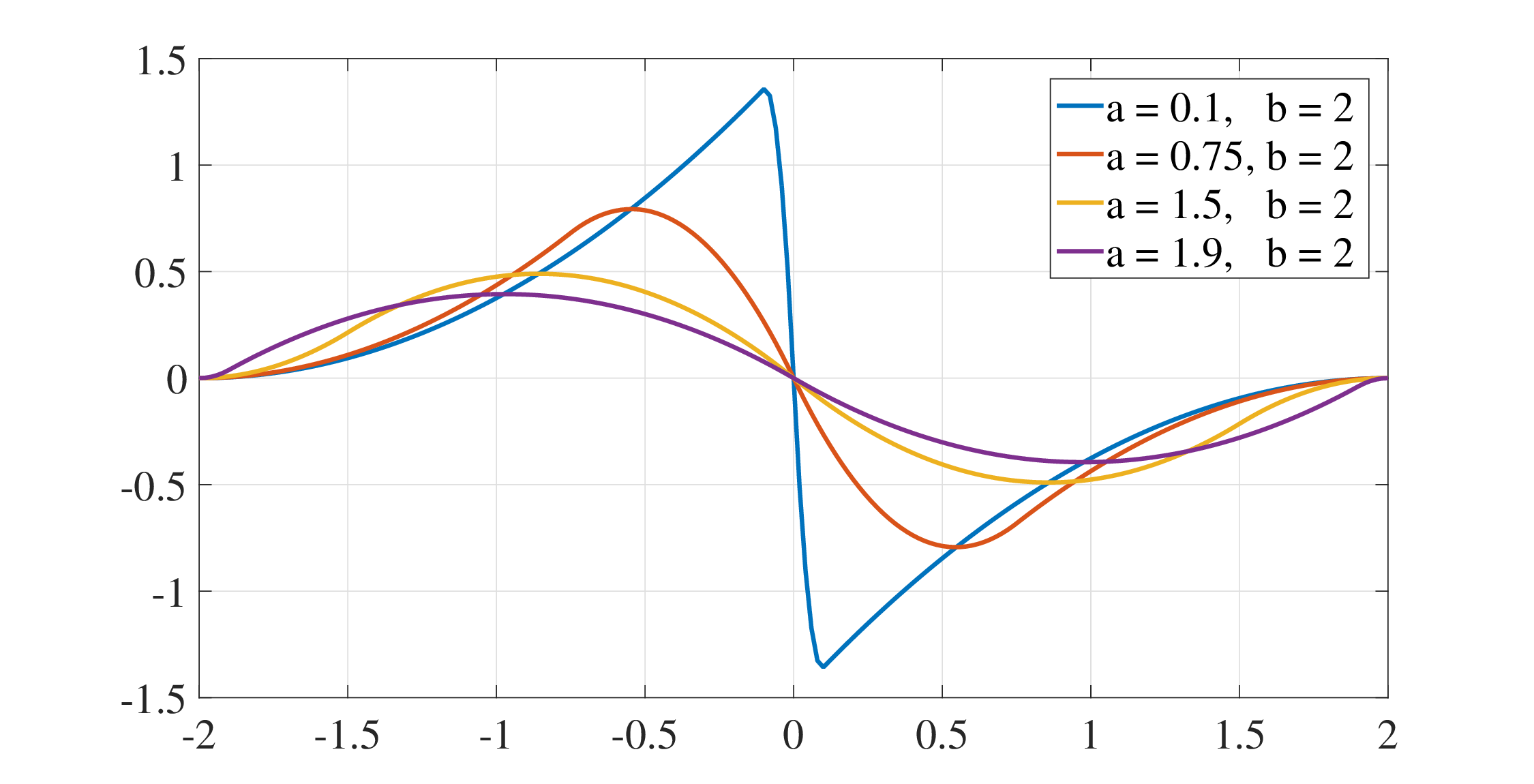}
\caption{}
\end{subfigure}
\caption{Cubic kernel over different intermediate knots; (a) Kernel, (b) first derivative of Kernel}
\label{figure 0}
\end{figure}

The local support property of the B-spline basis function gives $N_{I,P}(\zeta)>0~\forall ~ \zeta~\in~[\zeta_I,\zeta_{I+P+1})$. The  shape of $N_{I,P}(\zeta)$, within its support $[\zeta_I,\zeta_{I+P+1})$, can be modified by changing the position of intermediate knots $\lbrace \zeta_{I+1},...\zeta_{I+P} \rbrace$. The support of $N_{I,P}(\zeta)$ can be changed by changing the positions of the extreme knots $\lbrace \zeta_{I},\zeta_{I+P+1} \rbrace$. Herein, we take a symmetric knot vector $\Xi=\{-b,-a,0,a,b\}$ and the basis function $N_{0,3}$ to construct a symmetric cubic spline kernel. The resulting kernel we get is;
\begin{equation}
\label{kernel}
\begin{split}
  W(q,h)=\alpha_c \begin{cases}
    \frac{(a+b)q^3-3abq^2+a^2b^2}{a^2b(a+b)}, & \text{if $0\leq q<a$}\\
    \frac{(b-q)^3}{b(b^2-a^2)}, & \text{if $a\leq q<b$}\\
    0, & \text{if $b\leq q$}
  \end{cases}     
\end{split}
\end{equation}
where $\alpha_c$ is obtained from the normalising condition for the kernel, i.e. $\int_\Omega W(\boldsymbol{x}-\boldsymbol{x}',h)d\boldsymbol{x}'=1$. $\alpha_c=\frac{2}{bh}$ for 1D and $\alpha_c=\frac{10(a+b)}{\pi b(a^2+ab+b^2)h^2}$ for 2D. As shown in Figure \ref{figure 0}, changing the position of the knots results in a change in the shape of the kernel, which is the basis of the adaptive algorithm, as explained in the next section.

\subsection{Adaptive Algorithm}
\label{s5.2} 
In a 1D stability analysis, Swegle had shown that to remove \textit{tensile instability}, $W^{''}$ at the nearest neighbour should be less than zero for a state of tension and greater than zero for a state of compression. 
 
However, from most of the studies (\citep{monaghan2000sph},\citep{gray2001sph},\citep{morris1997modeling},\citep{marrone2013accurate}), one is led to understand that the instability in tension is more prominent and can severely pollute the solution. In \citet{monaghan2000sph}, the author provided an artificial pressure primarily when the material was under negative pressure (i.e., tension), and in \citet{gray2001sph}, the authors provided an artificial stress only along the principal direction in tension. In \citet{morris1997modeling}, and \citet{marrone2013accurate}, the authors provided a background pressure to ensure the pressure of the entire domain is positive at all times. In the simulations performed in this work too, it is shown that satisfying Swegle's condition for tension is sufficient to prevent instability.

\begin{figure}
\begin{subfigure}[b]{.32\textwidth}
\includegraphics[width=\textwidth]{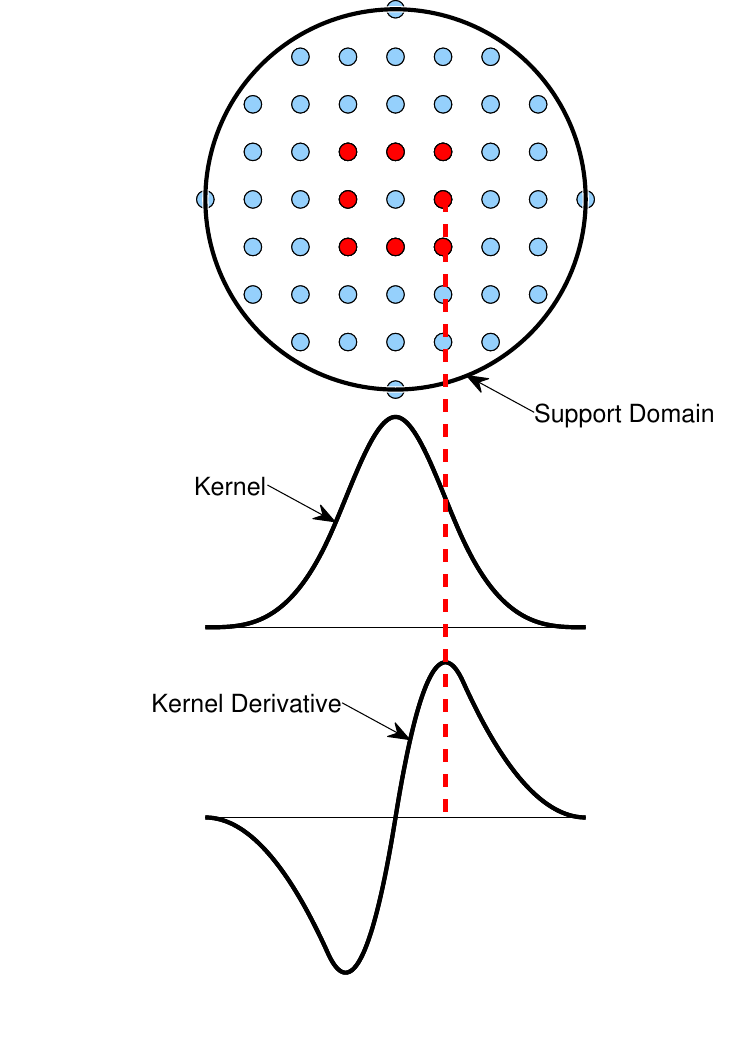}
\caption{$a = 0.7, b = 2.0$}
\label{figure 1a}
\end{subfigure}
\begin{subfigure}[b]{.32\textwidth}
\includegraphics[width=\textwidth]{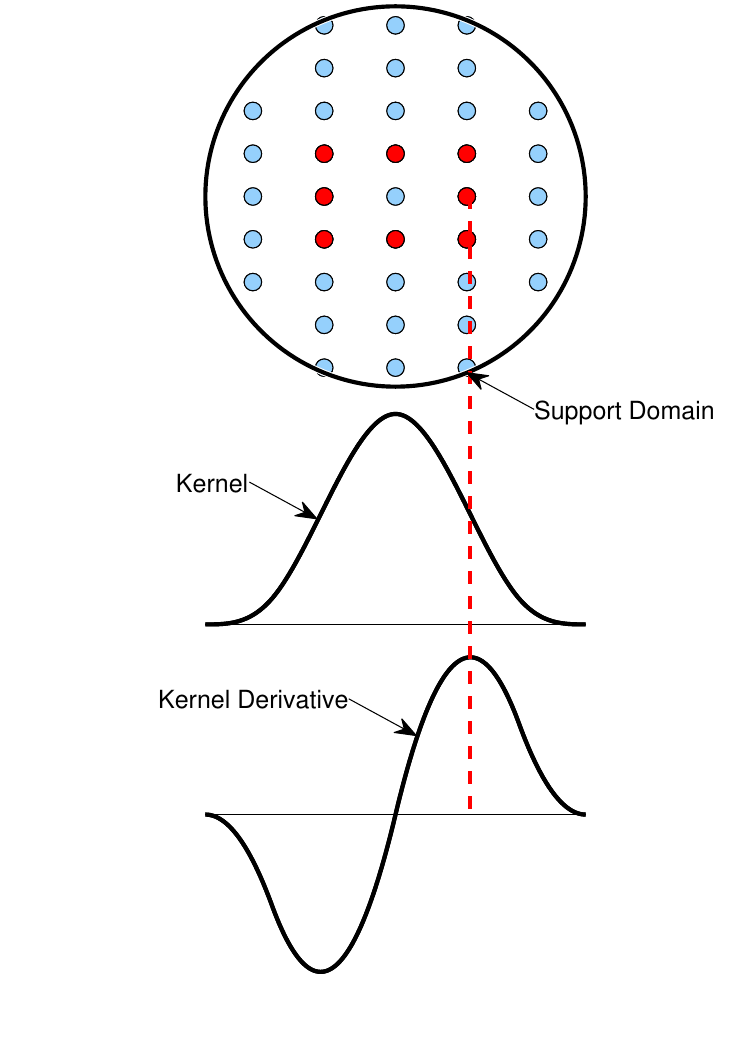}
\caption{$a = 1.3, b = 2.0$}
\label{figure 1b}
\end{subfigure}
\begin{subfigure}[b]{.32\textwidth}
\includegraphics[width=\textwidth]{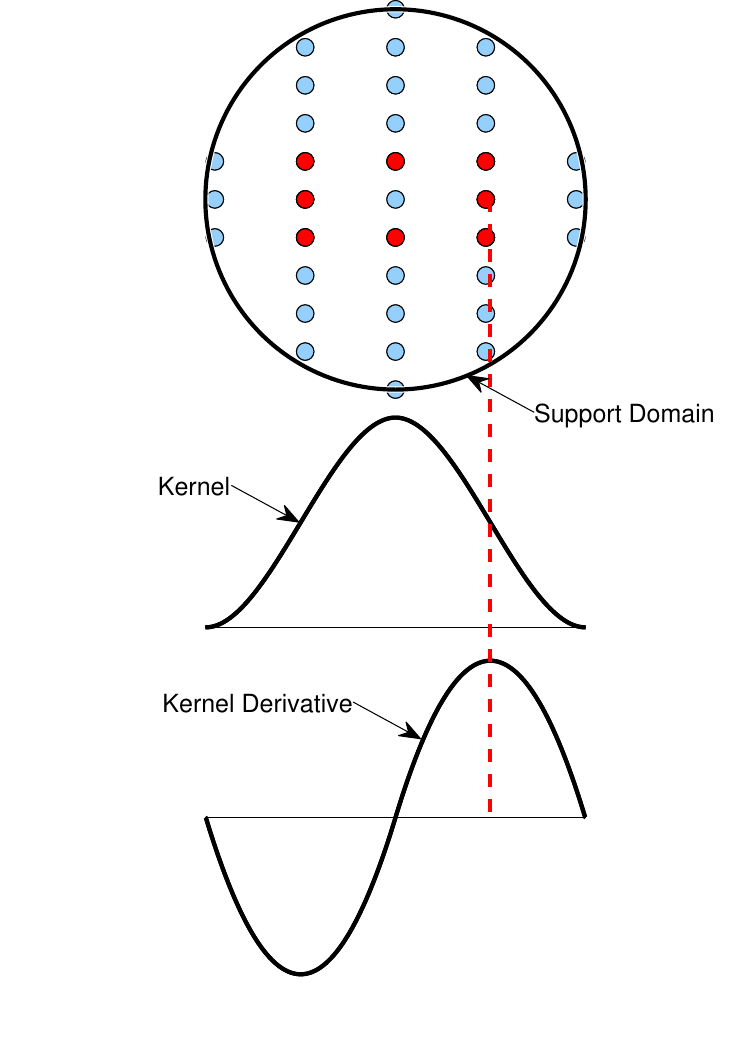}
\caption{$a = 1.99, b = 2.0$}
\label{figure 1c}
\end{subfigure}
\caption{Change in shape of the kernel and it's $1$-st derivative with the shifting of the \emph{farthest immediate neighbour}. The red points represent the nearest neighbours and the dotted red line represents the position of the extremum of $W^{'}$.}
\label{figure 1}
\end{figure}

Though the proposed adaptive algorithm is applicable for any quasi-uniform particle distribution, for a better comprehension, the steps involved in the method are demonstrated through a particle arrangement following a rectangular grid as shown in Figure \ref{figure 1}. The smoothing length $h$ is taken as $2\Delta p$, where $\Delta p$ is the particle spacing. The influence domain of a particle, say $i$-th particle with position $\boldsymbol{x}_i$, is defined as $\mathbb{N}^i = \lbrace j \in \mathbb{Z}^+ ~|~ ||\boldsymbol{x}_i-\boldsymbol{x}_j|| < bh ~\mathrm{and}~ i \neq j \rbrace$, with $b$ being the cutoff of the kernel as defined in Equation \eqref{kernel}. Let $\overline{\mathbb{N}^i} \subseteq \mathbb{N}^i$ be the set of immediate neighbours. For the given particle arrangement in Figure \ref{figure 1}, the immediate neighbours are highlighted in red. For simplicity, we are going to assume that tensile stress acts along the $x$ axis and compressive stress along the $y$ axis. To prevent the \textit{tensile instability} from arising at the $i$-th particle, we have to ensure that, in the direction of tension, $W^{''}_{ij}<0 ~ \forall j \in \overline{\mathbb{N}^i}$. Essentially, we have to track the \emph{farthest immediate neighbour} and ensure that $W^{''}<0$ at that position. The approach adopted in this work is described next.

\subsubsection{\textit{a}-adaptive}
\label{s5.2.1} 
For a cubic spline kernel (Equation \eqref{kernel}) with smoothing length $h$, the position of the extremum of $W^{'}$ is at $\frac{ab}{a+b}h$. Let $r_i=\max\limits_{j \in \overline{\mathbb{N}^i}}\lbrace \parallel \boldsymbol{x}_i-\boldsymbol{x}_j \parallel  \rbrace$ be the distance of the \emph{farthest immediate neighbour} (say $j$) from particle $i$. For the extremum of $W^{'}$ to be positioned at $j$, the value of knot $a$ should be: $a=\frac{br_i}{bh-r_i}$. Now, if the position of the extremum of $W^{'}$ is slightly beyond $j$, then the condition $W^{''}<0$ will be satisfied at all immediate neighbours. Hence the value of knot $a$ should be such that:
\begin{equation}
\label{adaptive_1}
\begin{split}
&a = \frac{br^*}{bh-r^*}, \textrm{where} \: r^* = Ar_i.  \\  
\end{split}
\end{equation}
In Equation \ref{adaptive_1}, $A>1$ is a multiplying constant which ensures that the stable zone of the kernel always covers the \emph{farthest immediate neighbour}. Equation \eqref{kernel}, with $a = 1$ and $b=2$, reproduces the commonly used Cubic B-spline kernel in the literature. In the present study, we also take $b=2$ unless large tensile strains occur, which is discussed in the next sub-section (\nameref{s5.2.2}). The intermediate knot $a \in (0,b)$ is adjusted according to Equation \ref{adaptive_1}. It is to be noted that $A$ in Equation \ref{adaptive_1} does not require any tuning or calibration. The sole purpose of taking a value of $A$ greater than 1 is to ensure that the extremum of $W^{'}$ is always slightly ahead of the \emph{farthest immediate neighbour} and thereby Swegle's criteria for preventing \textit{tensile instability} is effectively satisfied. It is observed in the simulations of this paper that values of $A$ from $1.05$ to $1.1$ serves the purpose. The concept is demonstrated in Figure \ref{figure 1}, where it can be observed how the extremum of $W^{'}$ is always slightly ahead of the \emph{farthest immediate neighbour} when $a$ is estimated from Equation \eqref{adaptive_1}. 

\subsubsection{\textit{ab}-adaptive}
\label{s5.2.2}
Now, consider a situation where the neighbourhood of a particle is under continuous tension. This causes the \emph{farthest immediate neighbour} to continuously move away from the centre particle, the $i$-th particle in this case. As $r_i$ increases, the intermediate knot $a$ is also increased as per Equation \ref{adaptive_1}. When $r^* \approx h$ or $r_i \approx 0.95h$ (for $A=1.05$), Equation \ref{adaptive_1} yields $a \approx 2$, which is also the value of $b$. This is the limiting situation beyond which the further shifting of $a$ is not possible as long as $b$ is fixed at 2. Now suppose the \emph{farthest immediate neighbour} further moves away due to continued tension. However, since $a$ has already reached its limiting value (i.e., $b$), the kernel shape cannot be further adjusted through $a$. This may cause the \emph{farthest immediate neighbour} to cross the extremum of $W^{'}$ and leave the stable zone of the kernel. In such a situation, to prohibit the instability from occurring, both the values of $a$ and $b$ are allowed to increase such that the position of the extremum can be shifted along with the \emph{farthest immediate neighbour}. 

\begin{figure}
\begin{subfigure}[b]{.32\textwidth}
\includegraphics[width=\textwidth]{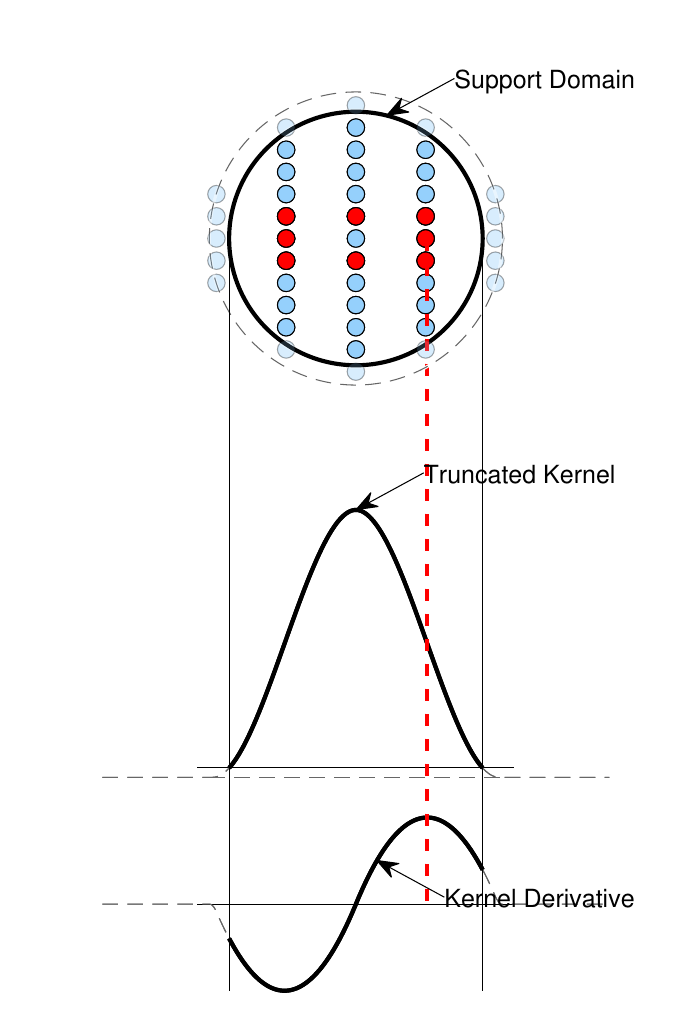}
\caption{$a = 2.19, b = 2.31$}
\label{figure a2}
\end{subfigure}
\begin{subfigure}[b]{.32\textwidth}
\includegraphics[width=\textwidth]{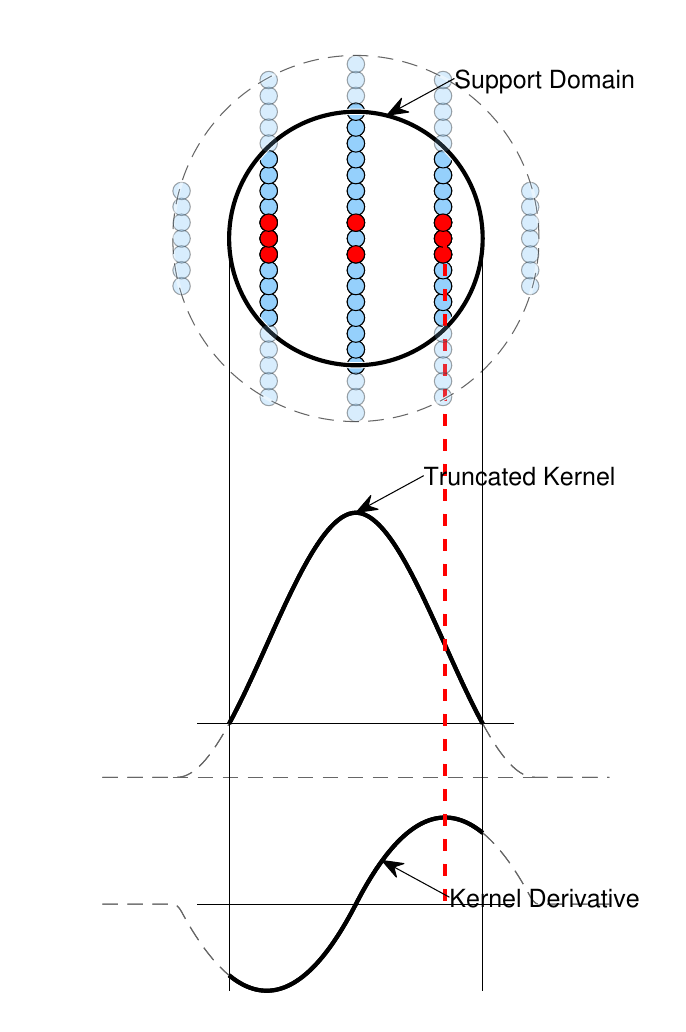}
\caption{$a = 2.74, b = 2.89$}
\label{figure 2b}
\end{subfigure}
\begin{subfigure}[b]{.32\textwidth}
\includegraphics[width=\textwidth]{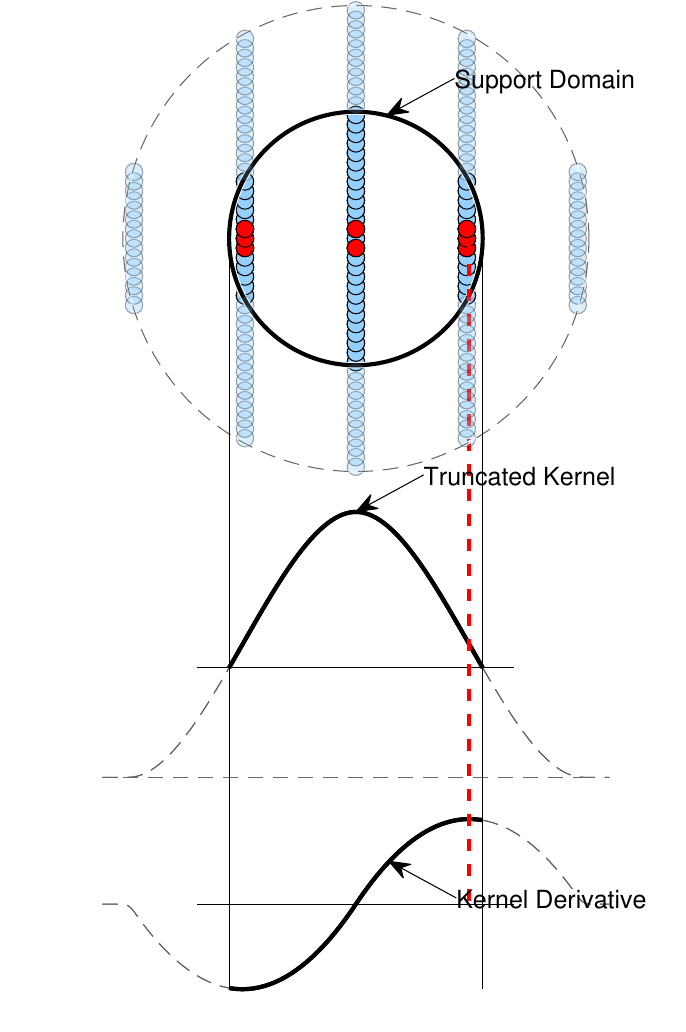}
\caption{$a = 3.49, b = 3.68$}
\label{figure 2c}
\end{subfigure}
\caption{Change in shape of the kernel and its $1$-st derivative with the shifting of the \emph{farthest immediate neighbour}. Both $a$ and $b$ are shifted, but the support domain is kept the same. The red points represent the nearest neighbours and the dotted red line represents the position of the extremum of $W^{'}$.}
\label{figure 2}
\end{figure}
\begin{figure}
\begin{subfigure}[b]{.5\textwidth}
\includegraphics[width=\textwidth]{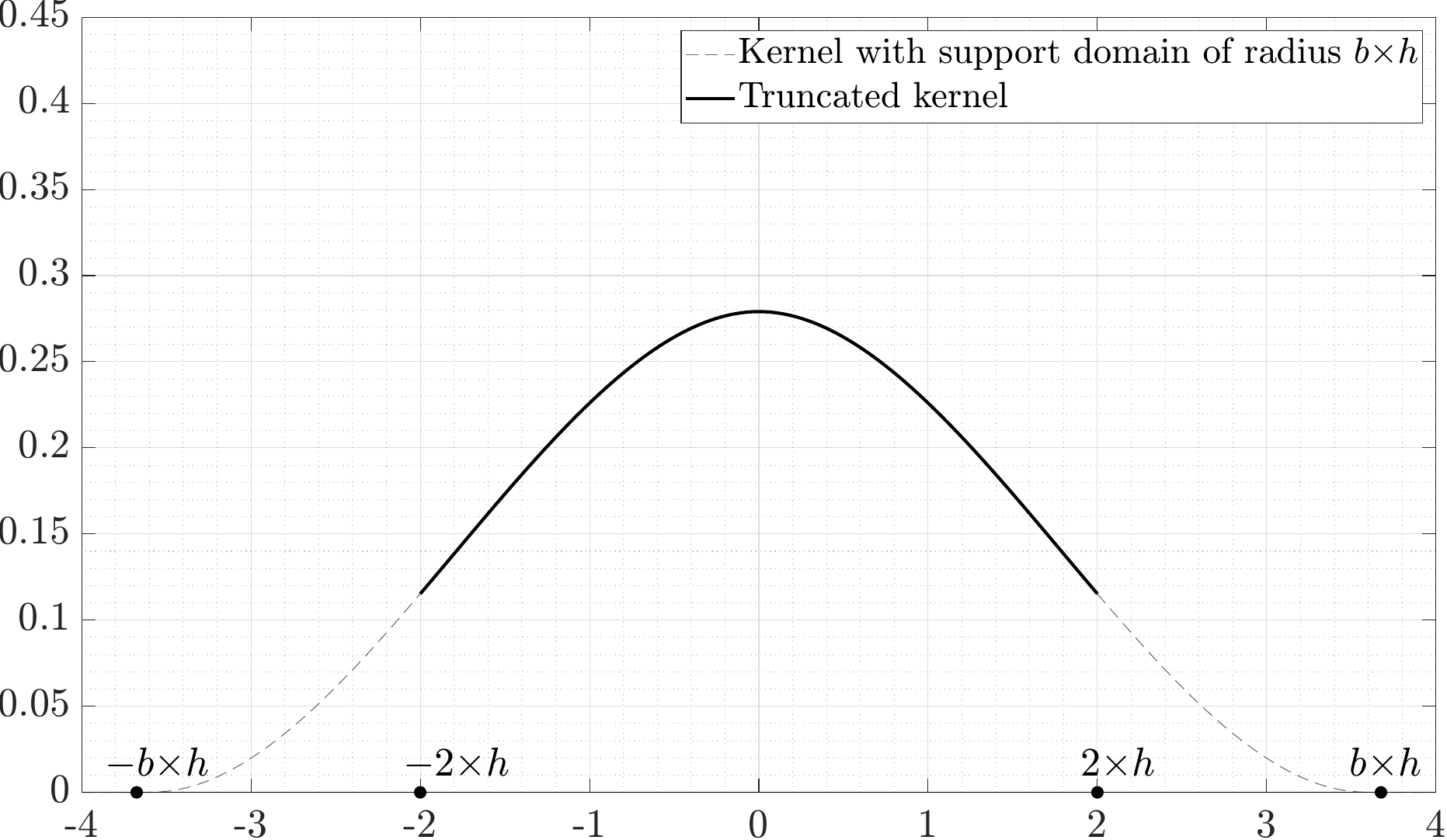}
\caption{}
\label{figure 3a}
\end{subfigure}
\begin{subfigure}[b]{.5\textwidth}
\includegraphics[width=\textwidth]{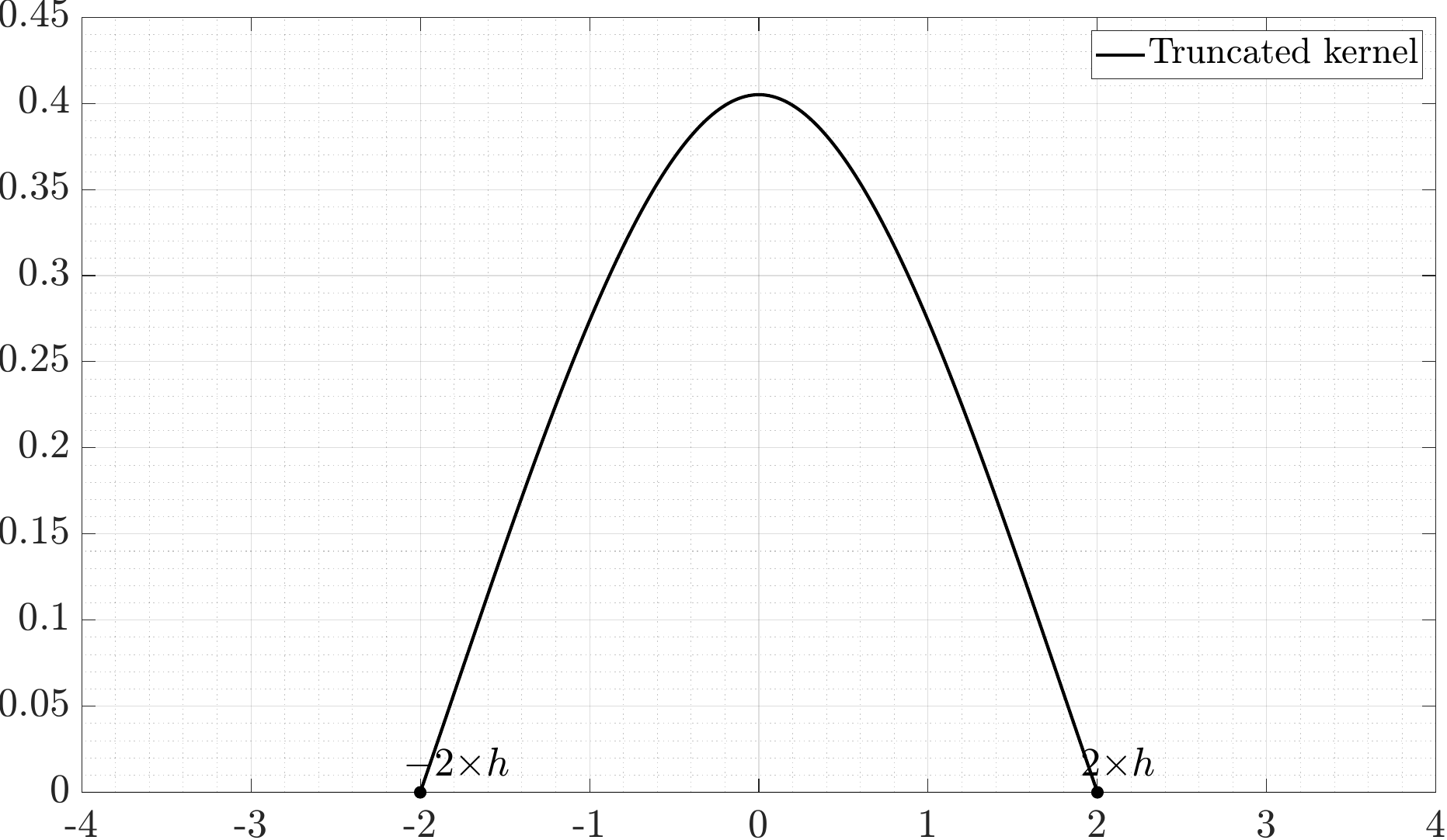}
\caption{}
\label{figure 3b}
\end{subfigure}
\begin{subfigure}[b]{.5\textwidth}
\includegraphics[width=\textwidth]{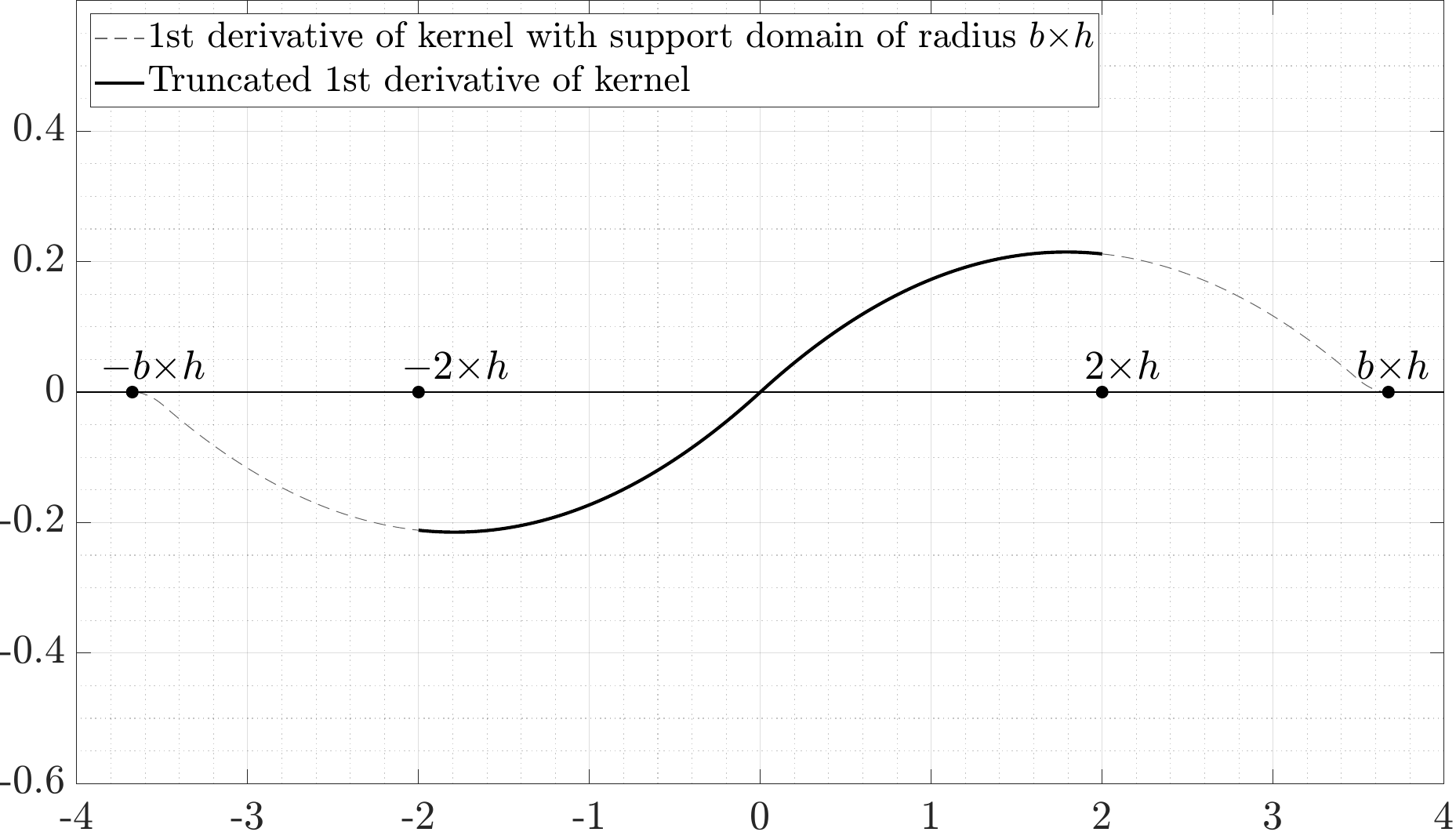}
\caption{}
\label{figure 3c}
\end{subfigure}
\begin{subfigure}[b]{.5\textwidth}
\includegraphics[width=\textwidth]{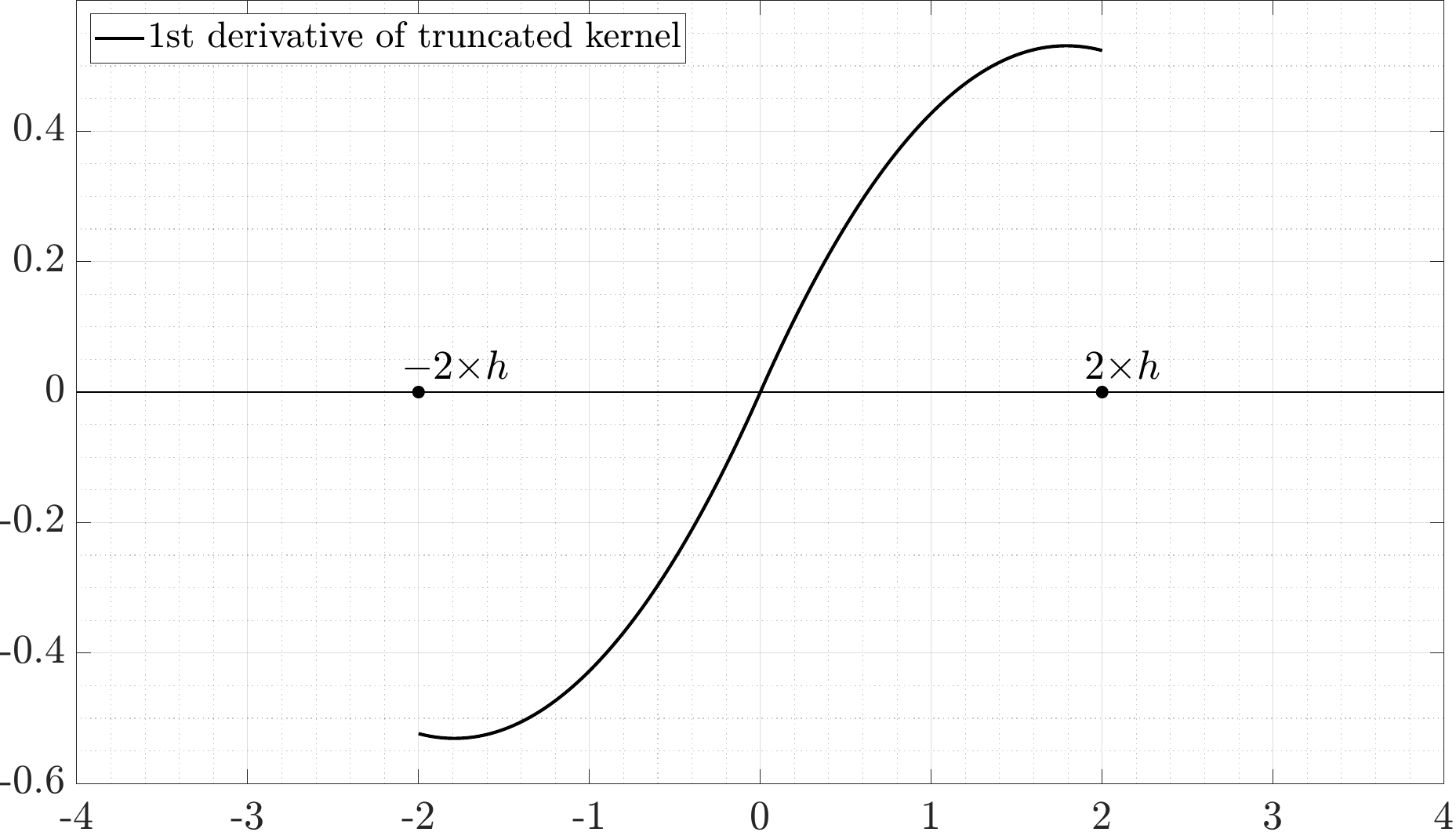}
\caption{}
\label{figure 3d}
\end{subfigure}
\caption{For $h=1, r_{i}=1.75$ and $A = 1.05$, which gives $a=3.58, b=3.77$ from Equation \eqref{adaptive_2}, (a) 1D kernel with support domain of radius $bh$ (dashed line) and truncated 1D kernel (solid line) ; (b) truncated 1D kernel normalised such that $\int_\Omega W(\textbf{x}-\textbf{x$'$},h)d\textbf{x$'$}=1$; (c) $1$-st derivative of 1D kernel with support domain of radius $bh$ (dashed line) and truncated $1$-st derivative (solid line); (d) $1$-st derivative of truncated 1D kernel of (b).}
\label{figure 3}
\end{figure}

Hence, if $a$ reaches a value close to $2$, the following algorithm is used: if $a>1.95$
\begin{equation}
\label{adaptive_2}
\begin{split}
&b = 2.05\times \frac{r^*}{h} ~~~[from ~Equation ~\ref{adaptive_1} ~with ~a = 0.95b], \\
&a = 0.95b.
\end{split}
\end{equation}
Increasing the value of $a$ and $b$ both allows the extremum of $W^{'}$ to shift along with $r_i$ (when $r_{i} > h$), as can be seen in Figure \ref{figure 2}. However, naively letting $a$ and $b$ increase with $r_i$ poses some problems. An increase in $b$ results in an increase of the support domain, thereby allowing more particles to interact with particle $i$. This not only results in an increased computational time but also leads to an artificial smoothening of results. But, a  more serious drawback is that the \textit{tensile instability} might not be eliminated. Swegle's condition says that in the case of tension, a positive value of the second derivative of the kernel contributes towards instability. Suppose the support domain is allowed to increase with increasing $b$. In that case, it can be understood from Figure \ref{figure 2c} that the particles in between the regions of radius $2h$ and $bh$ will have positive values of $W^{''}$. This will, in fact, result in \textit{tensile instability}. Therefore, in our approach, the kernel is truncated with the support domain having a constant radius of $2h$, as can be seen from Figure \ref{figure 2} and Figure \ref{figure 3}. Figure \ref{figure 3} shows the $1D$ kernel for a situation with $h=1$ and $r_i=1.75$. From Equation \eqref{adaptive_2} for $A=1.05$ we obtain $a=3.58$ and $b=3.77$. Figure \ref{figure 3a} shows the kernel with a support domain of radius $bh$ and also the truncated kernel whose support domain is of radius $2h$. The truncated kernel is again shown in Figure \ref{figure 3b} where it has been normalized such that $\int_\Omega W(\boldsymbol{x}-\boldsymbol{x}',h)d\boldsymbol{x}'=1$. Figure \ref{figure 3c} shows the $1$-st derivative of the kernel with support domain of radius $bh$ and also the truncated $1$-st derivative. Figure \ref{figure 3d} shows the $1$-st derivative of the normalized truncated kernel. Truncation of the kernel may cause inconsistency in the approximation. To ensure consistency, gradient correction is used where the first derivative of the kernel function is modified as $\frac{\partial W^C_{ij}}{\partial x^\alpha}=\mathbf{M}^{\alpha\beta}\frac{\partial W_{ij}}{\partial x^\beta}$, where $\mathbf{M}$ is a symmetric re-normalisation matrix obtained as $\mathbf{M}^{-1}_i = -\sum_{j\in N^i} \frac{m_j}{\rho_j} \boldsymbol{x}_{ij}  \otimes \mathbf{\bigtriangledown} W_{ij}$. 

\begin{figure}[htp]
\centering
\begin{tikzpicture}[font=\small,thick, auto, align=center]
 \tikzstyle{line} = [draw, -latex']
\node[draw,
    rounded rectangle,
    minimum width=2.5cm,
    minimum height=1cm] (block1) {  For any particle $i$  };
 
\node[draw,
    trapezium, 
    trapezium left angle = 65,
    trapezium right angle = 115,
    trapezium stretches,
    below=of block1,
    minimum width=3.5cm,
    minimum height=1cm
] (block2) { Estimate $r_i$ };
 
\node[draw,
    below=of block2,
    minimum width=3.5cm,
    minimum height=1cm
] (block3) { Calculate $a_i$ \\ $r^* = A\times r_i$ \\ $a_i = \frac{br^*}{bh-r^*}$ };
 
\node[draw,
    diamond,
    below=of block3,
    minimum width=2.5cm,
    inner sep=0] (block4) { $a_i>1.95$};
 
\node[draw,
    below left=of block4,
    minimum width=3.5cm,
    minimum height=1cm
] (block5) { $a_i = \frac{br^*}{bh-r^*}$ \\ $b_i = 2$ };
 
\node[draw,
    diamond,
    below right=1cm and 2cmof block4,
    minimum width=2.5cm,
    inner sep=0
] (block6) { Need to \\ extend $b_i$ \\ beyond 2};
 
\node[draw,
    below left=1cm and 0.01cm of block6,
    minimum width=2.5cm,
    minimum height=1cm] (block7) { $a_i = 1.95$ \\ $b_i = 2$ };
 
\node[draw,
    below right=1cm and 0.01cm of block6,
    minimum width=2.5cm,
    minimum height=1cm] (block8) { $b_i = 2.05\frac{r^*}{h}$ \\ $a_i = 0.95\times b_i$};
 
\node[draw,
    rounded rectangle,
    below=14cm of block1,
    minimum width=2.5cm,
    minimum height=1cm,] (block9) { Store Output };
 
\node[coordinate,below=4.85cm of block4] (block10) {};
\coordinate[left = 5cm of block9] (Empty1);
\coordinate[above = 15cm of Empty1] (Empty2); 
 
\draw[-latex] (block1) edge (block2)
    (block2) edge (block3)
    (block3) edge (block4);
 
\draw[-latex] (block4) -| (block5)
    node[pos=0.25,fill=white,inner sep=0,above]{No};
 
\draw[-latex] (block4) -| (block6)
    node[pos=0.25,fill=white,inner sep=0,above]{Yes};

\draw[-latex] (block6) -| (block7)
    node[pos=0.25,fill=white,inner sep=0,above]{No};
    
\draw[-latex] (block6) -| (block8)
    node[pos=0.25,fill=white,inner sep=0,above]{Yes};

\draw (block7) |- (block10);
\draw (block8) |- (block7|-block10);
\draw[-latex] (block10) -- (block9);
\draw (block5) |- (block10); 
\draw (block9) -| (Empty1);
\draw (Empty1)  edge  node[sloped, anchor=center, above] { Continue for particle $i+1$ } (Empty2);
\draw[-latex] (Empty2) -- (block1);
\end{tikzpicture}
\caption{Flow chart of the algorithm to estimate the knot values $a$ and $b$}
\label{figure 4}
\end{figure}
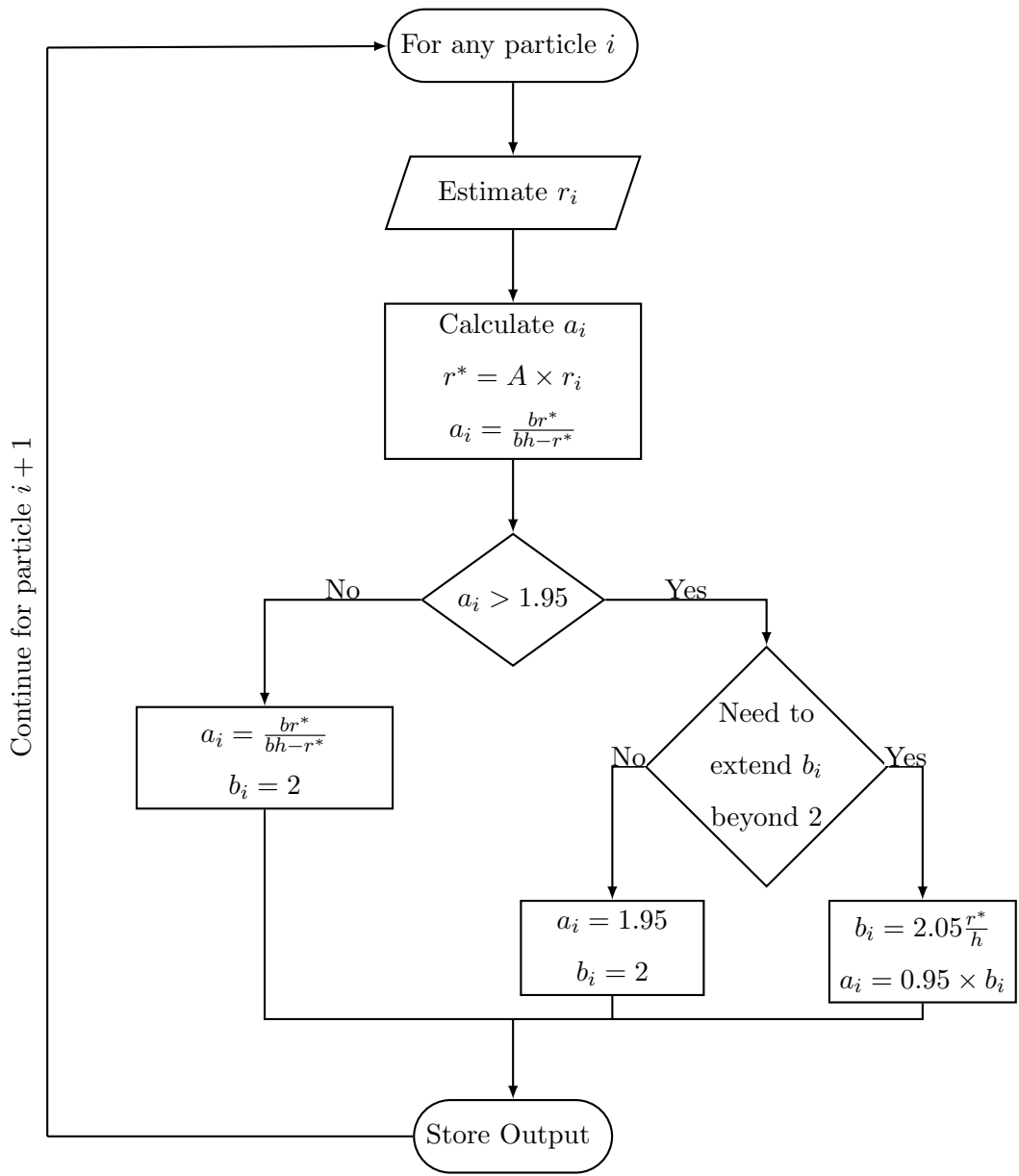

Figure \ref{figure 4} presents a flow chart of the algorithm to estimate the knot values of $a$ and $b$ as discussed in the previous paragraphs. It is shown in the flowchart that for particle $i$, if $a_i>1.95$, one might choose to extend $a_i$ and $b_i$ beyond 2 or one might assign $a_i=1.95$ and $b_i=2$. Of the two numerical simulations performed, in the Impacting drop problem in Section \ref{s6.2}, it was required to increase the values of $a$ and $b$ beyond 2 to prevent instability. But for the rotation of the fluid patch problem in Section \ref{s6.3}, the same was not required to prevent instability, as discussed in Section \ref{s6.3}.  

\subsubsection{Instability in compression and Hyperbolic kernel}
\label{s5.2.3}
\begin{figure}
\begin{subfigure}[b]{.5\textwidth}
\includegraphics[width=\textwidth]{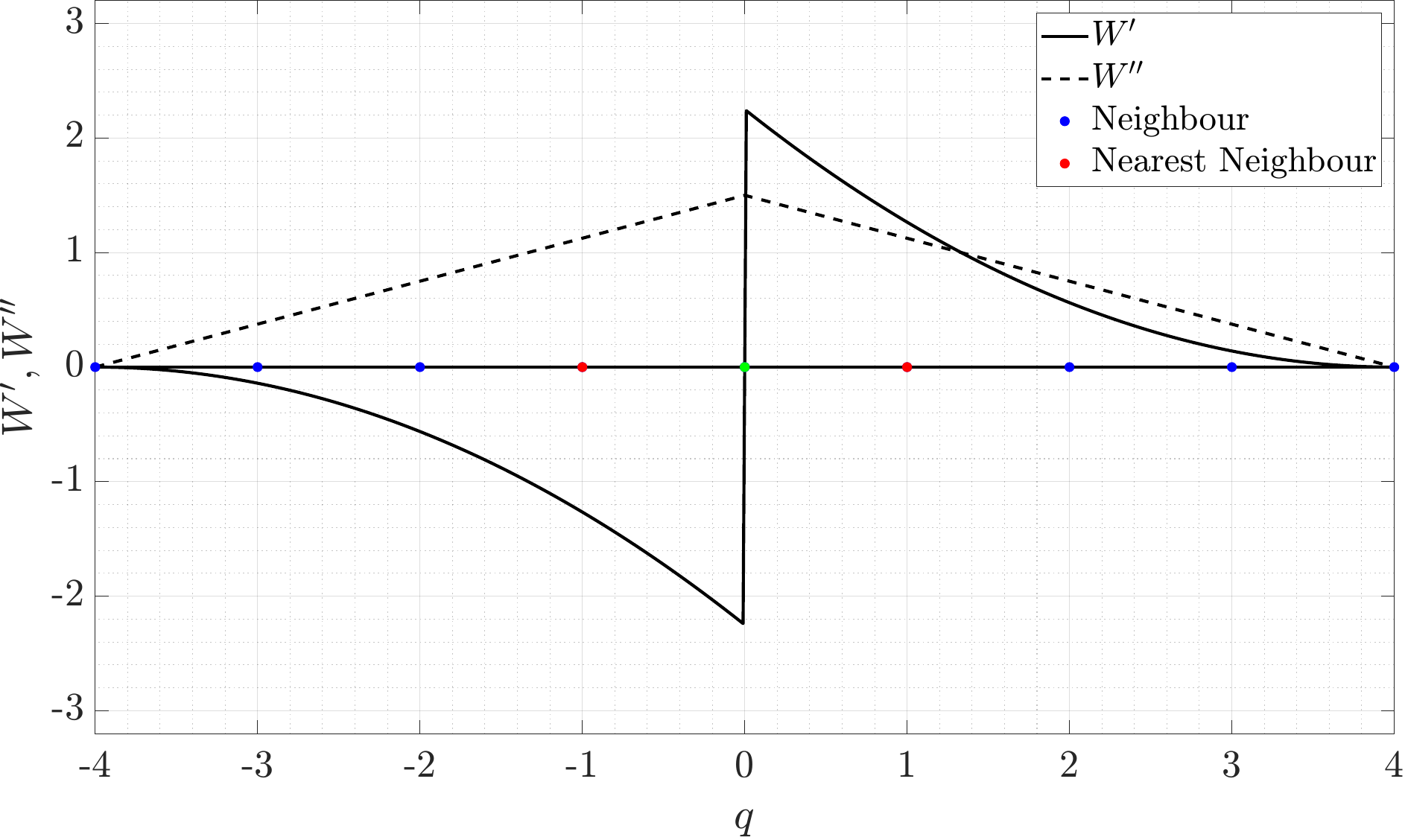}
\caption{a = 0, b = 2}
\label{figure 6c}
\end{subfigure}
\begin{subfigure}[b]{.5\textwidth}
\includegraphics[width=\textwidth]{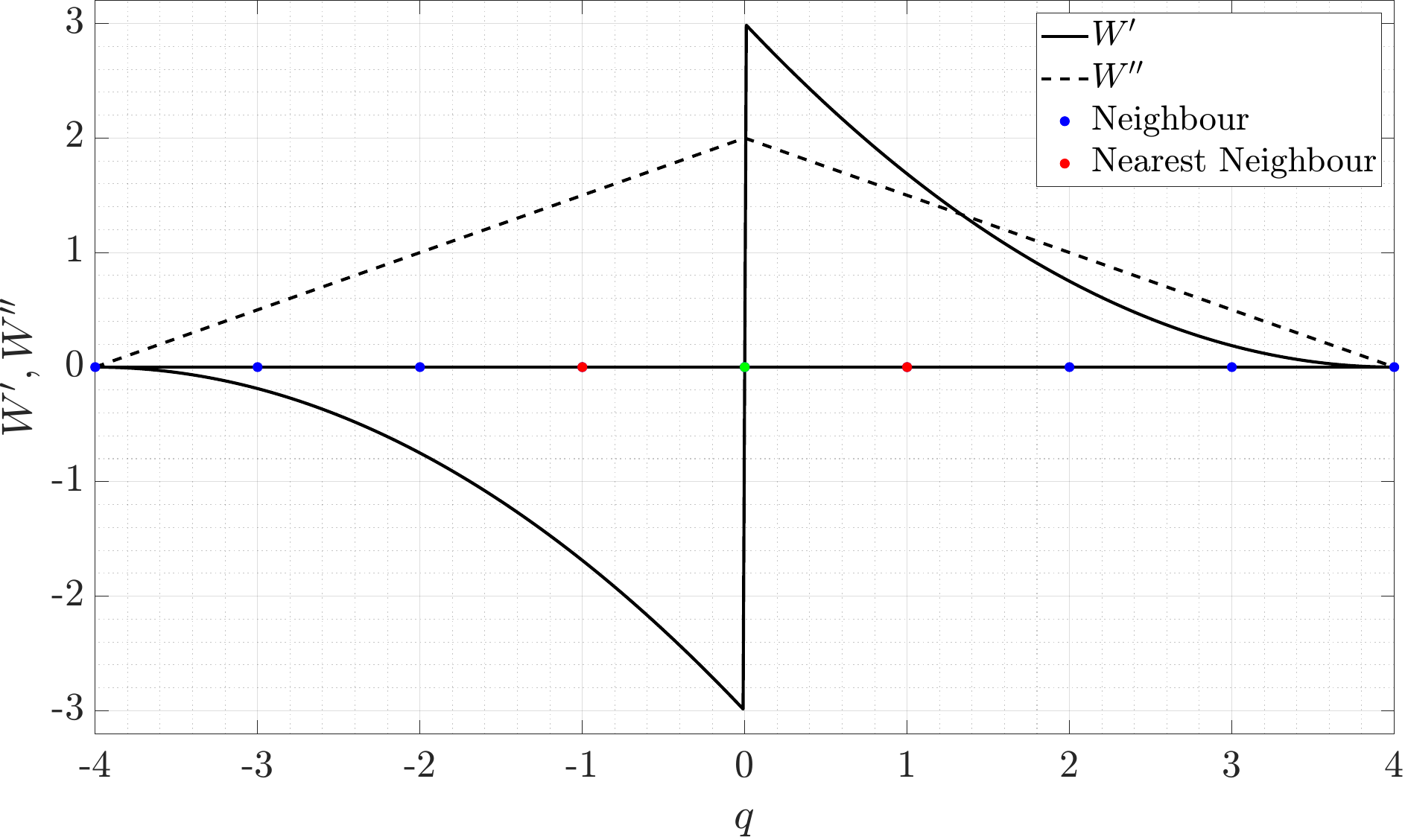}
\caption{a = -1, b = 2}
\label{figure 6d}
\end{subfigure}
\caption{Kernel with $a \le 0$ for compression dominated problems}
\label{figure 6}
\end{figure}

It is of relevance to mention the work by \citet{yang2014smoothed} wherein the authors used a hyperbolic kernel whose second derivative is non-negative over the full support and argued that the hyperbolic kernel can prevent the instability from occurring for positive pressure. This was demonstrated through an example of viscous liquid drops. Though the simulation performed in \cite{yang2014smoothed} is beyond the scope of the present work, it can be shown how in a situation of positive pressure, the kernel proposed in this work can be adapted to ensure that the second derivative is non-negative for all the particles within the support. 
In a 1D setting, as shown in Figure \ref{figure 6}, for values of $a \leq 0$, the second derivative is non-negative everywhere in the support domain.  

\subsection{Estimation of the farthest immediate neighbour}
\label{s5.3} 
One of the most vital steps in the algorithm is the estimation of $r_i$, i.e. the distance of the \emph{farthest immediate neighbour}. A possible method is to perform an inverse mapping of points in the candidate list and, in the mapped space, estimate the convex hull to get the nearest neighbours (\citet{randles2000normalized}). This exercise may make the adaptive algorithm computationally intensive, especially in problems involving a large number of particles. Development of a computationally efficient and easy to implement technique to determine $r_i$ requires further exploration. Nevertheless, to not deviate from the main focus of this work, which is to test the applicability of our approach in tackling \textit{tensile instability}, a simpler process to estimate $r_i$ based on strain is adopted here. This is found to work well, at least for the problems considered in the present study. The strategy is described below.

The measure of strain at a point should provide information about the position of points in its neighbourhood. Consider a 2D initially rectangular grid of particles with spacing $\Delta X^0$ and $\Delta Y^0$ in the $x$ and $y$ directions, respectively. At any subsequent time step $t$, the line segment $\Delta X^0$ changes its length to $\Delta X^t_i = \Delta X^0 + \sum_t\dot{\epsilon}_{i,t}^{xx} \Delta X^{t-1}_i dt$, and the relative displacement of one end of the line segment with respect to the other is $\sum_t\dot{\epsilon}_{i,t}^{xy} \Delta X^{t-1}_i dt$. Similarly, the length of the line segment $\Delta Y^0$ changes to $\Delta Y^t_i = \Delta Y^0 + \sum_t\dot{\epsilon}_{i,t}^{yy} \Delta Y^{t-1}_i dt$, and the relative displacement of one end of the line segment with respect to the other is $\sum_t\dot{\epsilon}_{i,t}^{yx} \Delta Y^{t-1}_i dt$, where, $\dot{\epsilon}_{i,t}^{xx}$, $\dot{\epsilon}_{i,t}^{yy}$ and $\dot{\epsilon}_{i,t}^{xy}$ are the components of strain rate at particle $i$ at the $t$-th time step. It is to be noted that because SPH is an updated lagrangian method, the strain rates at the $t$-th step are calculated with respect to the configuration at time $t-1$ as,  
\begin{equation}
\label{Estimate_ri_2}
\begin{split}
&\dot{\epsilon}_{i,t}^{xx} =\sum_j \frac{m_j}{\rho^{t-1}_j}(u^{t-1}_j-u^{t-1}_i)\frac{\partial W_{ij}}{\partial x_i}, \\
&\dot{\epsilon}_{i,t}^{yy} =\sum_j \frac{m_j}{\rho^{t-1}_j}(v^{t-1}_j-v^{t-1}_i)\frac{\partial W_{ij}}{\partial y_i}, \\
&\dot{\epsilon}_{i,t}^{xy} =\frac{1}{2} \sum_j \frac{m_j}{\rho^{t-1}_j} \left[ (v^{t-1}_j-v^{t-1}_i)\frac{\partial W_{ij}}{\partial x_i} + (u^{t-1}_j-u^{t-1}_i)\frac{\partial W_{ij}}{\partial y_i} \right].
\end{split}
\end{equation}
Now, the rectangle of sides $\Delta X^0$ and $\Delta Y^0$ becomes a rhombus whose diagonals ($S_1$ and $S_2$) may be determined as, 
\begin{equation}
\label{Estimate_ri_1}
\begin{split}
&S_1 = \sqrt{(\Delta X^t _i + \sum\dot{\epsilon}_{i,t}^{xy} \Delta Y^{t-1} dt)^2 + (\Delta Y^t_i + \sum\dot{\epsilon}_{i,t}^{yx} \Delta X^{t-1} dt)^2}\\
&S_2 = \sqrt{(\Delta X^t_i  - \sum\dot{\epsilon}_{i,t}^{xy} \Delta Y^{t-1} dt)^2 + (\Delta Y^t_i - \sum\dot{\epsilon}_{i,t}^{yx} \Delta X^{t-1} dt)^2}.
\end{split}
\end{equation}
The \emph{farthest immediate neighbour} of the $i$-th particle at time step $t$ is considered to be at a distance $\mathrm{max}(S_1, S_2)$ away.

For inviscid fluids, as in the case of the second example (Section \ref{s6.3}), the contribution of shear strain is ignored, and the distance of the \emph{farthest immediate neighbour} is determined as, 
\begin{equation}
\label{Estimate_ri_2}
\begin{split}
&S = \sqrt{(\Delta X^t _i)^2 + (\Delta Y^t_i)^2}.
\end{split}
\end{equation}

\begin{figure}
\begin{subfigure}[b]{.5\textwidth}
\includegraphics[trim={4cm 0 4cm 0}, width=\textwidth]{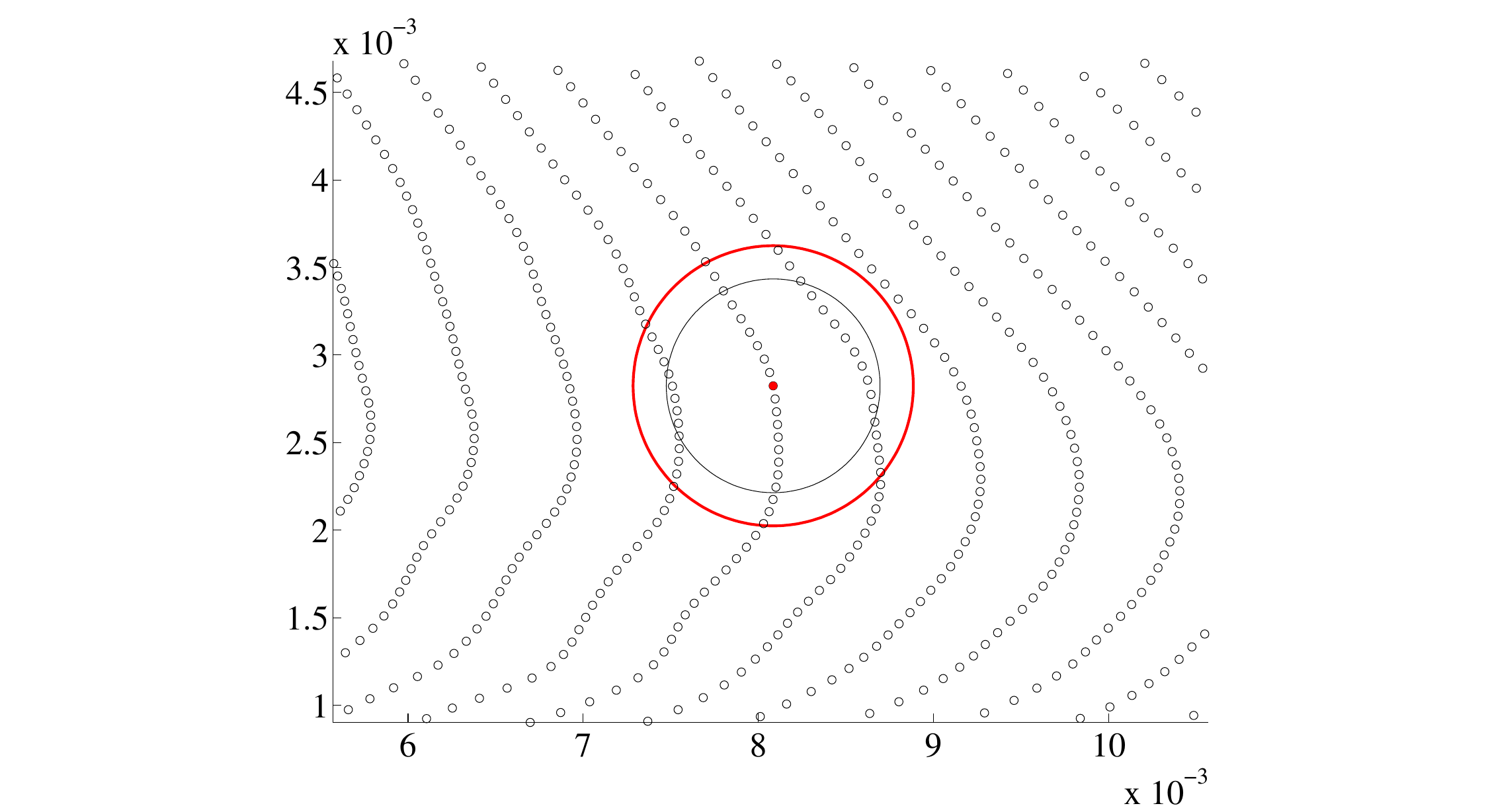}
\caption{}
\end{subfigure}
\begin{subfigure}[b]{.5\textwidth}
\includegraphics[trim={4cm 0 4cm 0}, width=\textwidth]{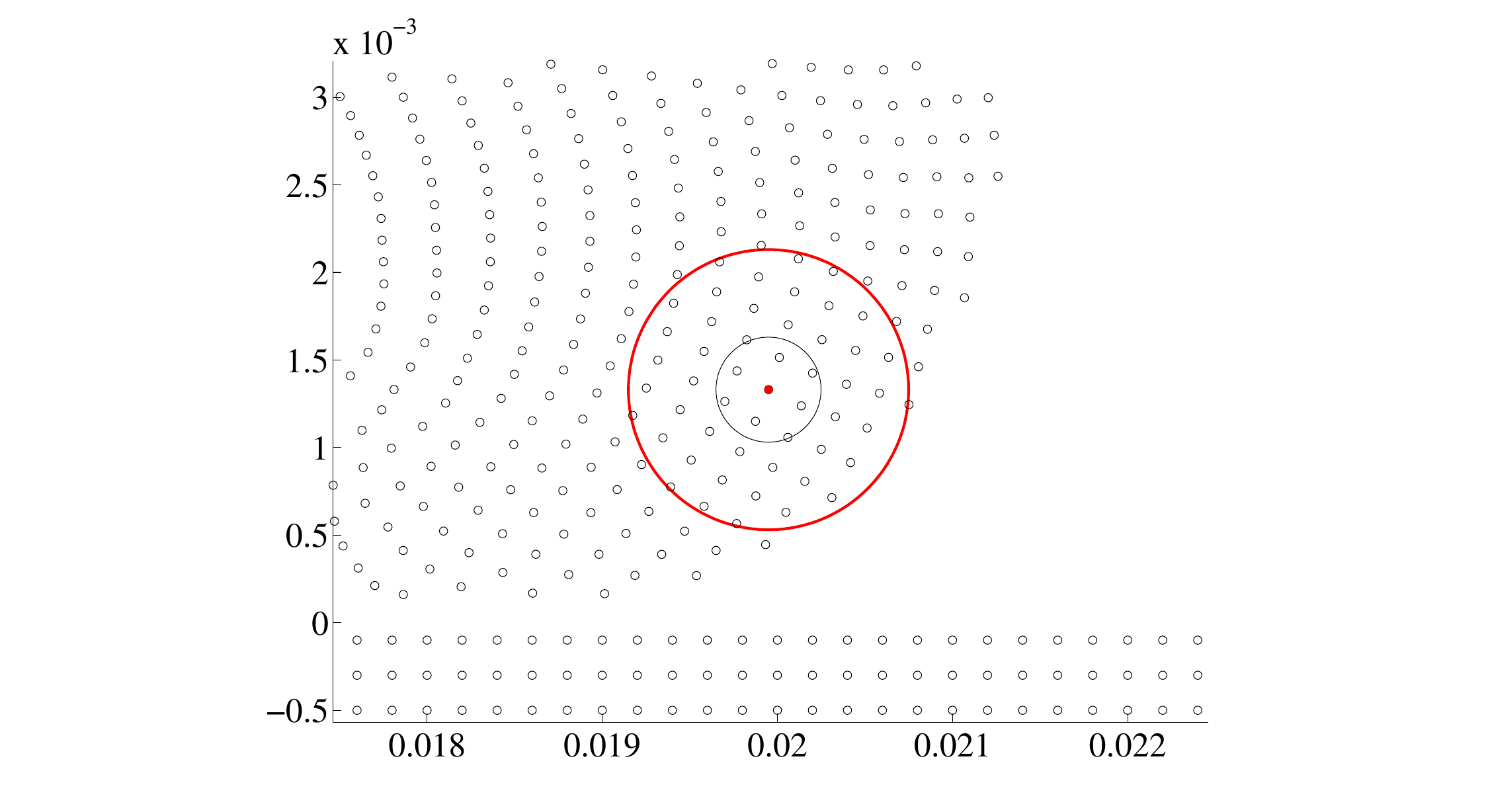}
\caption{}
\end{subfigure}
\caption{Plot showing the horizon size (red circle) and the value of $r^*$ (black circle) calculated from Equation \eqref{Estimate_ri_1} at different positions for the drop impact problem.}
\label{figure 7}
\end{figure}

To verify that the above approach calculates $r_i$ reasonably well, consider Figure \ref{figure 7}, which shows two views of the impacting drop problem from Section \ref{s6.2}. The red circle shows the SPH horizon, and the black circle has a radius $r^*$. $r_i$ is calculated using Equation \ref{Estimate_ri_1} and $r^*=Ar_i$, where $A$ is taken as $1.05$. Figure \ref{figure 7} shows how the $r_i$ obtained is able to track the \emph{farthest immediate neighbour} with reasonable accuracy.

\subsection{Plot of 1D Dispersion Relation}
\label{s5.4} 
\begin{figure}
\begin{subfigure}[b]{.33\textwidth}
\includegraphics[width=\textwidth]{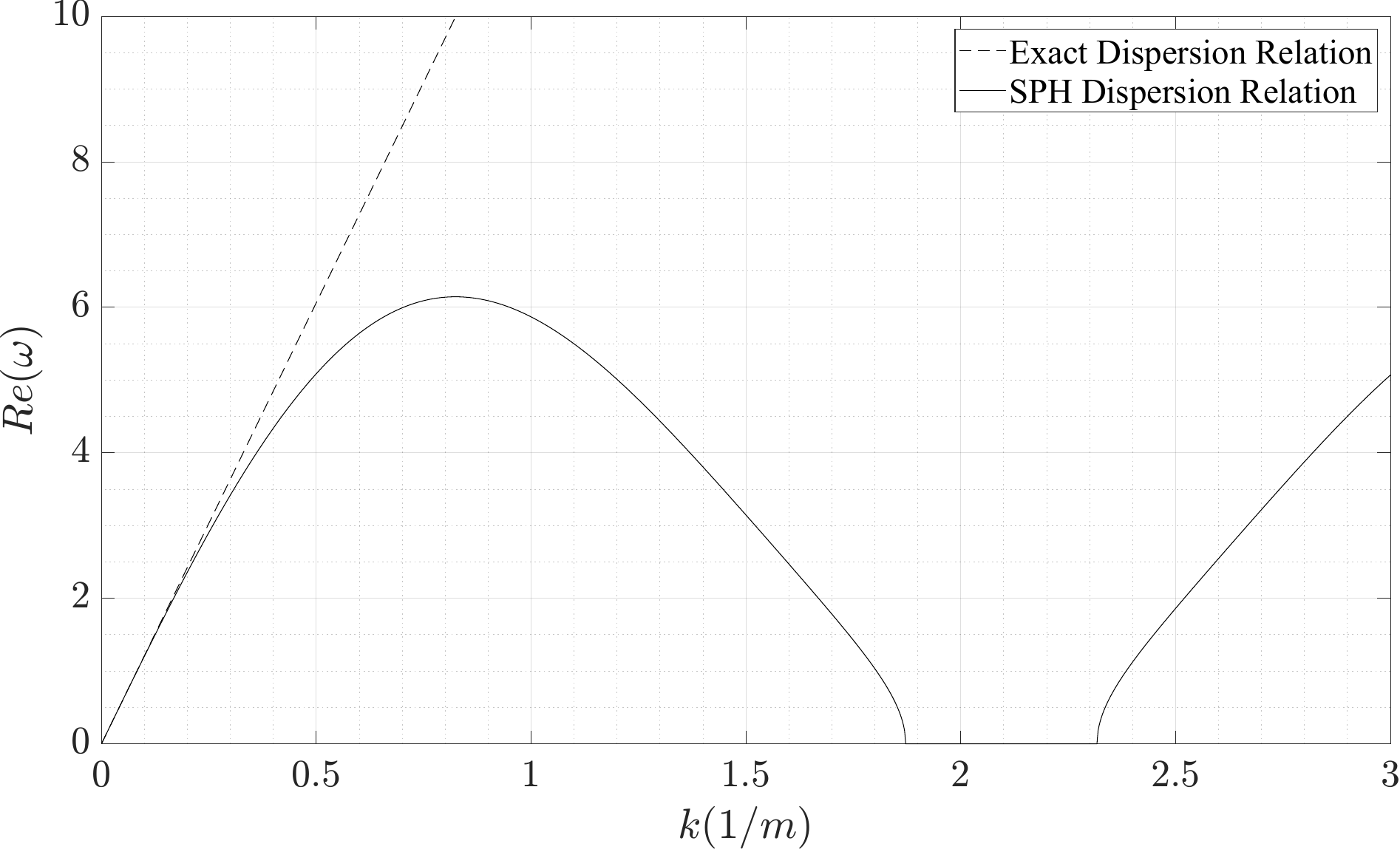}
\caption{$r=1.5,a=1,b=2$}
\label{figure8a}
\end{subfigure}
\begin{subfigure}[b]{.33\textwidth}
\includegraphics[width=\textwidth]{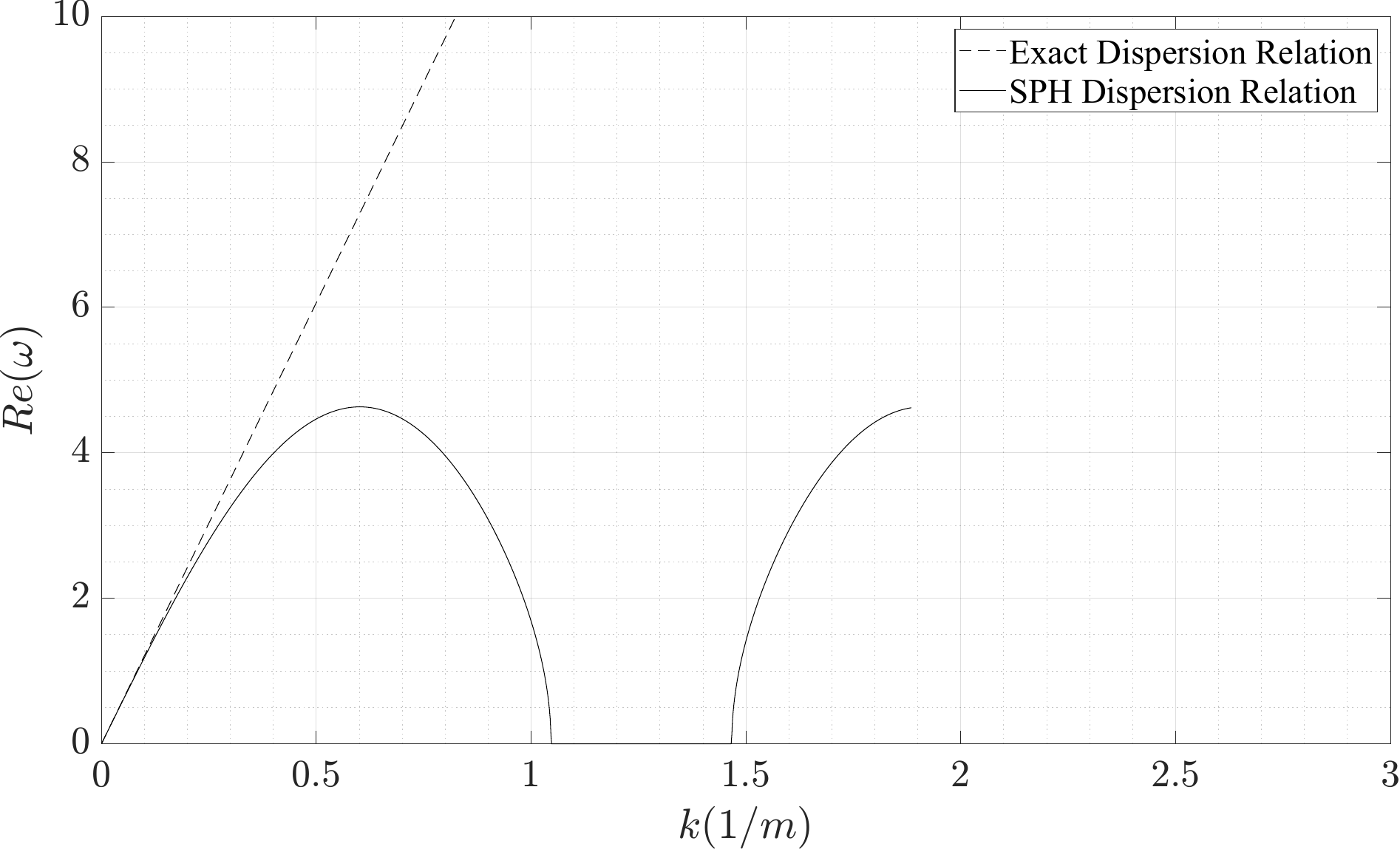}
\caption{$r=2.5,a=1,b=2$}
\label{figure8b}
\end{subfigure}
\begin{subfigure}[b]{.32\textwidth}
\includegraphics[width=\textwidth]{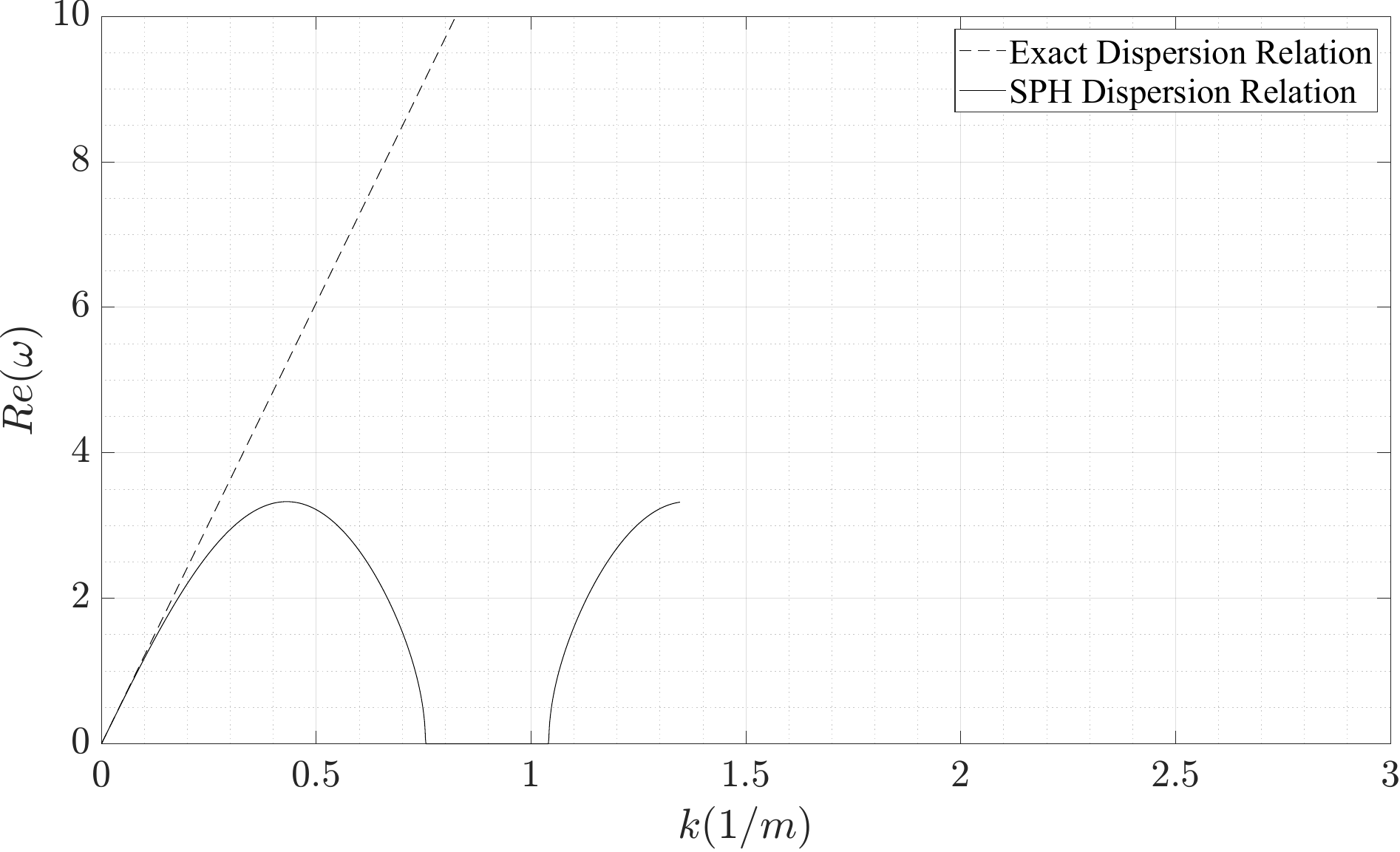}
\caption{$r=3.5,a=1,b=2$}
\label{figure8c}
\end{subfigure}
\begin{subfigure}[b]{.32\textwidth}
\includegraphics[width=\textwidth]{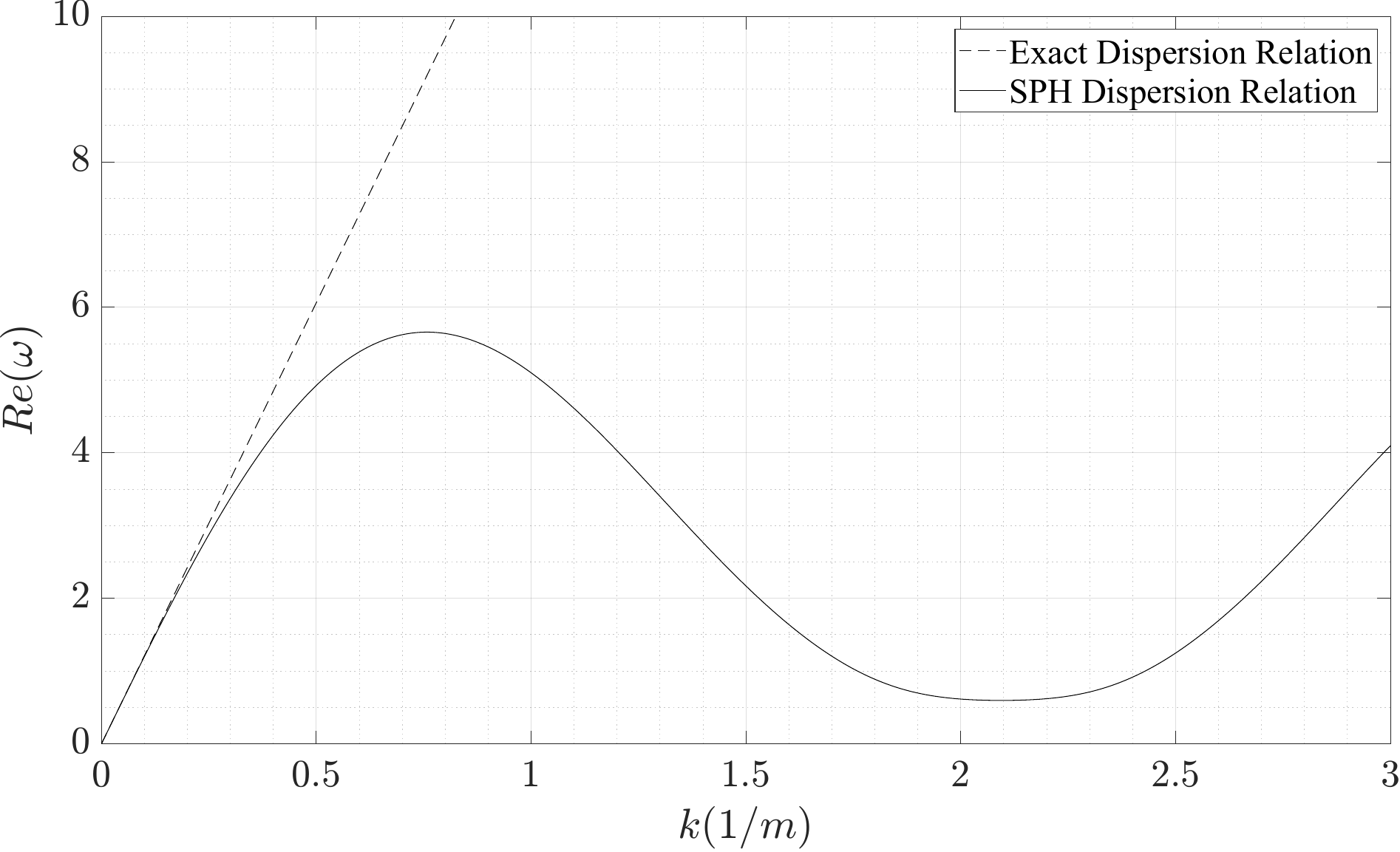}
\caption{$r=1.5,a=1.3,b=2$}
\label{figure8d}
\end{subfigure}
\begin{subfigure}[b]{.32\textwidth}
\includegraphics[width=\textwidth]{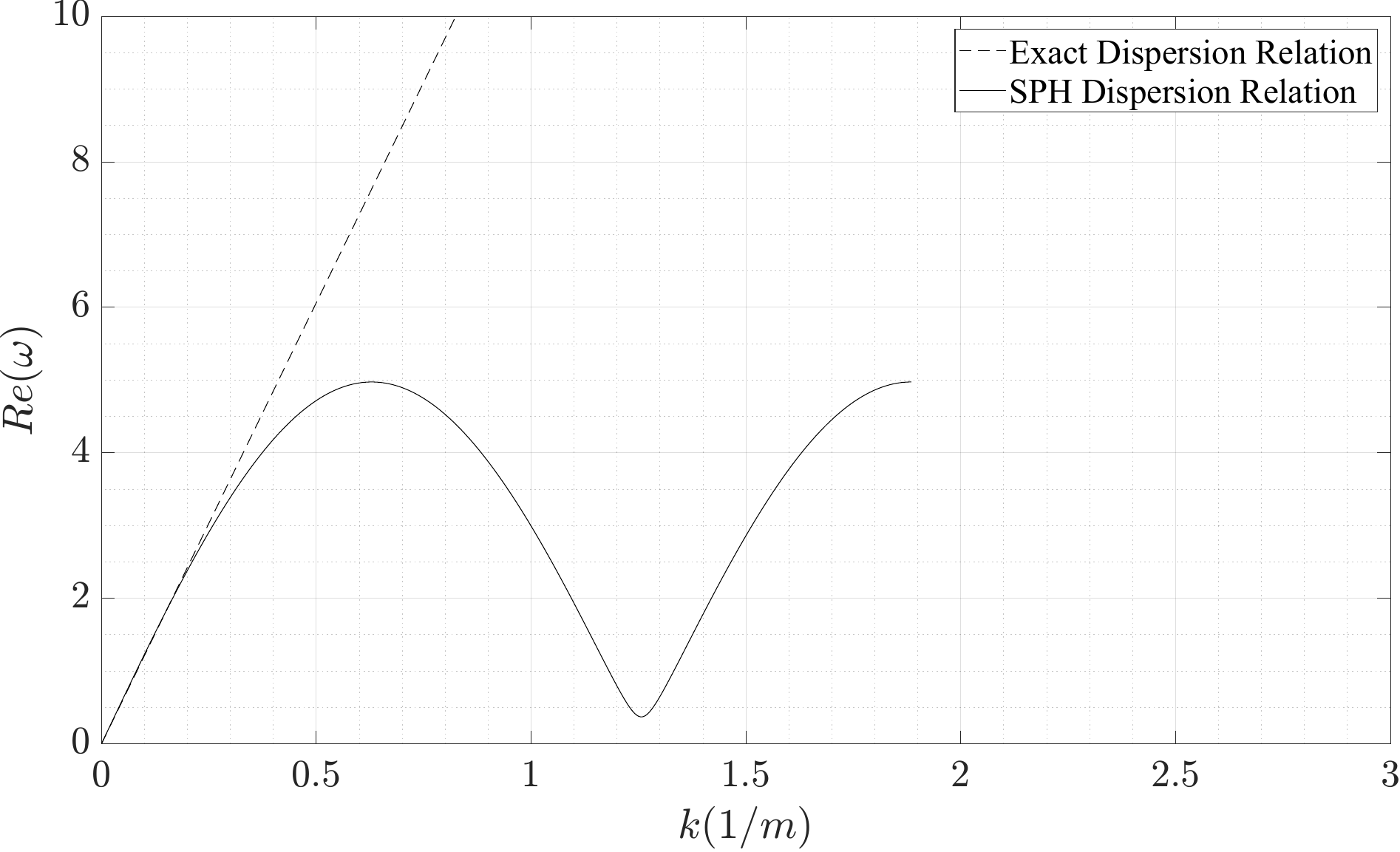}
\caption{$r=2.5,a=2.56,b=2.69$}
\label{figure8e}
\end{subfigure}
\begin{subfigure}[b]{.32\textwidth}
\includegraphics[width=\textwidth]{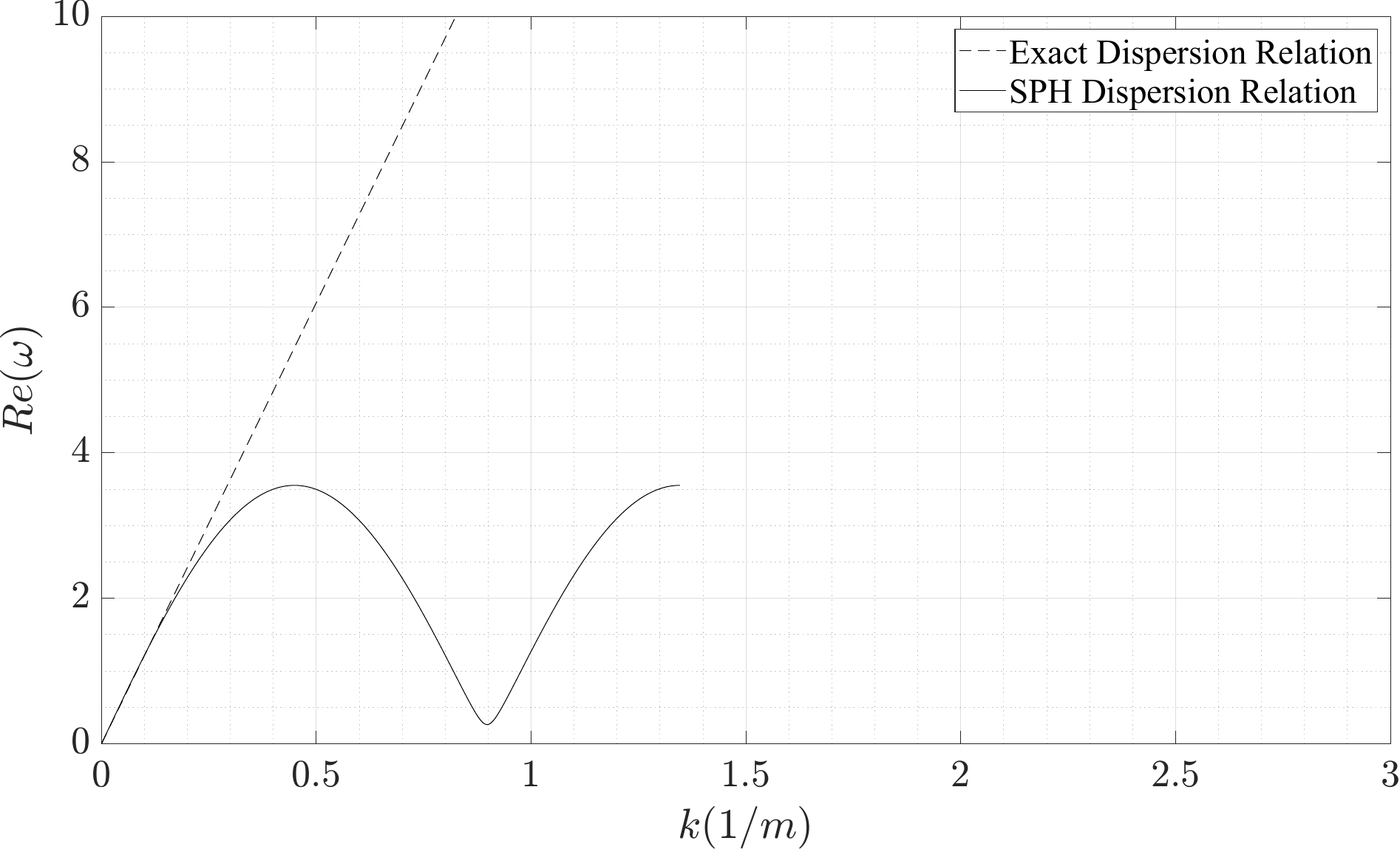}
\caption{$r=3.5,a=3.58,b=3.77$}
\label{figure8f}
\end{subfigure}
\caption{Plot of $Re(\omega)$ versus wave number $(k)$ from the 1D dispersion relation for $r=1.5,2.5,3.5$ for the standard SPH (a), (b) and (c); and for SPH with the adaptive kernel (d), (e) and (f). For large wavelengths, comparison is made with the exact dispersion relation (dashed line).}
\label{figure 8}
\end{figure}

In this section, by plotting the dispersion relations, we show how the adaptive algorithm helps in alleviating the \textit{tensile instability}. Towards this, a 1D bar is considered, and perturbations are provided to obtain the wave speeds for the exact Oldroyd B material (Equation \ref{exact_wavespd}) and for the SPH approximations (Equation \ref{SPH_wavespd}) in Section \ref{s4}. For the 1D SPH bar, the horizon size is kept constant at 2 units. Three different particle spacings are considered; 1.5, 2.5 and 3.5 units. $\frac{\overline{\rho}}{\rho_0}$ is taken as 0.99, as the stiff equation of state (Equation \ref{Pressure_EOS}) controls the density variation to $1\%$. $\overline{\uptau_p}$ is taken as $2000 ~ \mathrm{Pa}$ (the maximum $\overline{\uptau_p}$ obtained in the Impact drop simulation (Section \ref{s6.2}) is around $800 ~ \mathrm{Pa}$). In Figure \ref{figure8a}, \ref{figure8b} and \ref{figure8c}, are plotted the exact dispersion relation and the SPH dispersion relation with the standard cubic B-spline kernel for the three particle spacings. The wave numbers for which $Re(\omega)=0$ represent instability, i.e. zero energy modes. Now, if the adaptive approach proposed in Section \ref{s5.2} is used, then the instabilities are eliminated. Figure \ref{figure8d}, \ref{figure8e}, and \ref{figure8f} show the SPH dispersion relation using the adaptive algorithm, with $A=1.05$. For spacing of 1.5 units, the value of $a$ is $1.3$, and of $b$ is $2$. For spacing of $2.5$ and $3.5$, the values of $a$ and $b$ need to be greater than $2$ to prevent instability. For spacing of $2.5$, $a = 2.56$, $b = 2.69$; and for a spacing of $3.5$, $a = 3.58$ and $b = 3.77$. From Figure \ref{figure 8}, it can be seen that for long wavelength modes, i.e. $k \rightarrow 0$, there is a good accuracy between the exact and the SPH dispersion relations. 

\section{Numerical Simulations}\label{s6} 
In the literature, the \textit{tensile instability} of SPH in the simulation of weakly compressible fluids is discussed through several examples. Among them, two benchmark problems, viz., a liquid drop impacting a rigid surface and the rotation of a fluid patch, are taken in this section to demonstrate the proposed algorithm's efficacy in alleviating \textit{tensile instability}. For the first example, i.e., the impacting liquid drop problem, a no-slip boundary condition needs to be enforced between the liquid drop and the rigid surface. The approach adopted for the no-slip boundary condition is discussed in the following sub-section.  

After obtaining the values of $a$ and $b$ for all the particles using the algorithm described in Section \ref{s5.2}, two different approaches are adopted in this study to solve Equations \eqref{SPH_discret_1}-\eqref{SPH_discret_4}. 
The shape of the kernel centred at a given particle, say $i$-th, depends on the value of $a_i$ and $b_i$, which in turn depend on the value of $r^*$. Since particles in a domain will have different values of $r^*$, the knot positions and, consequently, the shape of the kernel will not be the same for all the particles. If differently shaped kernels at particles $i$ and $j$ are used for estimating the interaction between the particle pair, energy conservation will be violated. So, in one approach, the interaction between any particle pair $i-j$ is performed through a kernel function constructed with average values of $a_{ij}=(a_i+a_j)/2$ and $b_{ij}=(b_i+b_j)/2$, which ensures that the conservativeness of the method is restored. However, consider another particle $k$ in the vicinity of $i$. $a_{ik}$ and $b_{ik}$ will, in general, be different from $a_{ij}$ and $b_{ij}$, which makes the kernel centred at $i$, lose its symmetry. 
Hence in the other approach, to ensure that the kernel at any particle $i$ is symmetric, Equations \eqref{SPH_discret_1}-\eqref{SPH_discret_4} at particle $i$ are solved by constructing the kernel from $a_i$ and $b_i$, instead of considering averaged knot values. Though there will be a non-conservation of energy, it is quite small, as shown in the numerical simulations. The results from the numerical simulations show a minor difference between the two approaches.

\subsection{No-slip boundary Conditions}
\label{s6.1} 
\begin{figure}[htp]
\centering
\includegraphics[scale=0.4]{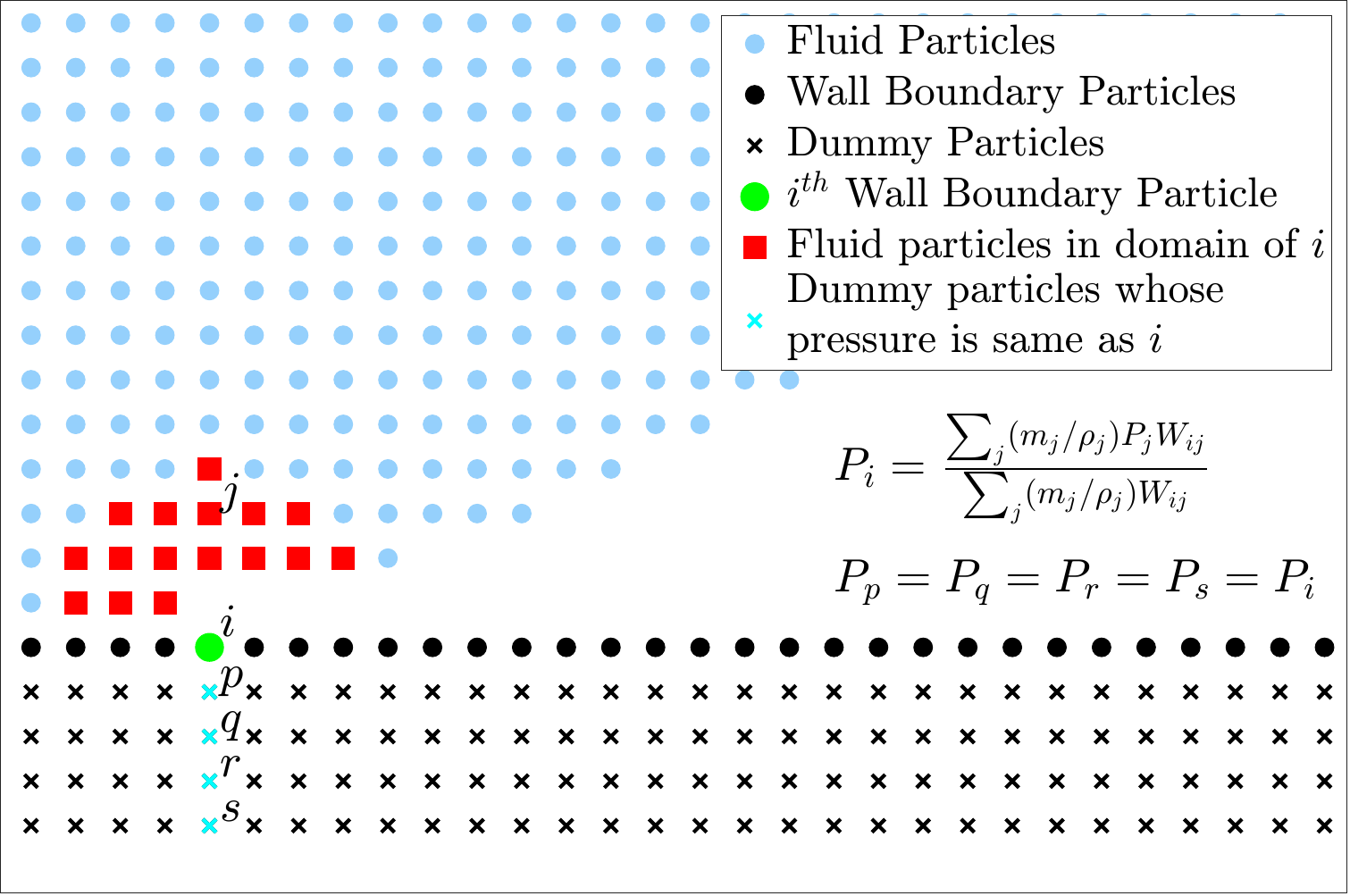}
\caption{Implementing a no - slip boundary condition}
\label{figure 10}
\end{figure}

Implementing boundary conditions in SPH sometimes poses problems because of particle deficiency at or near the boundaries. In the $1$-st numerical simulation in Section \ref{s6.2}, the boundary to be simulated is a solid wall. Following the approach of \citet{xu2013sph}, two types of boundary particles are considered, wall boundary particles and dummy particles. The wall boundary particles are placed along the solid wall, as shown in Figure \ref{figure 10}. The dummy particles are arranged in a grid just outside the solid wall, and they fill a domain with a depth of at least $2h$. Because this is a fixed wall, the wall boundary particles are fixed in position, and the pressure of each of these particles ($i$) is calculated as
$P_i=\frac{\sum_j(m_j/\rho_j)P_jW_{ij}}{\sum_j(m_j/\rho_j)W_{ij}}$, where $j$ is the index of liquid particles in the support domain of $i$. The dummy particles are also stationary, and the pressure of each particle is set as the pressure of the closest wall particle. Instead of providing repulsive forces to the fluid particles (\citet{monaghan1994simulating},                  \citet{monaghan2009sph}), the wall boundary and the dummy particles interact with the fluid particles by contributing to the expressions of continuity (Equation \eqref{SPH_discret_1}) and the conservation of linear momentum (Equation \eqref{SPH_discret_2}). A combination of these wall boundary and dummy particles solves the problem of particle deficiency of the fluid particles near the boundaries and also enforces a no-slip boundary condition. 

\subsection{Example 1: Impacting Liquid Drop}
\label{s6.2} 
As the first example, the numerical simulation of a visco-elastic drop impacting a rigid surface is investigated. In 2D, a disc of initial radius $R = 1$ cm is dropped onto a rigid surface from a height of 4 cm. The drop is given an initial downward speed of $V = 1$ m/s. The acceleration due to gravity is taken as -9.81 $\mathrm{m/s^2}$. For the discretisation of the drop, a rectangular grid of particles with spacing 0.02 cm in both the $x$ and $y$ directions is considered, and then a circle of radius 1 cm is drawn. The particles within and on the periphery of the circle are retained, and the rest are deleted. This gives a total of 7957 particles. The particles within the circle have volume $4\times 10^{-8}$ $\mathrm{m^3}$, while for the particles on the periphery of the circle, volumes are distributed such that the total volume of the drop is $\pi\times 0.01^2$ $\mathrm{m^3}$. The impacting surface is modelled as a solid wall, and no-slip boundary conditions are implemented, as explained in Section \ref{s6.1}. The speed of sound $c_0$ in the equation of state (Equation \ref{Pressure_EOS}) is taken as 12.5  $\mathrm{m/s}$ \cite{fang2006numerical}. The smoothing length $h$ is taken as $2\Delta p$, where $\Delta p$ is the initial interparticle spacing (0.02 cm in this case). A time step of $4\times 10^{-6}$ $\mathrm{s}$ is considered for numerical stability. It will be shown subsequently that the artificial viscosity alone cannot prevent the \textit{tensile instability} from occurring. Nevertheless, as mentioned in \citet{fang2006numerical} and \citet{rafiee2007incompressible}, without the artificial viscosity, the solution may blow up and diverge. Thus some amount of artificial viscosity with parameters $\gamma_1=0.5$ and $\gamma_2=0.5$ is used in all the simulations. Results are plotted against a non-dimensionless time $T=tV/2R$. In the following simulations, averaged knot values $a_{ij}$ and $b_{ij}$ are considered when estimating the interaction between a particle pair $i-j$.

\begin{figure}[htp]
\centering
\includegraphics[scale=0.4]{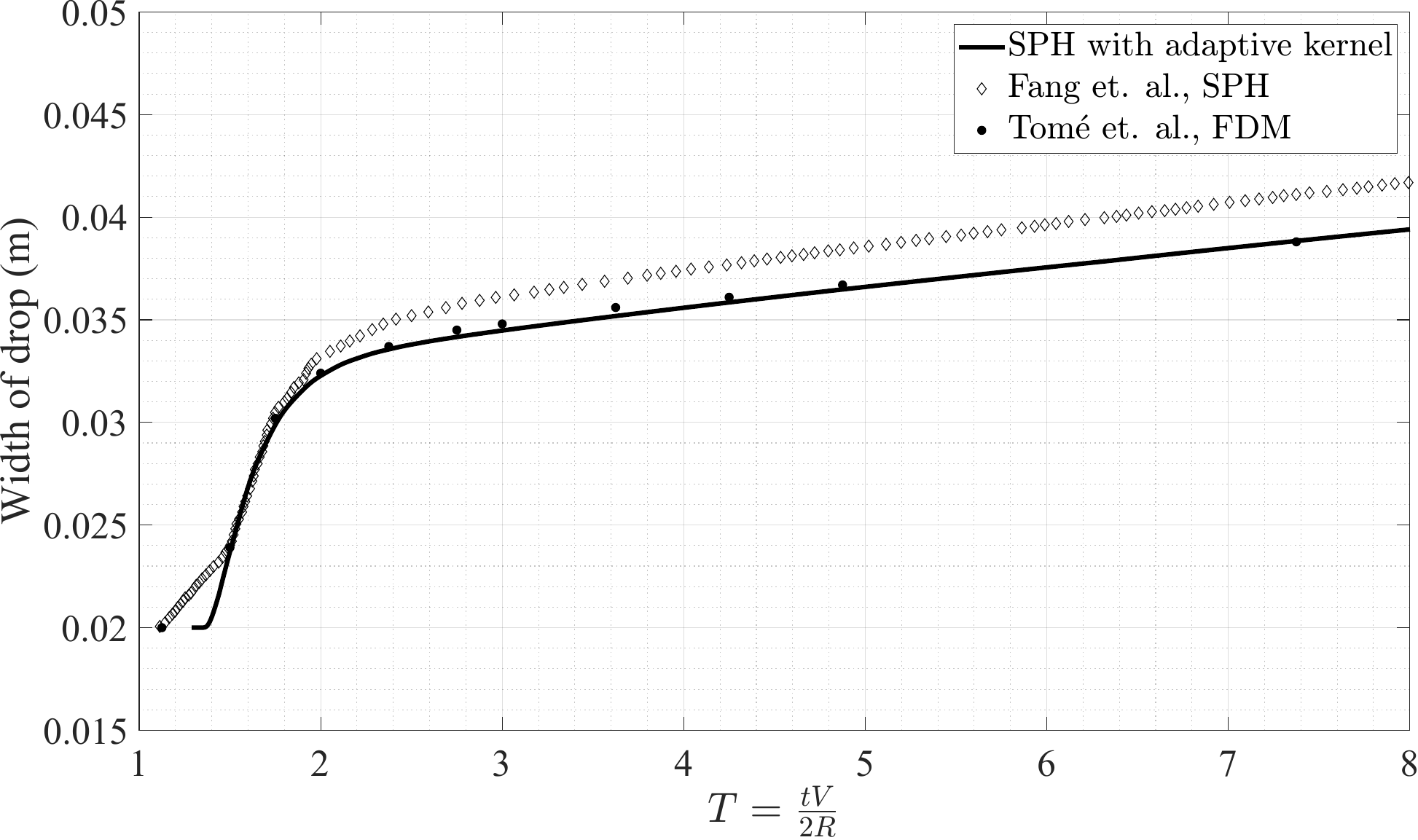}
\caption{Time history of width of drop for a Newtonian fluid using SPH with the adaptive kernel approach. Comparison is made with SPH simulations from \cite{fang2006numerical} and FDM simulations from \cite{tome2002finite}.}
\label{figure 11}
\end{figure}

First, the simulation is performed for a Newtonian fluid with $\theta = 0$ and $\eta_s = 4.0$ $\mathrm{Pa \, s}$. As reported in \cite{fang2006numerical}, our simulations also show that the Newtonian drop does not exhibit any \textit{tensile instability}. Hence, even though the adaptive kernel is used in the simulation, it is not required that $a$ be increased beyond $1.95$ once it reaches that value. The plot of the time history of the width of the fluid drop is presented in Figure \ref{figure 11}, where the result is compared with the SPH simulations of \citet{fang2006numerical} and the FDM simulations of \citet{tome2002finite}. It can be seen that the present SPH results match the results from the literature, and particularly, there is a decent agreement with the FDM result. It can be understood that after impact, the Newtonian drop spreads uniformly, and it continues to spread with time. Because the \textit{tensile instability} is not visibly evident even in standard SPH, the profiles of the drop at different times are not shown.

\begin{figure}
\begin{subfigure}[b]{.5\textwidth}
\includegraphics[width=\textwidth]{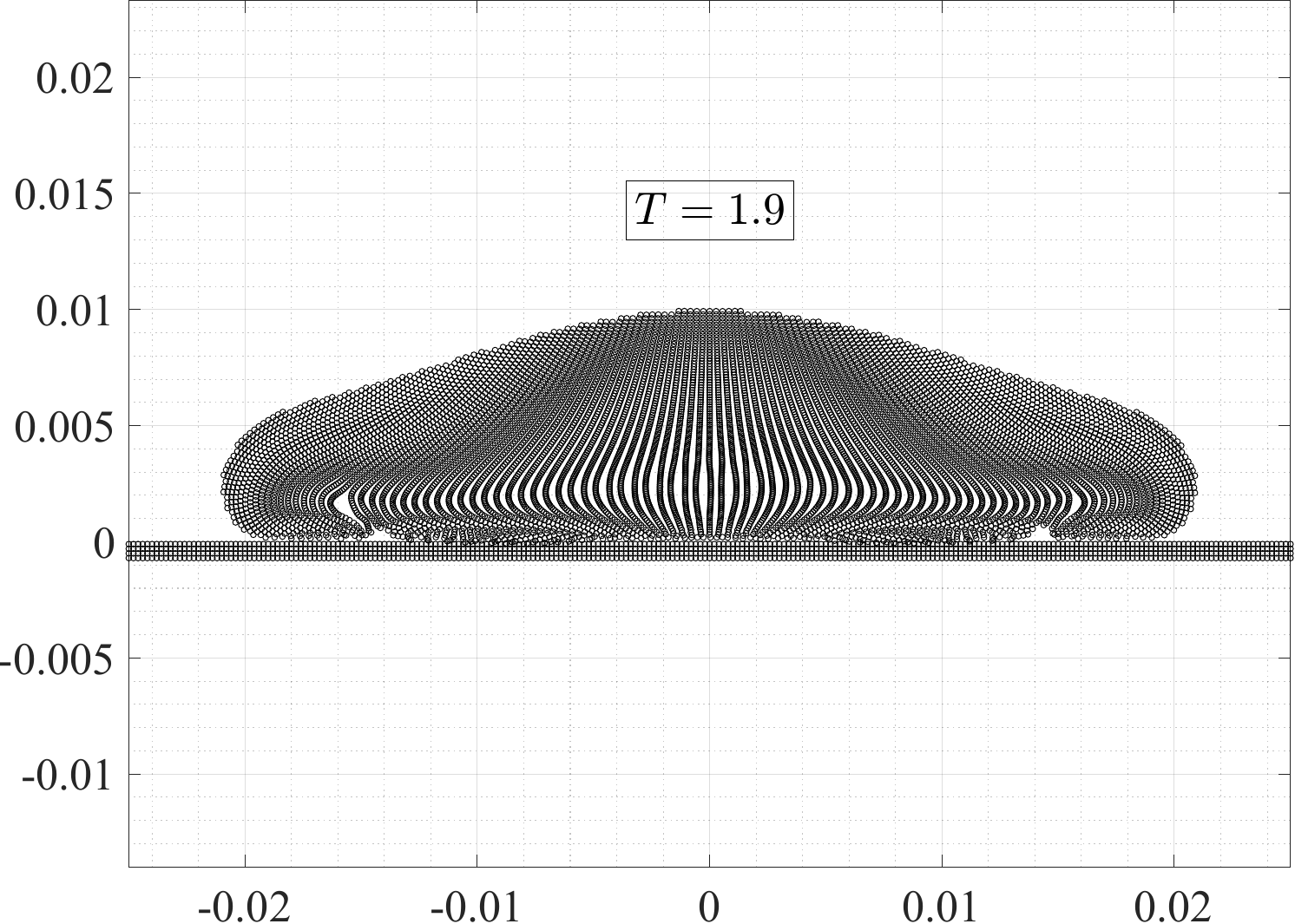}
\caption{}
\label{figure 12a}
\end{subfigure}
\begin{subfigure}[b]{.5\textwidth}
\includegraphics[width=\textwidth]{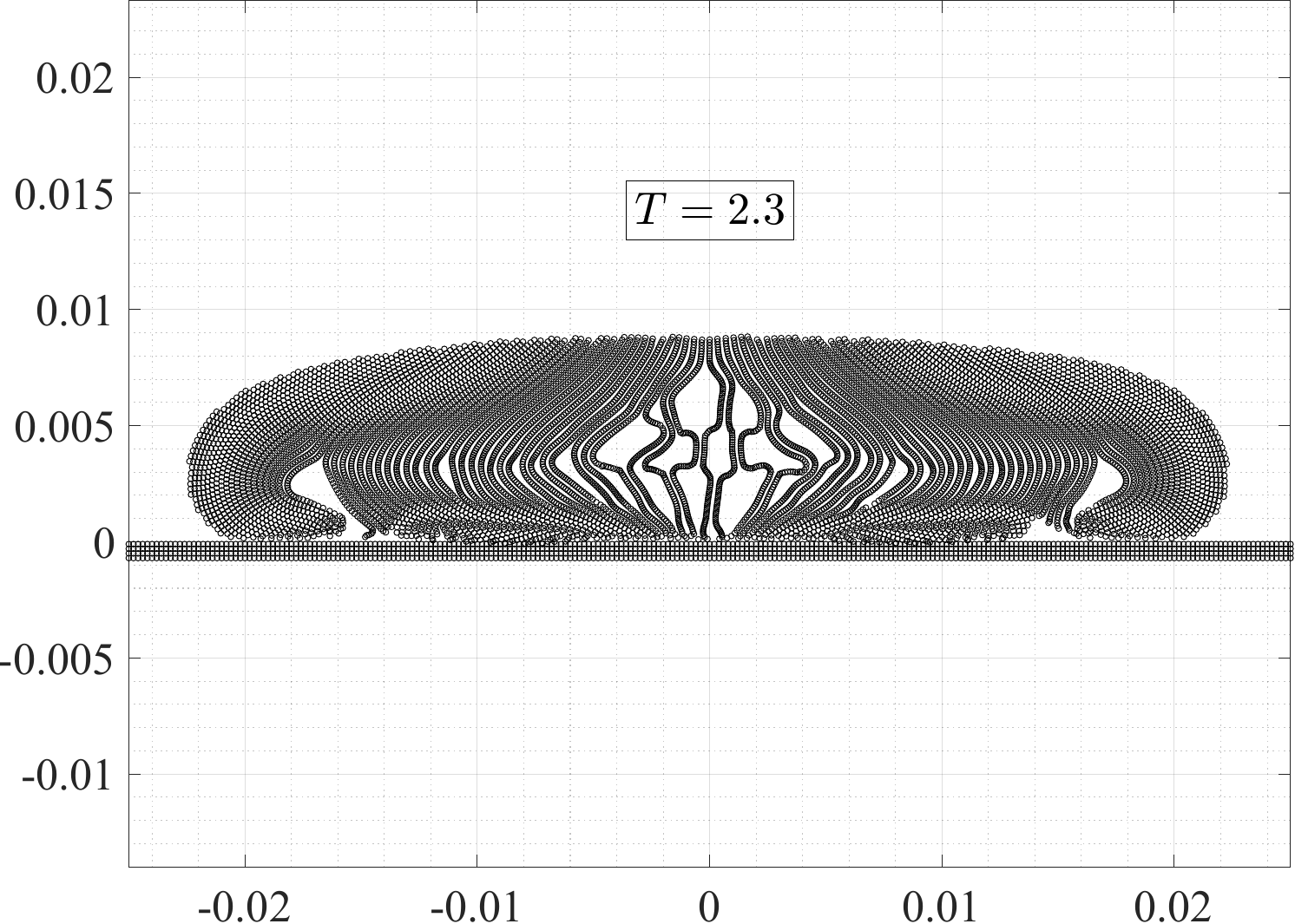}
\caption{}
\label{figure 12b}
\end{subfigure}
\begin{subfigure}[b]{.5\textwidth}
\includegraphics[width=\textwidth]{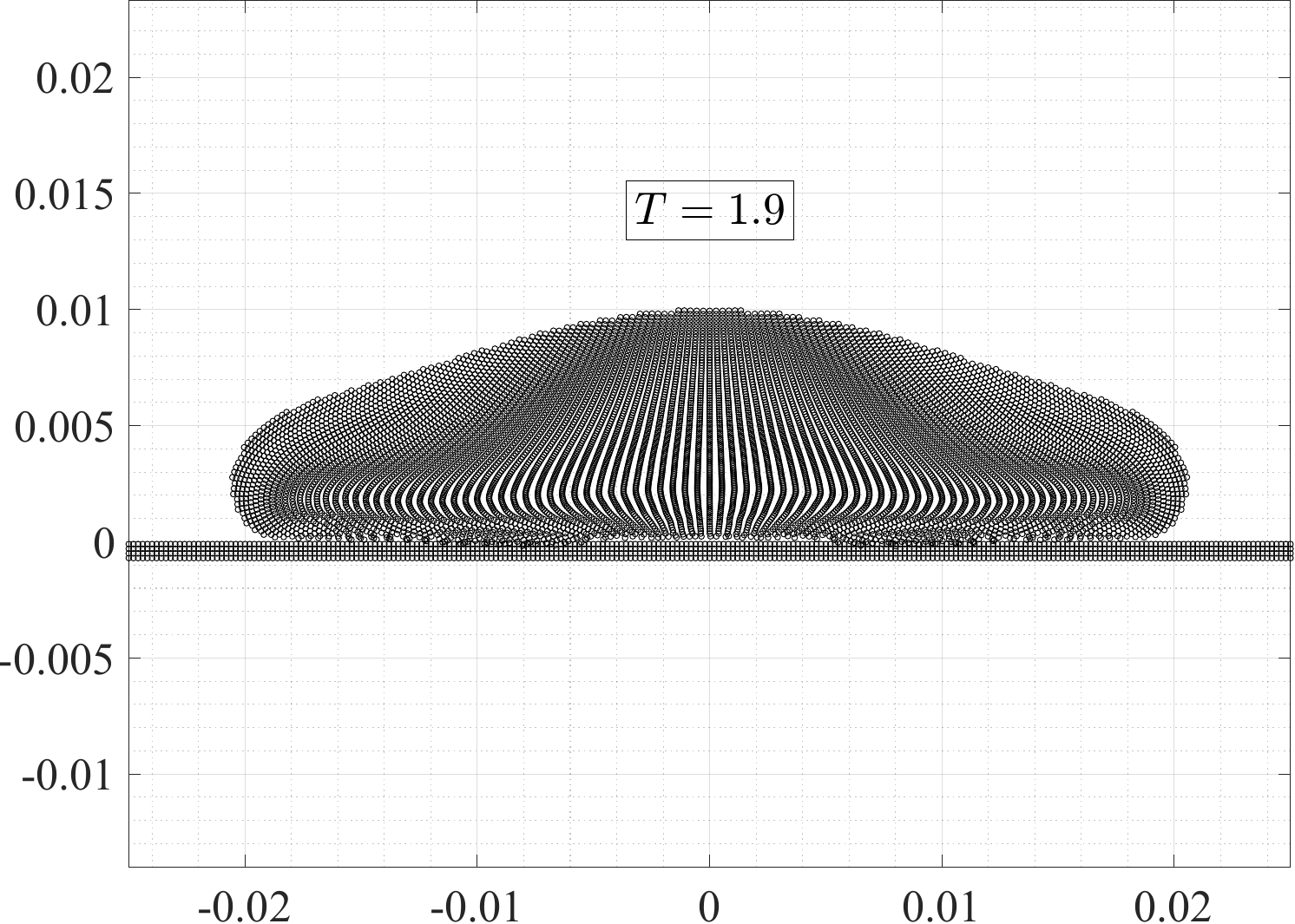}
\caption{}
\label{figure 12c}
\end{subfigure}
\begin{subfigure}[b]{.5\textwidth}
\includegraphics[width=\textwidth]{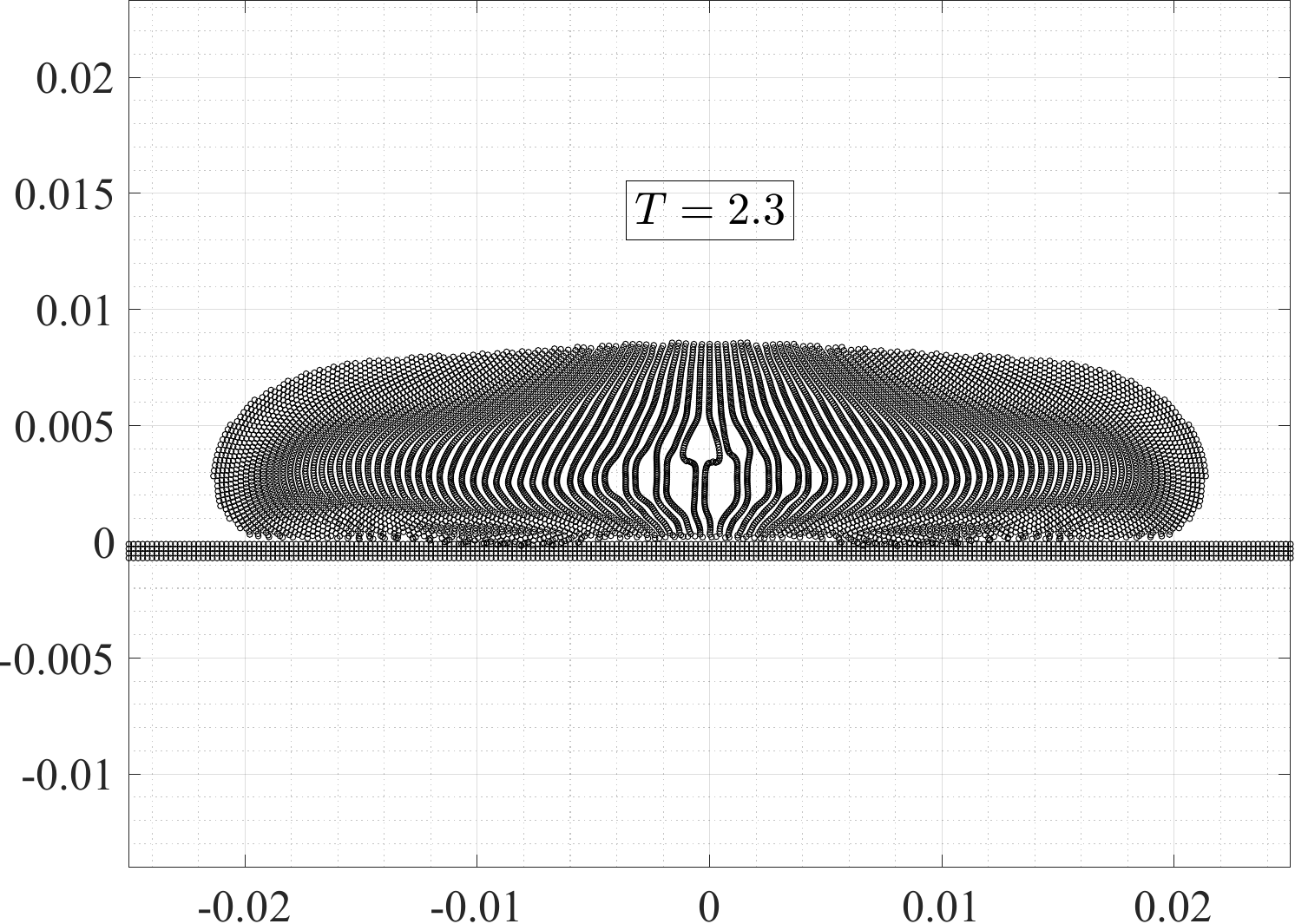}
\caption{}
\label{figure 12d}
\end{subfigure}
\begin{subfigure}[b]{.5\textwidth}
\includegraphics[width=\textwidth]{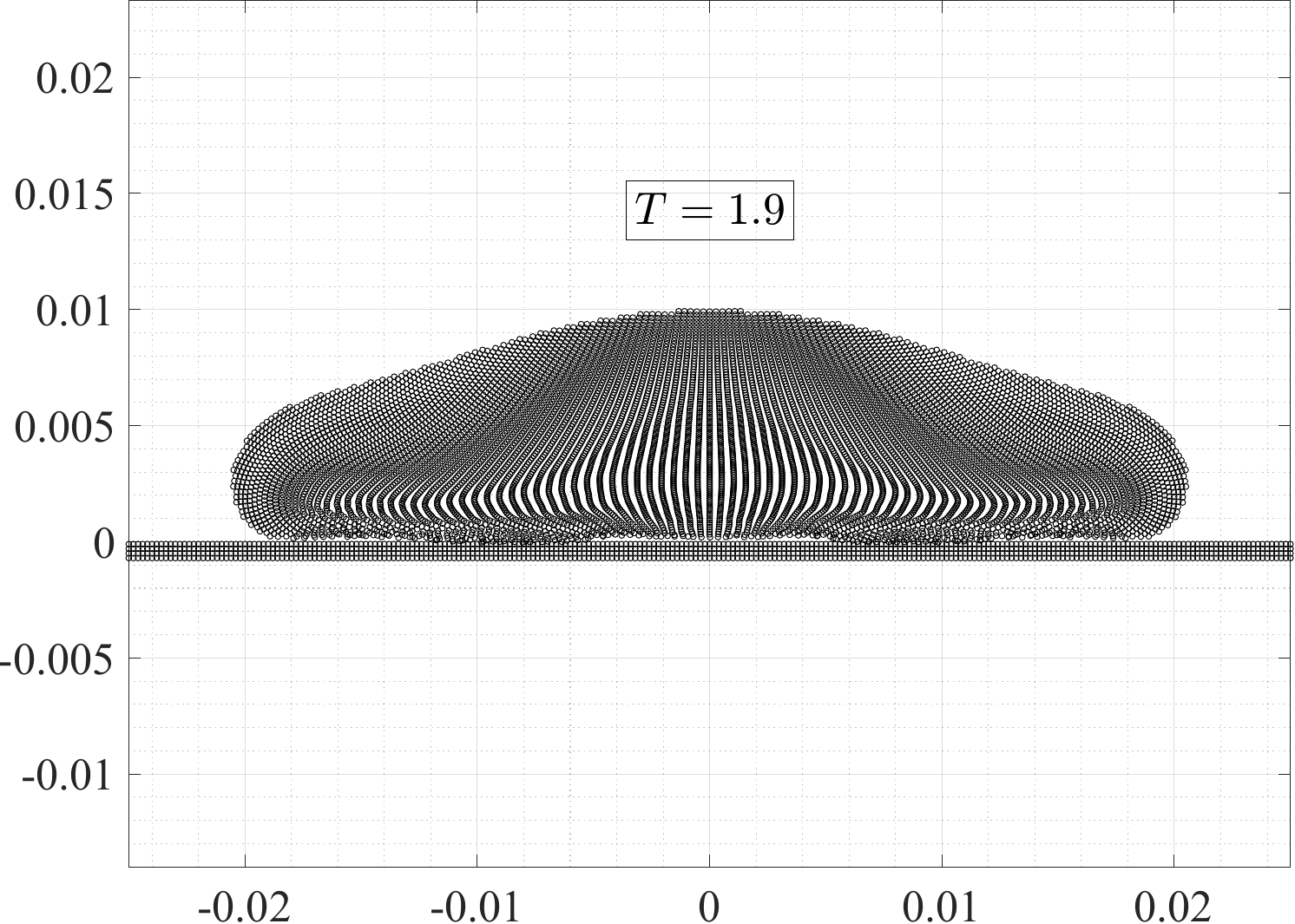}
\caption{}
\label{figure 12e}
\end{subfigure}
\begin{subfigure}[b]{.5\textwidth}
\includegraphics[width=\textwidth]{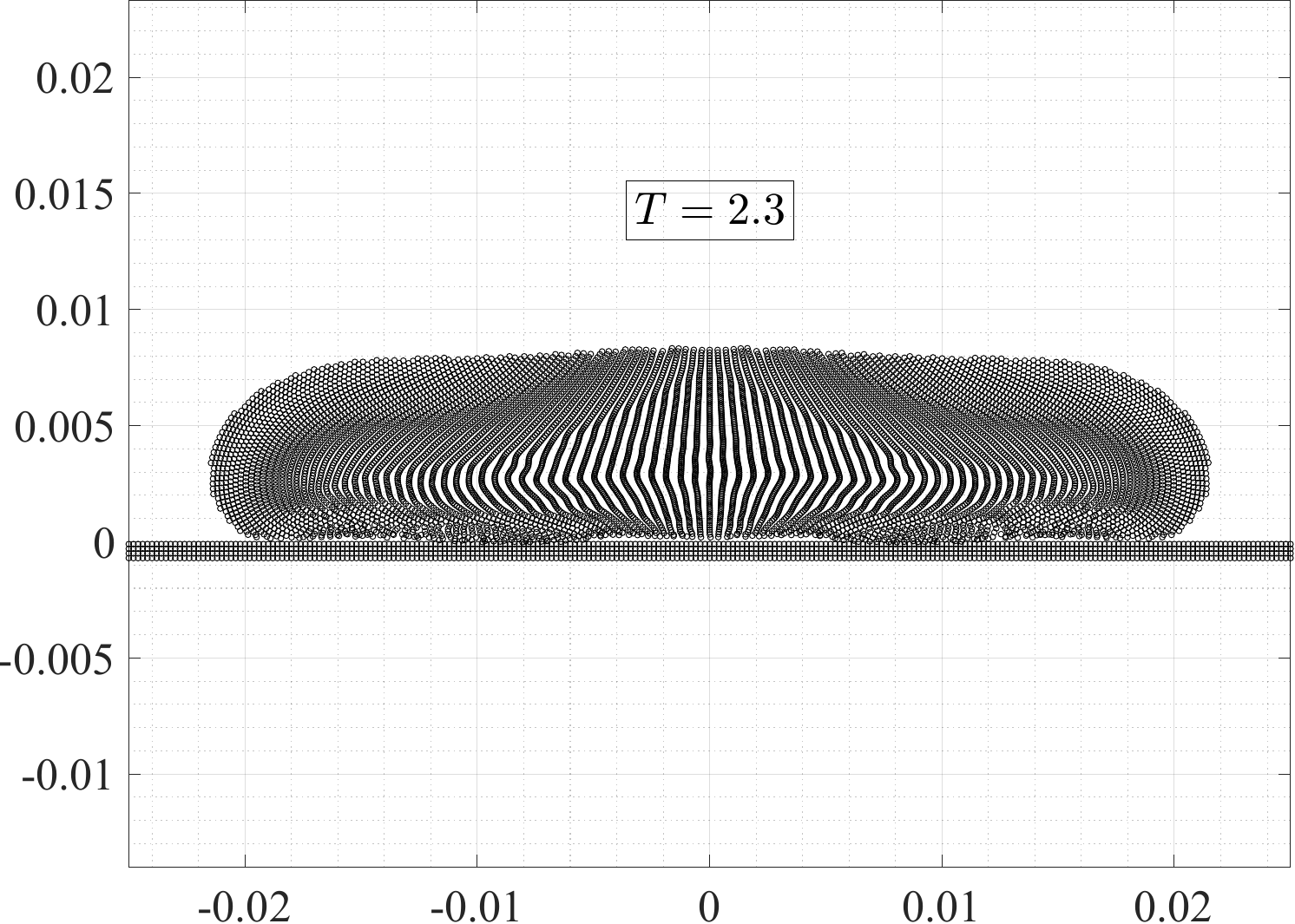}
\caption{}
\label{figure 12f}
\end{subfigure}
\caption{SPH simulation of an Oldroyd B drop using cubic B spline kernel, (a) $T=1.9$, (b) $T=2.3$; simulation using adaptive kernel with maximum value of $a$ as $1.95$, (c) $T=1.9$, (d) $T=2.3$; simulation using adaptive kernel with $a$ and $b$ calculated from Equation \eqref{adaptive_2}, (e) $T=1.9$, (f) $T=2.3$.}
\label{figure 12}
\end{figure}

Next, the simulation is performed for an Oldroyd B fluid. In this case, the solvent viscosity is $\eta_s = 0.4$ $\mathrm{Pa \, s}$, the polymer contribution to the dynamic viscosity is $\eta_p = 3.6$ $\mathrm{Pa \, s}$, and the relaxation time of the fluid is $\lambda_1 = 0.02$ $\mathrm{s}$. Figure \ref{figure 12} shows the drop profiles at two different times ($T=1.94$ and $T=2.29$) for three simulations, one with the standard SPH and two with the adaptive approach presented here. Figures \ref{figure 12a} and \ref{figure 12b} show the drop profiles when the standard B spline kernel is used. The \textit{tensile instability} in the solution is clearly evident. As mentioned previously, even though artificial viscosity with $\gamma_1 = 0.5$ and $\gamma_2 = 0.5$ is used in these simulations, \textit{tensile instability} prevails. In the next set of simulations, the adaptive kernel proposed in this work is used, and the value of knot $a$ is obtained from Equation \ref{adaptive_1}. However, once the value of $a$ has reached $1.95$, $a$ and $b$ are not modified according to Equation \ref{adaptive_2}. Hence the maximum value that $a$ can take in this simulation is $1.95$. The drop profiles are shown in Figures \ref{figure 12c} and \ref{figure 12d}. Though the instability is prevented at the initial stages (Figure \ref{figure 12c}), at later stages of the simulation, the instability appears (Figure \ref{figure 12d}). This is because at the later stages of the simulation, $r_i$ becomes greater than $h$, but the values of $a$ and $b$ are not increased in order that the extremum of $W^{'}$ shifts along with $r_i$. Hence, for the last simulation, the adaptive algorithm is again used, but now once $a$ exceeds the value of $1.95$, $a$ and $b$ are modified according to Equation \ref{adaptive_2}. 
It can be seen from Figures \ref{figure 12e} and \ref{figure 12f} how the problem of \textit{tensile instability} is completely alleviated in this simulation.

\begin{figure}
\centering
\begin{subfigure}[b]{.4\textwidth}
\includegraphics[width=\textwidth]{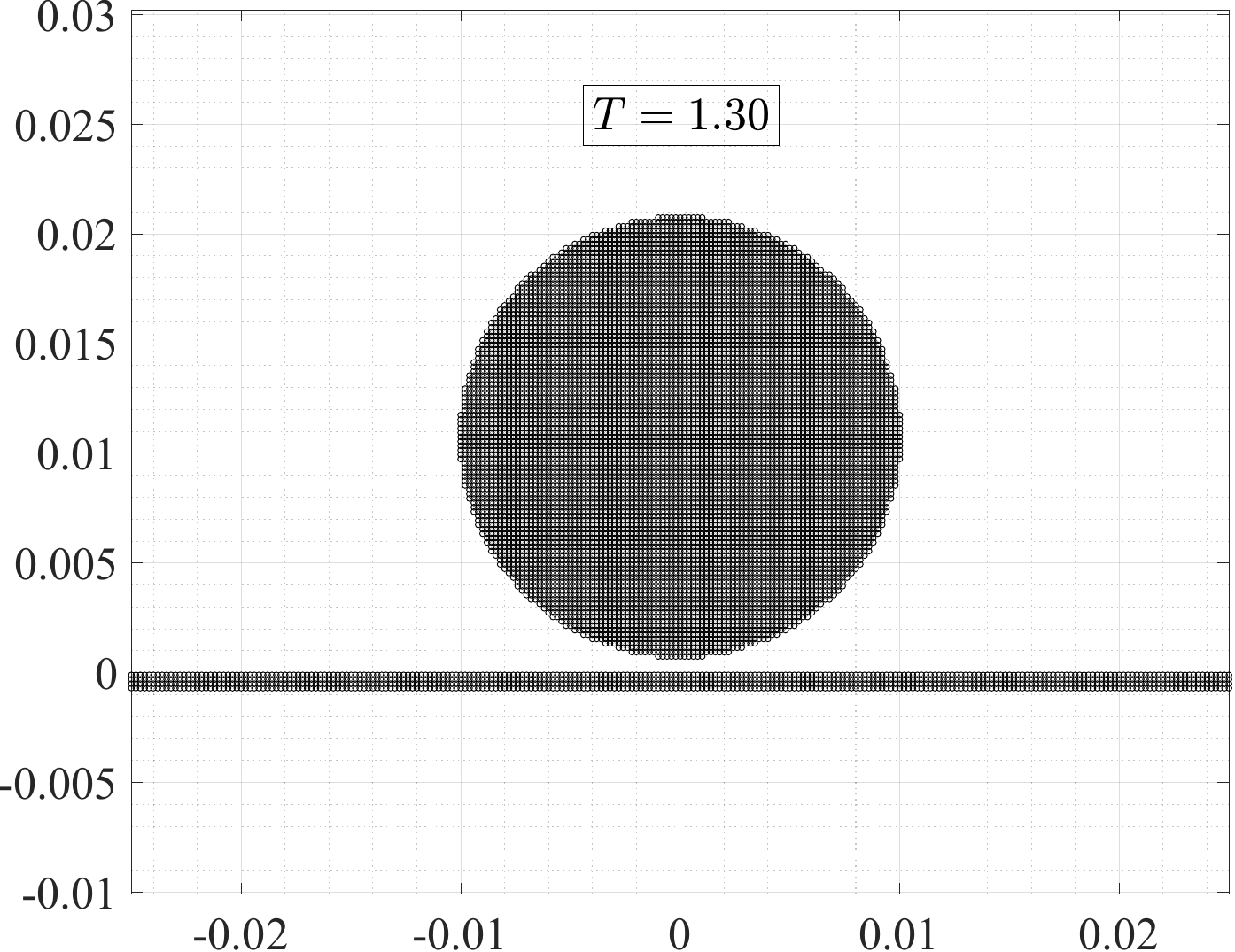}
\caption{}
\end{subfigure}
\begin{subfigure}[b]{.4\textwidth}
\includegraphics[width=\textwidth]{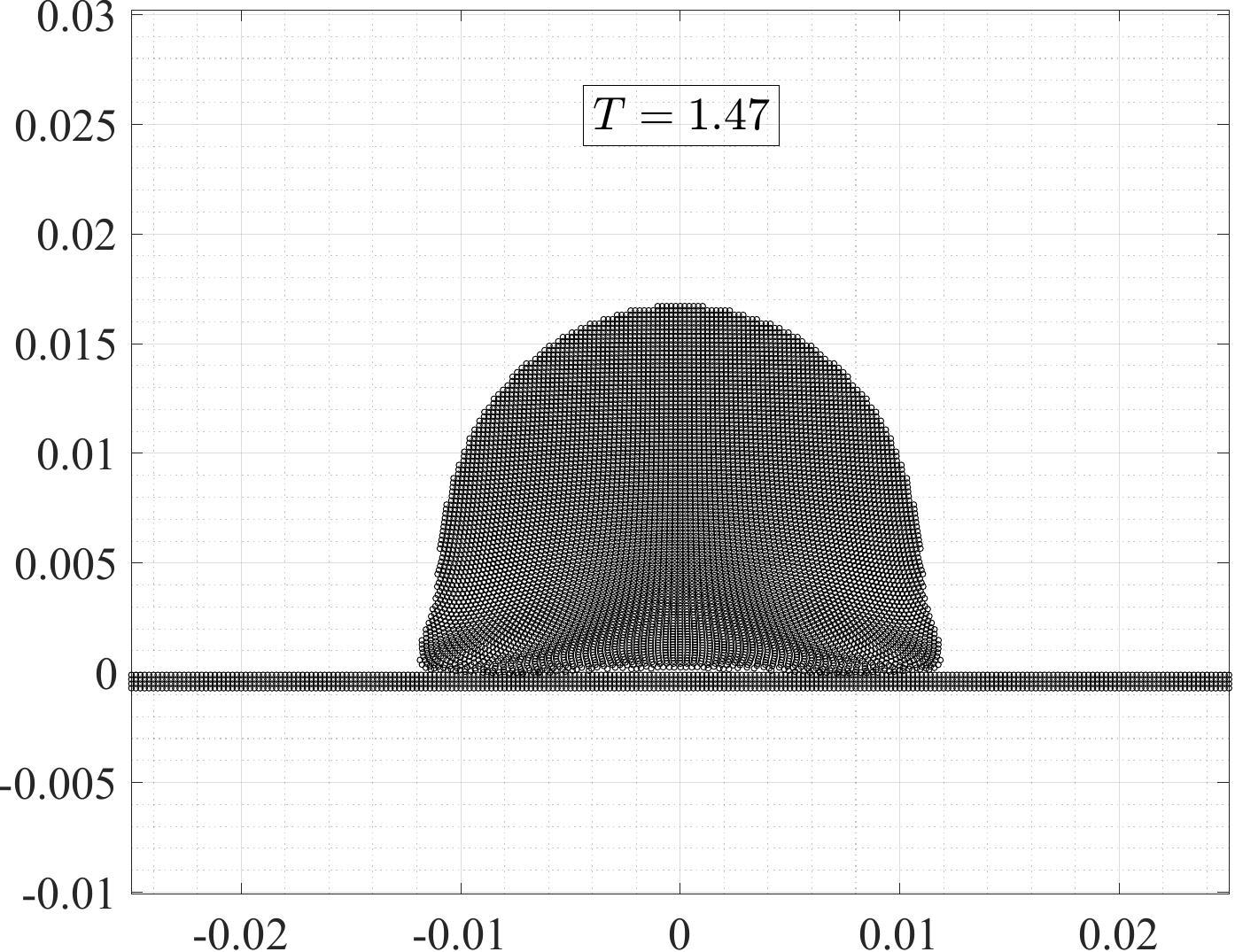}
\caption{}
\end{subfigure}
\begin{subfigure}[b]{.4\textwidth}
\includegraphics[width=\textwidth]{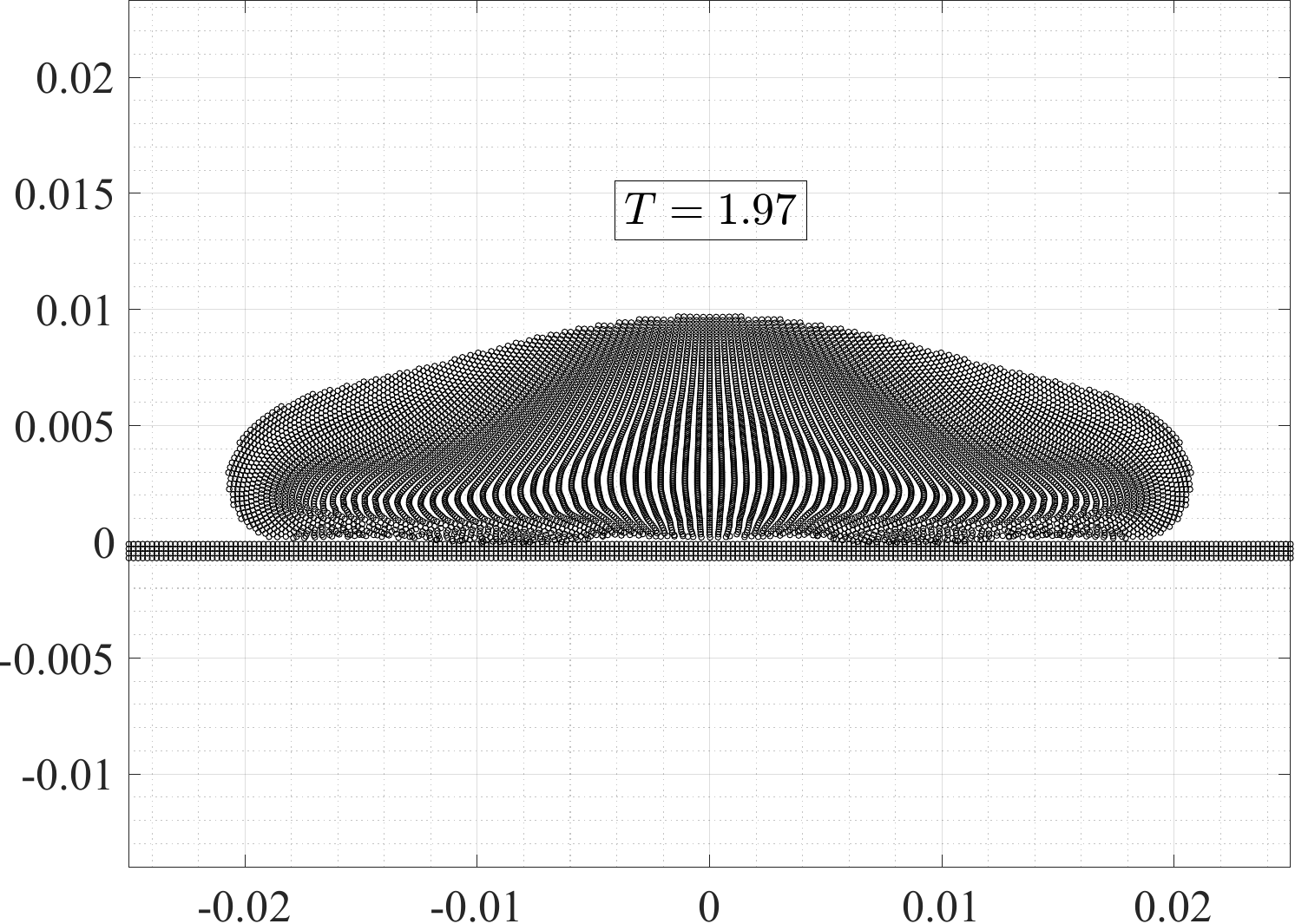}
\caption{}
\end{subfigure}
\begin{subfigure}[b]{.4\textwidth}
\includegraphics[width=\textwidth]{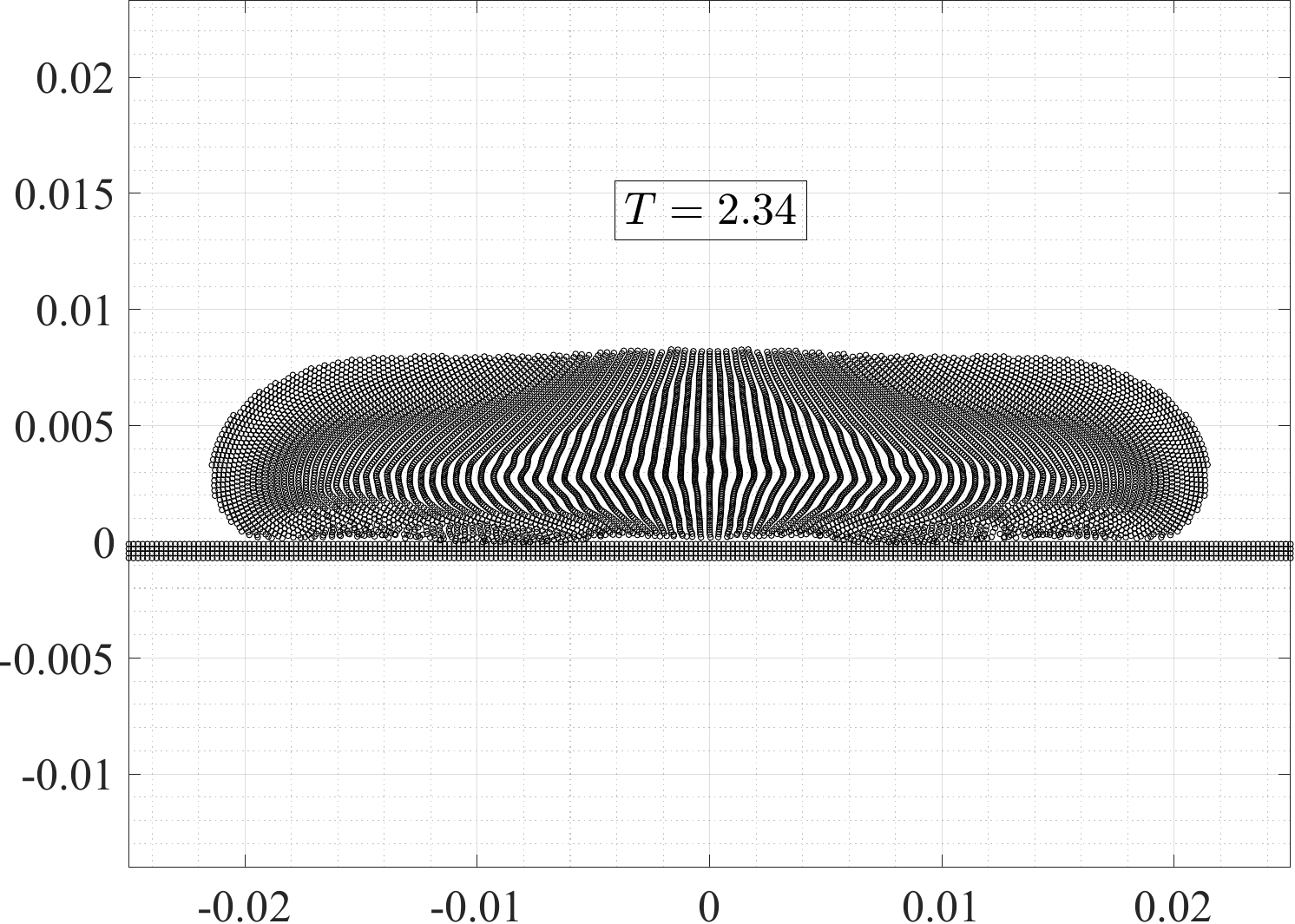}
\caption{}
\end{subfigure}
\begin{subfigure}[b]{.4\textwidth}
\includegraphics[width=\textwidth]{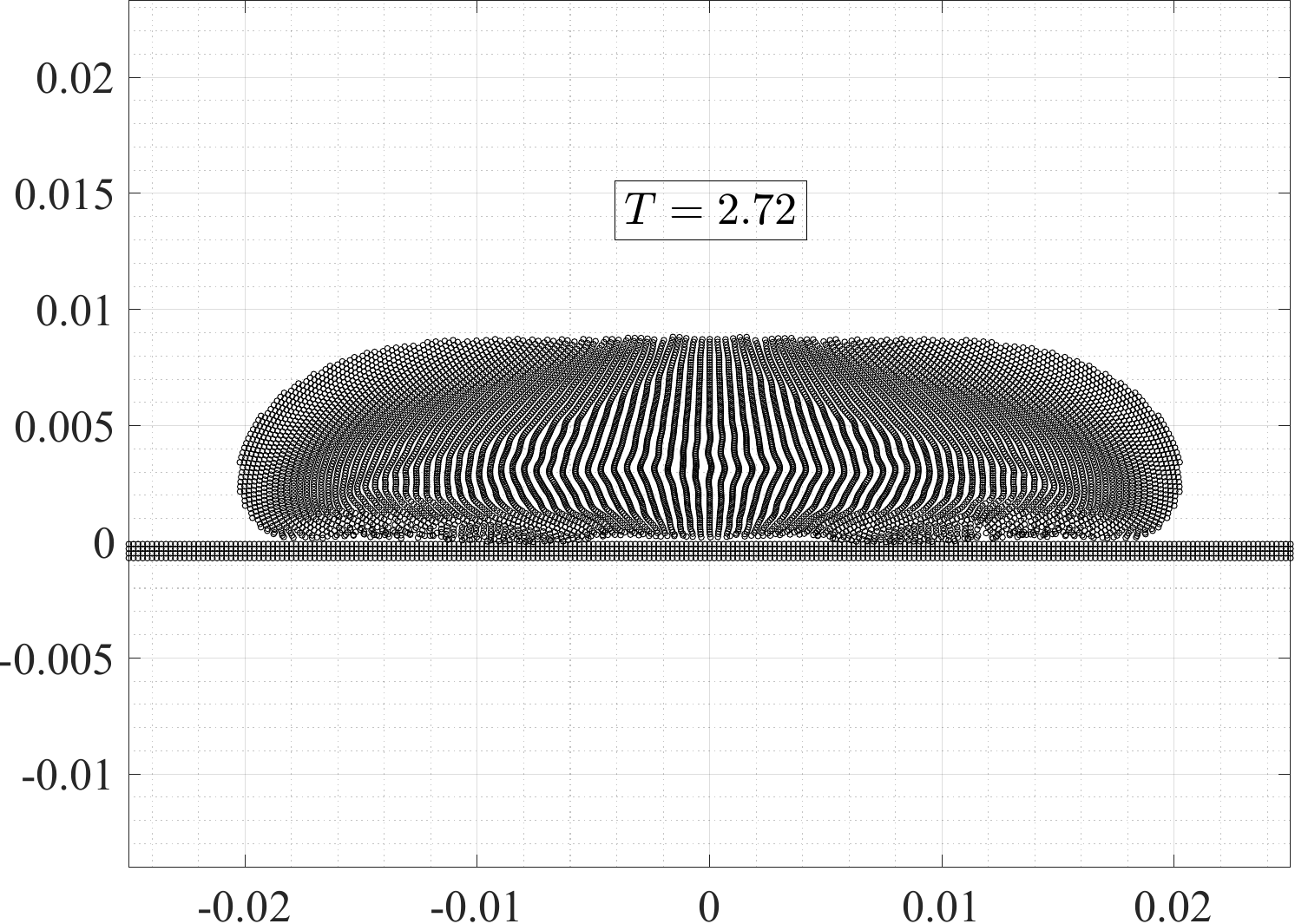}
\caption{}
\end{subfigure}
\begin{subfigure}[b]{.4\textwidth}
\includegraphics[width=\textwidth]{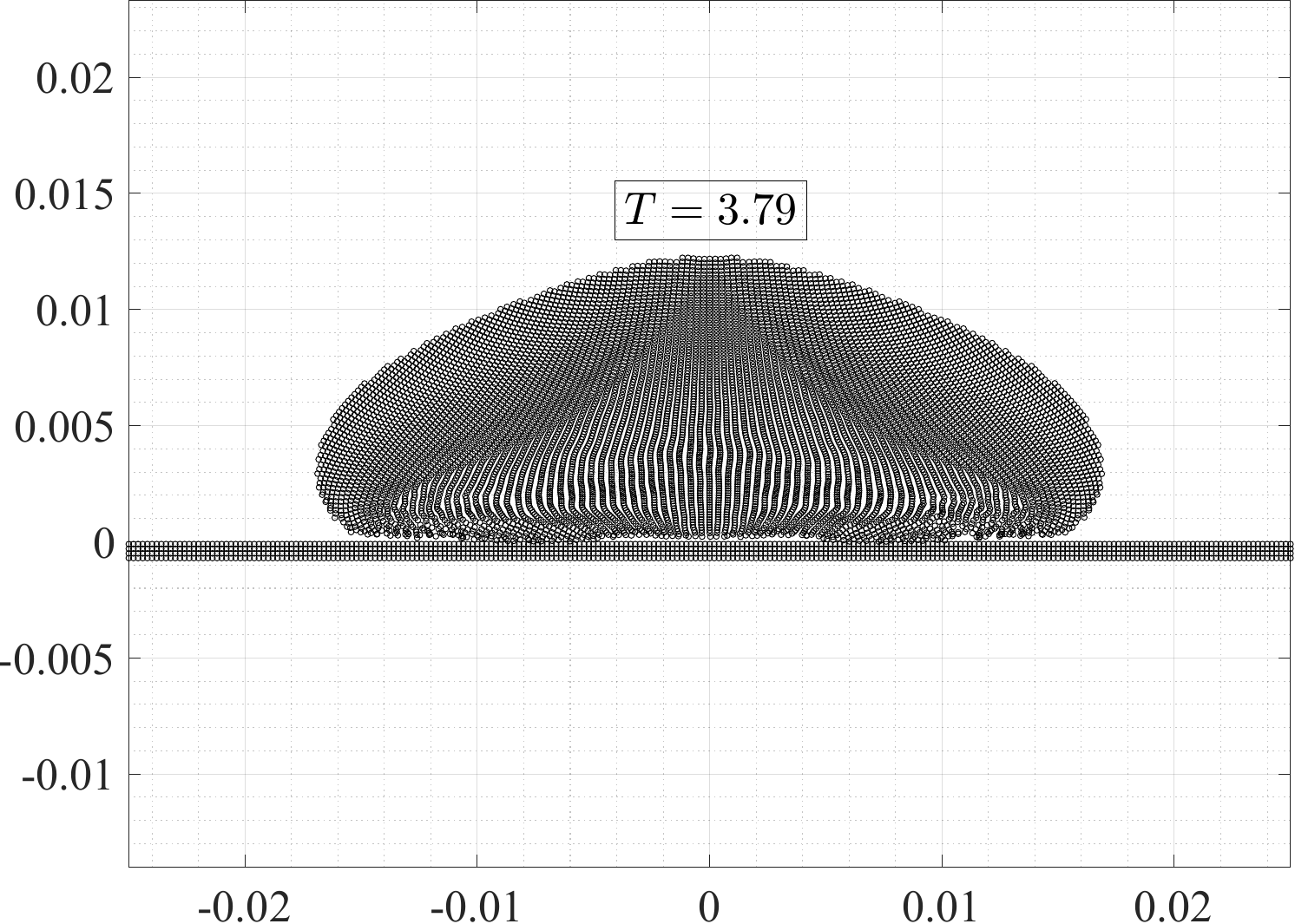}
\caption{}
\end{subfigure}
\begin{subfigure}[b]{.4\textwidth}
\includegraphics[width=\textwidth]{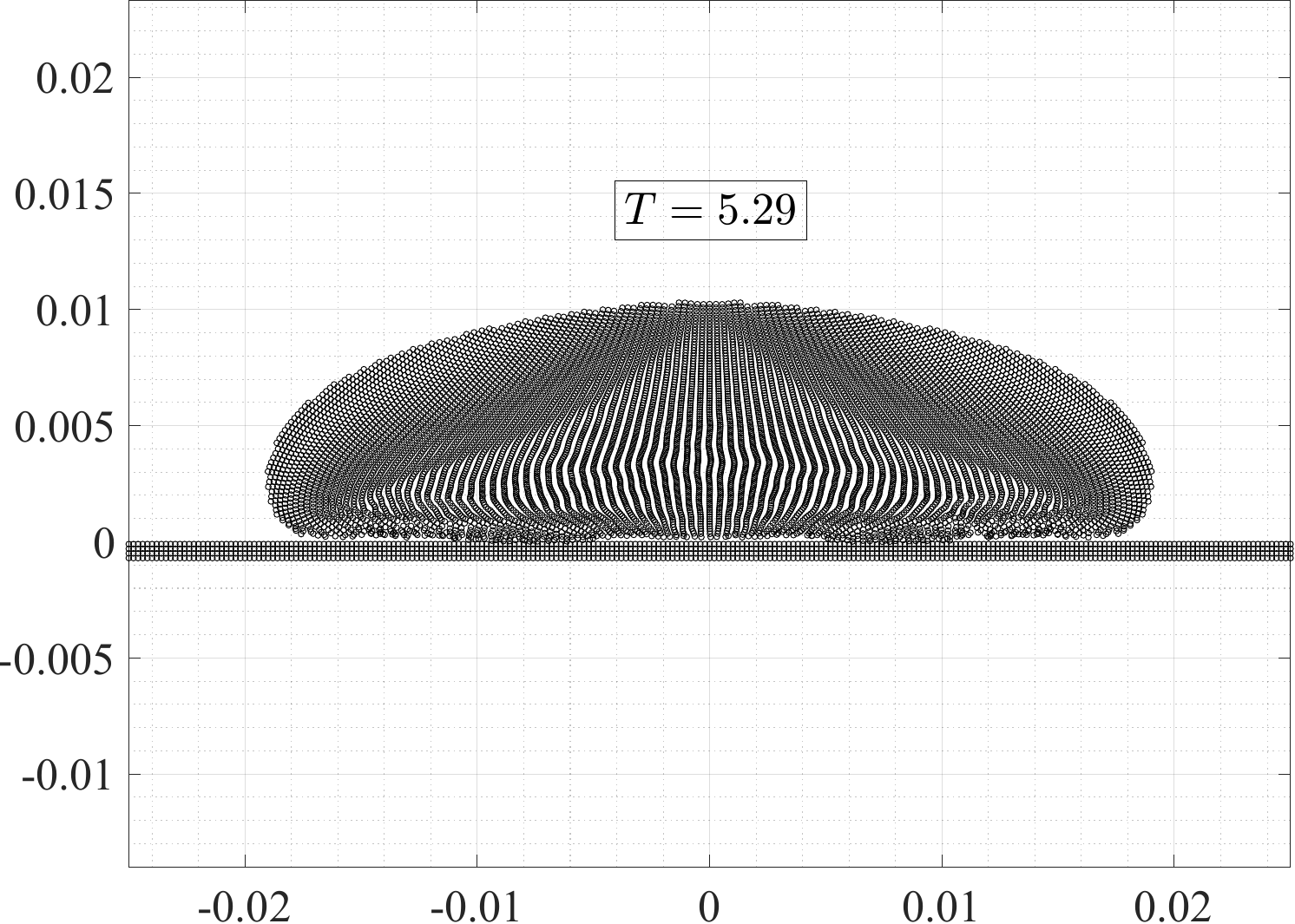}
\caption{}
\end{subfigure}
\begin{subfigure}[b]{.4\textwidth}
\includegraphics[width=\textwidth]{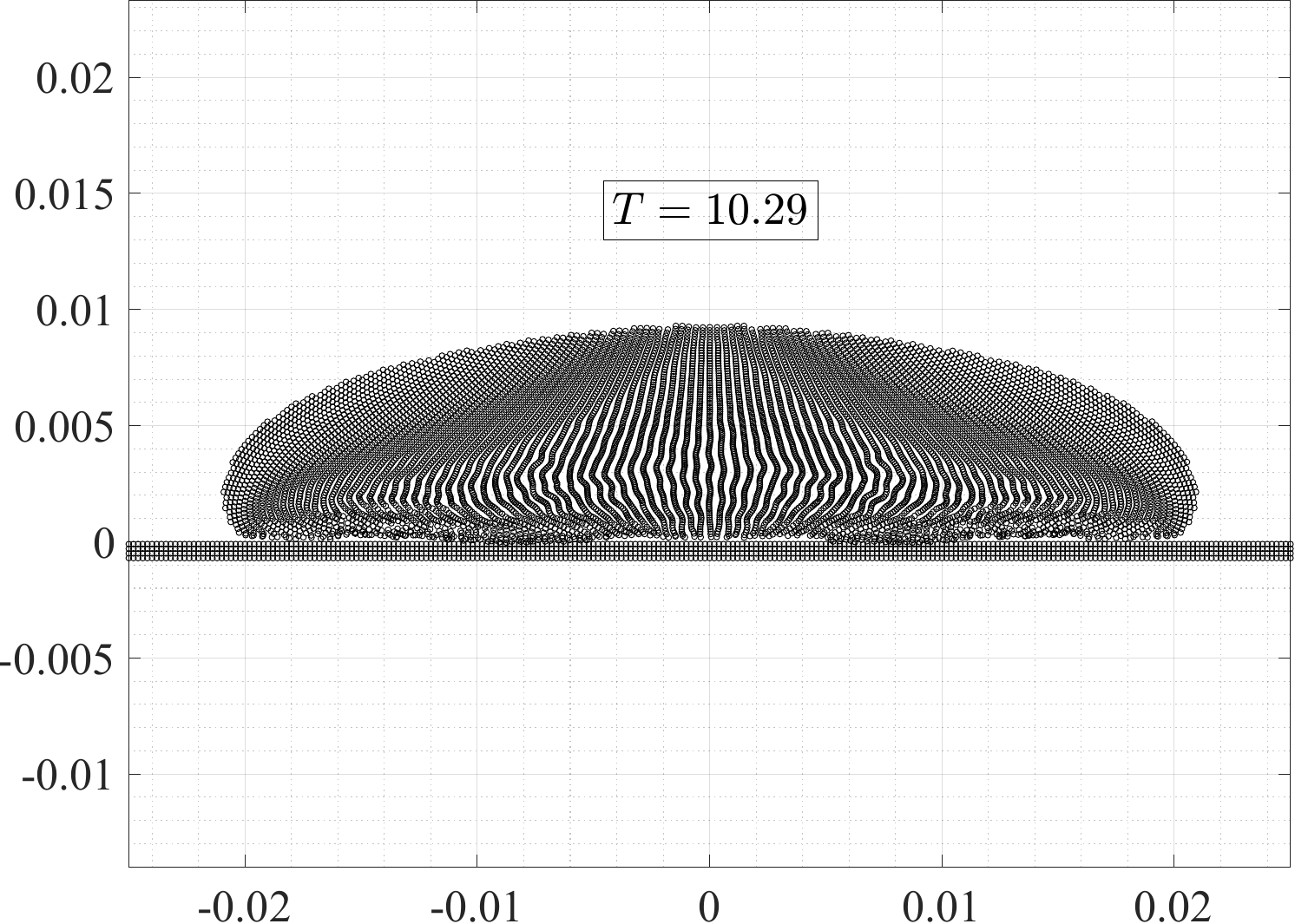}
\caption{}
\end{subfigure}
\caption{SPH simulation of an Oldroyd B drop impacting a rigid surface, using the adaptive B-spline kernel, at different non-dimensionless times as indicated.}
\label{figure 13}
\end{figure}

\begin{figure}[htp]
\centering
\includegraphics[scale=0.4]{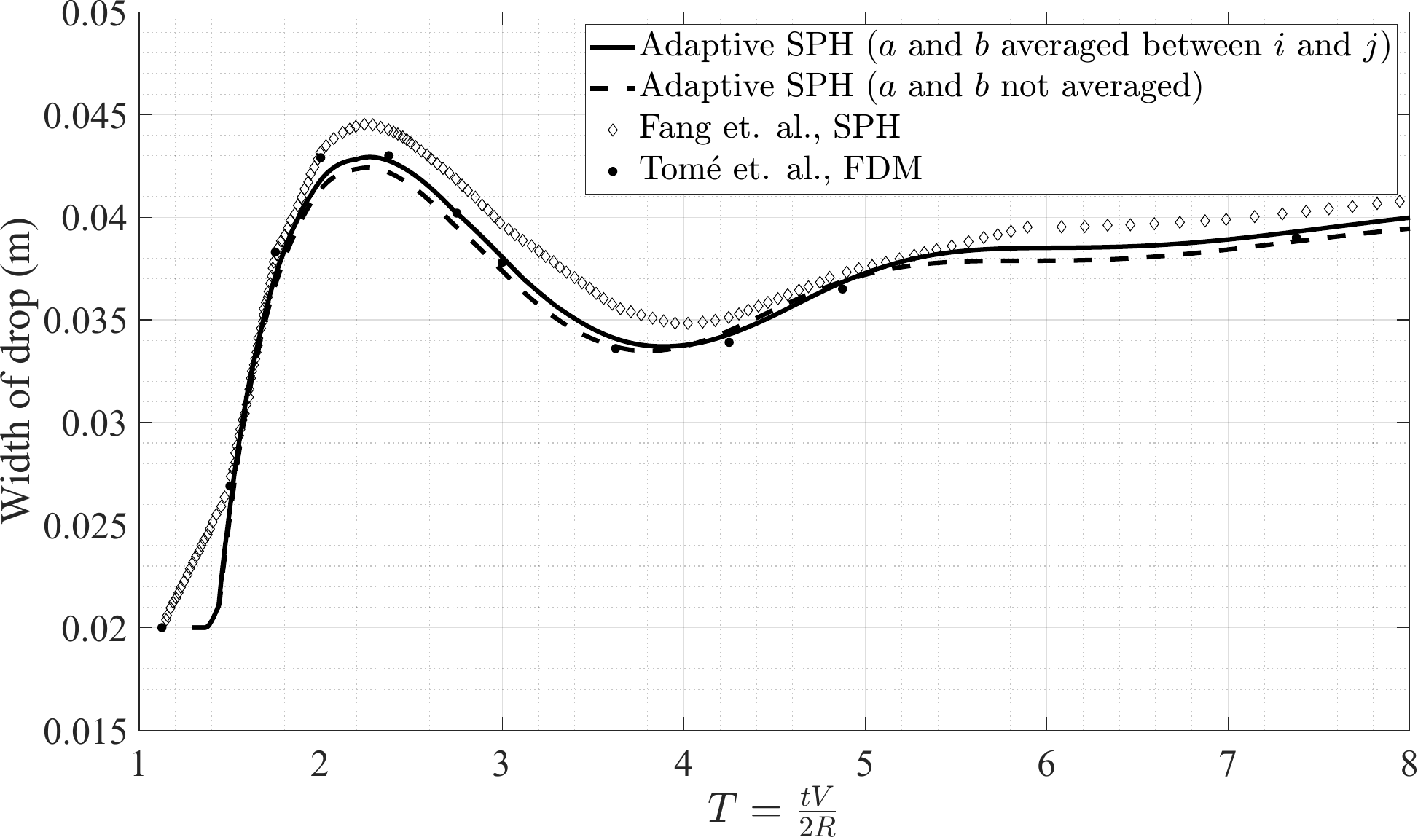}
\caption{Time history of width of drop for an Oldroyd B fluid using SPH with the adaptive kernel approach. Comparison is made with SPH simulations from \citet{fang2006numerical} and FDM simulations from \citet{tome2002finite}.}
\label{figure 14}
\end{figure}

\begin{figure}
\centering
\begin{subfigure}[b]{.45\textwidth}
\includegraphics[width=\textwidth]{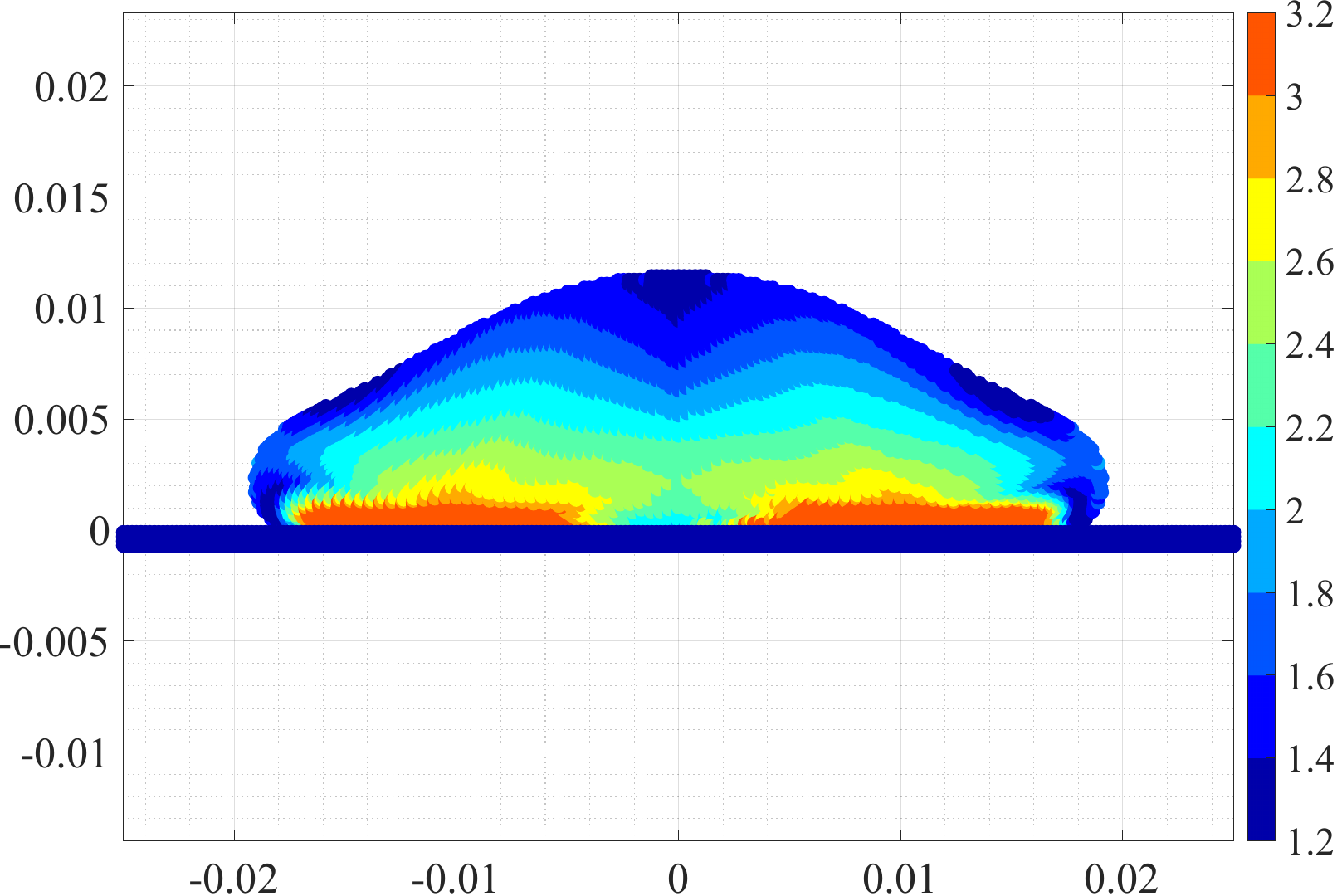}
\caption{$a$ at T = $1.8$}
\end{subfigure}
\begin{subfigure}[b]{.45\textwidth}
\includegraphics[width=\textwidth]{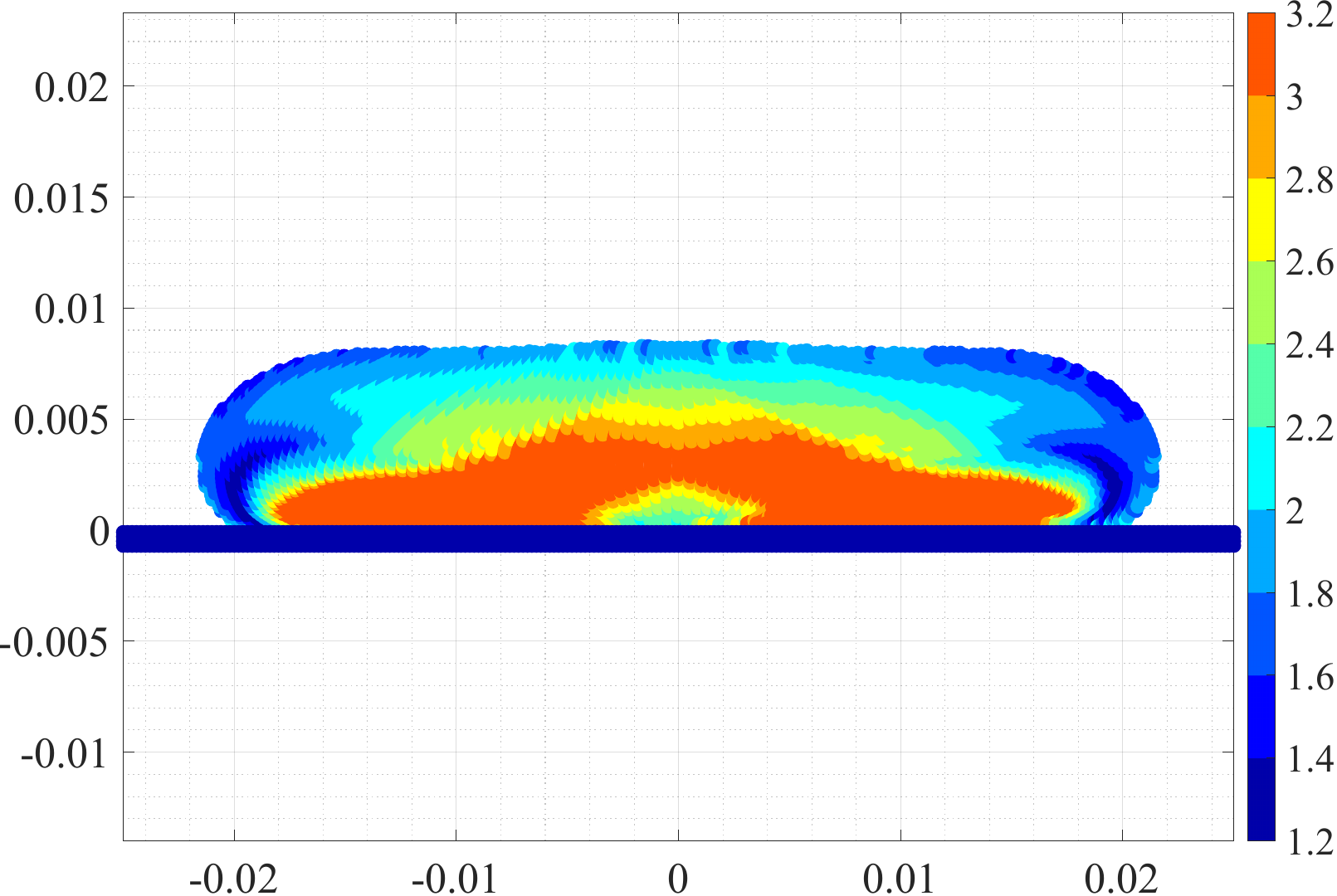}
\caption{$a$ at T = $2.4$}
\end{subfigure}
\begin{subfigure}[b]{.45\textwidth}
\includegraphics[width=\textwidth]{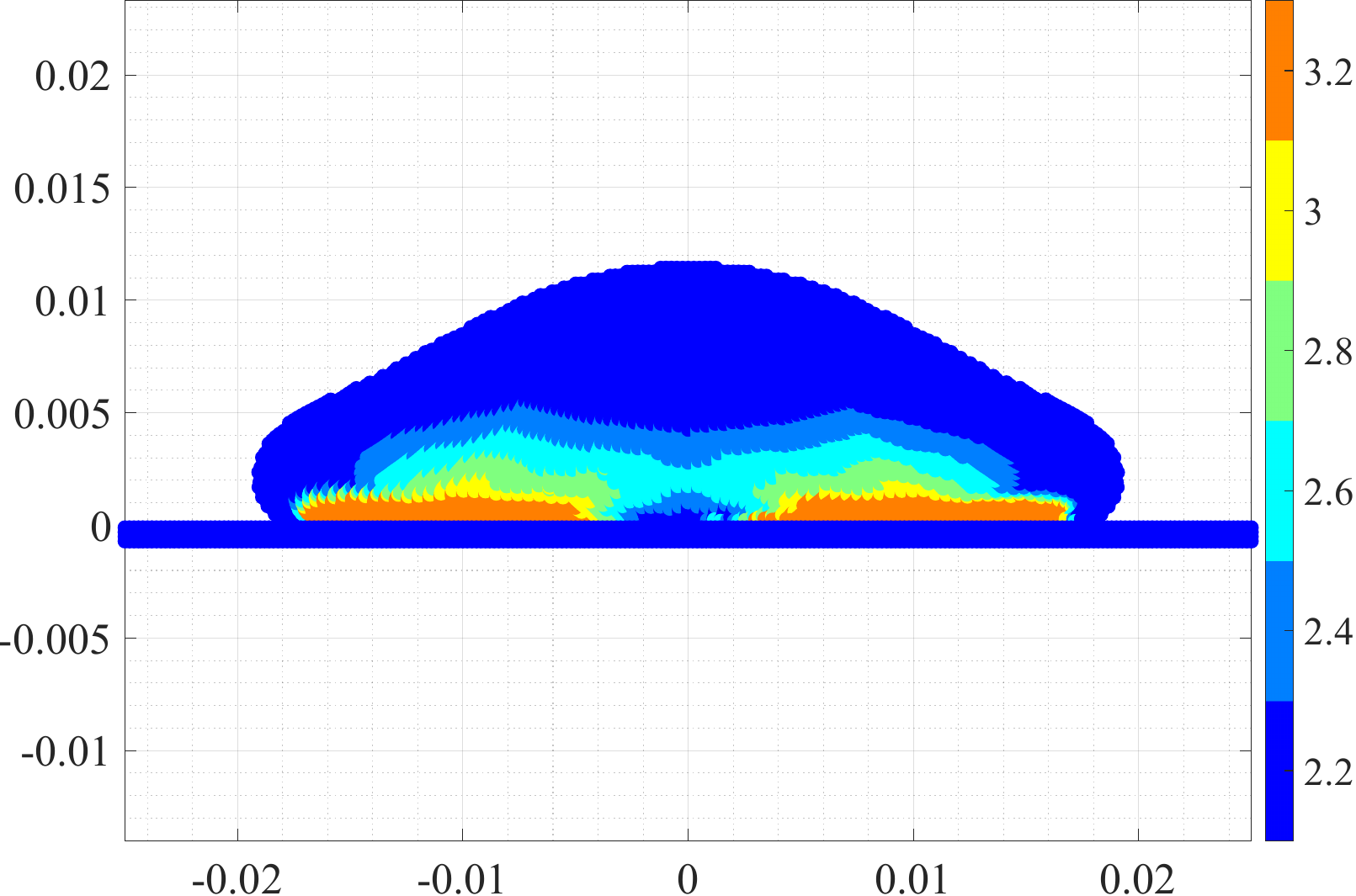}
\caption{$b$ at T = $1.8$}
\end{subfigure}
\begin{subfigure}[b]{.45\textwidth}
\includegraphics[width=\textwidth]{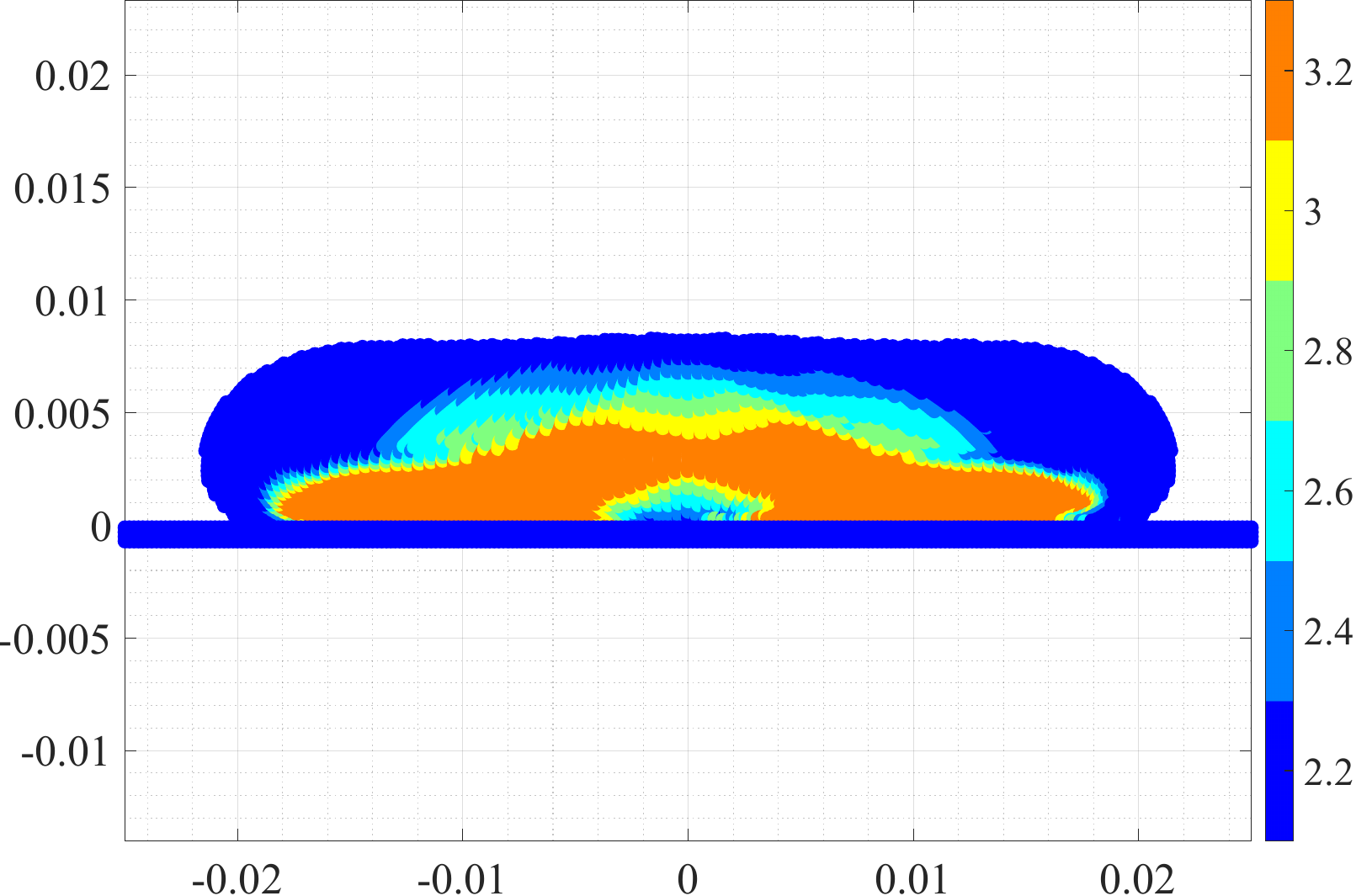}
\caption{$b$ at T = $2.4$}
\end{subfigure}
\caption{Distribution of $a$ and $b$ at different time instants.}
\label{figure 14a}
\end{figure}

For the simulation with the adaptive approach, where $a$ and $b$ are modified according to Equation \ref{adaptive_2}, the drop profiles for the Oldroyd B drop for different times are shown in Figure \ref{figure 13}. It can be observed how the instability is prevented from occurring at all times. In Figure \ref{figure 14}, the evolution of the width of the drop with time is shown, and the result is compared with that given in the literature (\cite{fang2006numerical, tome2002finite}). The present SPH results are comparable with the results from the literature, and furthermore, there is a decent agreement with the FDM result in \cite{tome2002finite}. The spatial distribution of $a$ and $b$ at different time instants are shown in Figure \ref{figure 14a}. Finally, a simulation is performed without the averaging of knot values between an interacting particle pair $i-j$ to compare with the previous simulations where the averaging has been considered. From the evolution of the drop width shown in Figure \ref{figure 14}, it can be seen that there is a marginal difference in the results of the two approaches.   

\subsection{Example 2: Rotation of a fluid patch}
\label{s6.3} 
\begin{figure}[htp]
\centering
\includegraphics[scale=0.4]{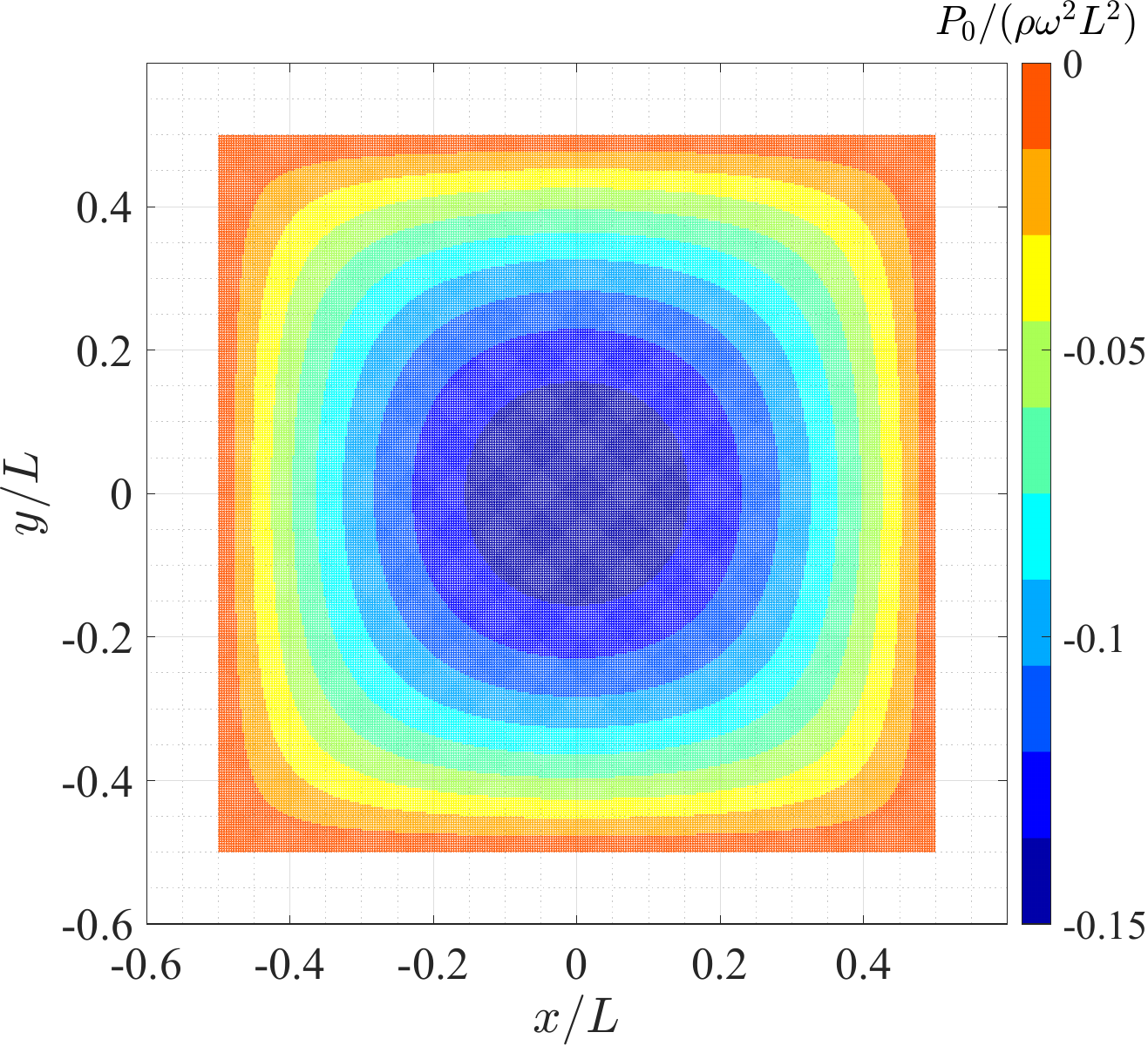}
\caption{Initial pressure distribution in the rectangular patch of fluid}
\label{figure 15}
\end{figure}

In this section, the example of a rotating, initially square patch (side length L) of fluid is considered. This test is considered a benchmark for demonstrating instability in particle-based methods (\citet{le2013critical}). Figure \ref{figure 15} shows the initial configuration of the fluid patch, which is given an initial velocity field as,
\begin{equation}
\label{vel_patch}
\begin{split}
u(x,y,t=0) = +\omega y; \qquad v(x,y,t=0) = -\omega x,
\end{split}
\end{equation}
where $\omega$ is the angular velocity. The initial pressure field consistent with the initial velocity field (Equation \ref{vel_patch}) is given by,
\begin{equation}
\label{pr_patch}
\begin{split}
P_0(x,y) = \rho \sum_m^{\infty} \sum_n^{\infty} -\frac{32\omega^2/(mn\pi^2)}{(n\pi/L)^2+(m\pi/L)^2}\sin(\frac{m\pi x^*}{L})\sin(\frac{n\pi y^*}{L}) ~~~~~ m,n \in \mathbb{N}^{odd},
\end{split}
\end{equation}
where $x^*=x+L/2$ and $y^*=y+L/2$. The initial pressure is calculated from Equation \ref{pr_patch} and applied to the fluid patch as shown in Figure \ref{figure 15}. In the literature, the simulation of this problem is performed without any gradient correction for consistency, and therefore, here also, the gradient correction is not implemented. 
The value of $c_0$ is taken as $7\omega L$ \cite{le2013critical}. The density is periodically reinitialised as is done in \cite{le2013critical, colagrossi2003numerical}. At every $20^{th}$ time step, the density is recalculated as,
\begin{equation}
\label{dens_reinitialisation_1}
\begin{split}
\rho_{i}=\sum_{j}m_j W_{j}^{MLS}(\boldsymbol{x}_i),
\end{split}
\end{equation}
where the moving least square kernel $W_{j}^{MLS}$ is calculated through the following equations;
\begin{equation}
\label{dens_reinitialisation_2}
\begin{split}
&W_{j}^{MLS}(\boldsymbol{x}_i)=[\beta_0(\boldsymbol{x}_i)+\beta_1(\boldsymbol{x}_i)(x_i-x_j)+\beta_2(\boldsymbol{x}_i)(y_i-y_j)]W_{ij},\\
&\boldsymbol{\beta}(\boldsymbol{x}_i)=\begin{bmatrix}
           \beta_{0} \\
           \beta_{1} \\
           \beta_{2} 
         \end{bmatrix}=\boldsymbol{A}^{-1}(\boldsymbol{x}_i)\begin{bmatrix}
           1 \\
           0 \\
           0, 
           \end{bmatrix},\\
&\boldsymbol{A}(\boldsymbol{x}_i)=\sum_{j}W_{j}(\boldsymbol{x}_i)\tilde{\boldsymbol{A}}_{ij},\\
&\tilde{\boldsymbol{A}}_{ij}=\begin{bmatrix}
      1&{(x_i-x_j)}&{(y_i-y_j)}\\
      {(x_i-x_j)}& (x_i-x_j)^2 & {(y_i-y_j)(x_i-x_j)} \\
       {(y_i-y_j)}&{(y_i-y_j)(x_i-x_j)}& (y_i-y_j)^2 
    \end{bmatrix}. 
\end{split}
\end{equation}
The advantage of using density reinitialisation is that a more regular pressure distribution can be obtained.

In the following simulations, the averaging of the knot values between a pair of interacting particles is not considered when solving Equations \eqref{SPH_discret_1}-\eqref{SPH_discret_4}. The number of SPH particles considered is $90,000$. The smoothing length $h$ is taken as $2\Delta p$, where $\Delta p$ is the initial interparticle spacing. First, the simulations are performed with standard SPH with artificial viscosity coefficients $\gamma_1 = 0.8$ and $\gamma_2 = 0.8$. Because of the -ve pressures in the fluid patch (Figure \ref{figure 15}), \textit{tensile instability} occurs in standard SPH. This results in unphysical fragmentation, as can be seen from the deformed shapes in Figure \ref{figure 16}. Even for higher values of $\gamma_1$ and $\gamma_2$, the instability could not be prevented. Next, the simulations are performed using the adaptive kernel approach proposed in this work. As mentioned in Section \ref{s5.3}, in this case, $r_i$ is calculated via Equation \ref{Estimate_ri_2} where the contribution of shear strains is not considered. Furthermore, in this simulation, the maximum value of $a$ in the unstable zone (i.e., the central part of the patch) is found to be always less than 1.95, and therefore the \textit{a-adaptive} approach is sufficient to tackle the \textit{tensile instability}. The value of $a$ reaches close to $1.95$ only near the legs, but the pressure in these regions is close to zero; hence no \textit{tensile instability} is observed in the legs. 

In this simulation, the adaptive kernel approach alone could not prevent instability; artificial viscosity is needed in conjunction with the adaptive approach in controlling the \textit{tensile instability}. It is documented in \citet{morris1996study} that artificial viscosity cannot eliminate the instability, but it can reduce its growth rate. In this problem, we can observe how it aids the adaptive approach in eliminating instability, but artificial viscosity alone cannot eliminate it. The minimum values of the artificial viscosity parameters required for the simulation are $\gamma_1 = 0.8$ and $\gamma_2 = 0.8$. The deformed shapes of the rotating patch at times $t\omega = 2$ and $t\omega = 4$ are shown in Figure \ref{figure 17}. The deformed shapes are compared with the numerical solutions obtained by the Boundary Element Method (\citet{oger2016sph}, \citet{sun2017deltaplus}). {Though the \textit{tensile instability} is prevented, slight deviations of the deformed shapes with respect to the reference solution can be observed. The reason for this is the comparatively high values of $\gamma_1$ and $\gamma_2$ adopted. The \textit{tensile instability} occurs in regions of high negative pressure; hence it is in those regions where the high artificial viscosity is needed. In the regions where the pressure is positive or where the magnitude of negative pressure is low, the values of $\gamma_1$ and $\gamma_2$ can be reduced. Hence in the next set of simulations, $\gamma_1$ and $\gamma_2$ are varied according to the distribution of pressure. 
At the beginning of the simulation, the artificial viscosity parameters are set such that the points with the maximum tensile pressure are assigned $(\gamma_1, \gamma_2)=(0.8,0.8)$, and the points where the pressure is compressive the $(\gamma_1, \gamma_2)$ are reduced to $(0.1,0.1)$. For points with a pressure between zero and the maximum tensile pressure, the $(\gamma_1, \gamma_2)$ is linearly varied in between $(0.1,0.1)$ and $(0.8,0.8)$. Using this approach, we can see from Figure \ref{figure 18a} that there is a reasonably good agreement with the reference solutions.
\begin{figure}[htp]
\begin{subfigure}[b]{.5\textwidth}
\includegraphics[width=\textwidth]{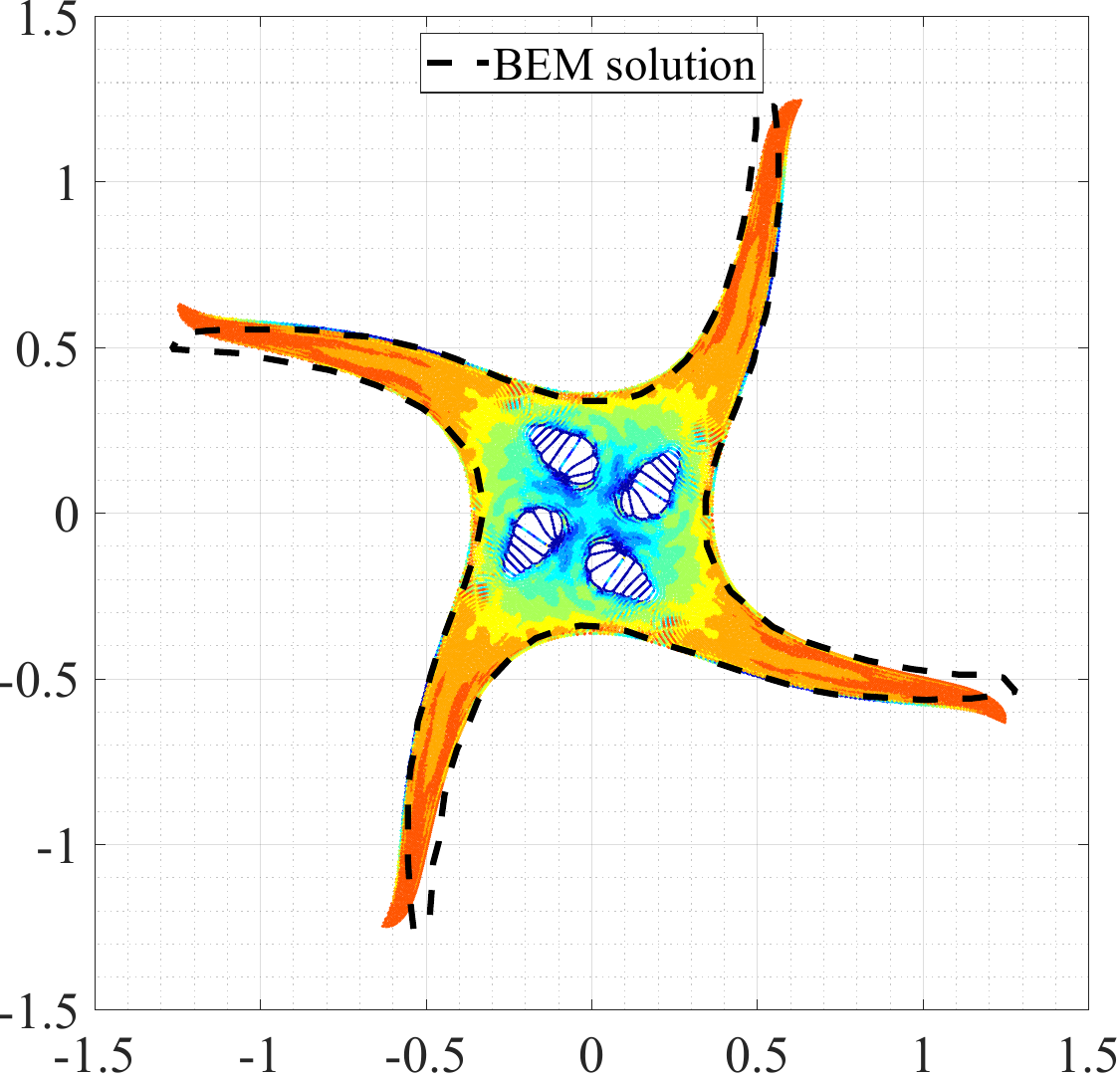}
\caption{$t\omega = 2$}
\end{subfigure}
\begin{subfigure}[b]{.5\textwidth}
\includegraphics[width=\textwidth]{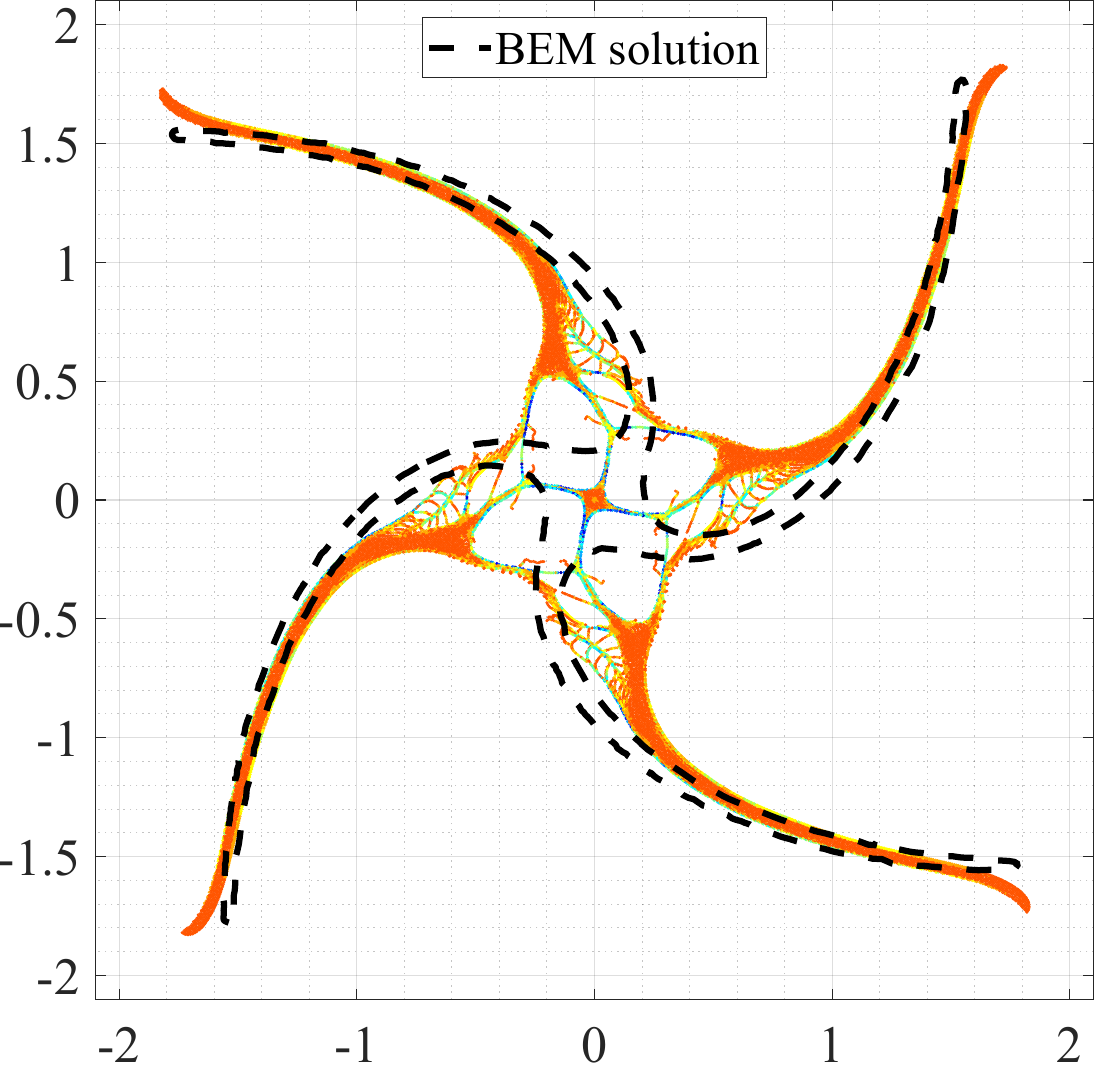}
\caption{$t\omega = 4$}
\end{subfigure}
\caption{Plot of the deformed shapes of the fluid patch without any \textit{tensile instability} correction. The dashed lines represent the solution obtained by BEM (\citep{oger2016sph} and \citep{sun2017deltaplus}). Particles are colored with pressure contours.}
\label{figure 16}
\end{figure}
\begin{figure}[htp]
\begin{subfigure}[b]{.5\textwidth}
\includegraphics[width=\textwidth]{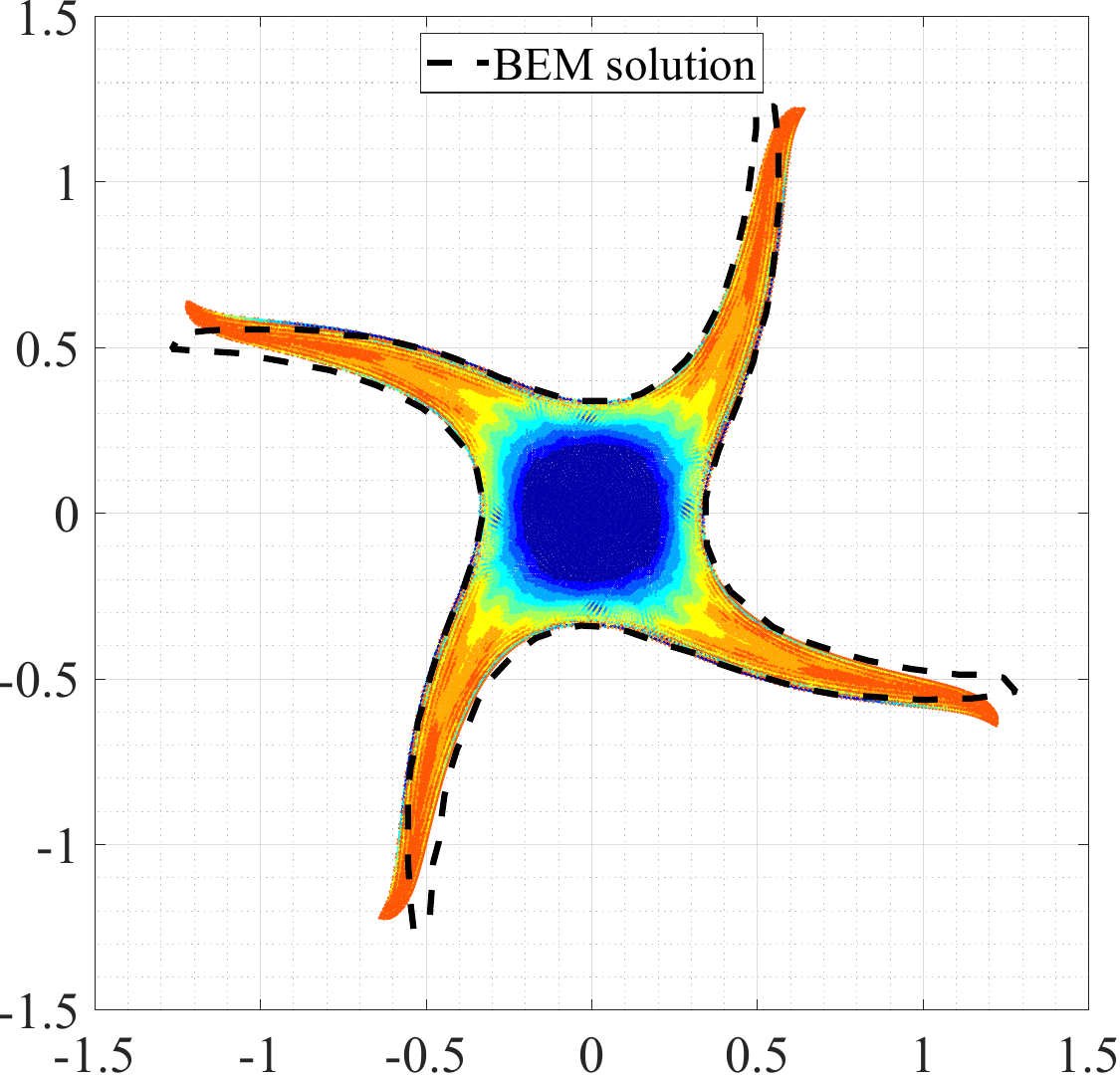}
\caption{$t\omega = 2$}
\end{subfigure}
\begin{subfigure}[b]{.5\textwidth}
\includegraphics[width=\textwidth]{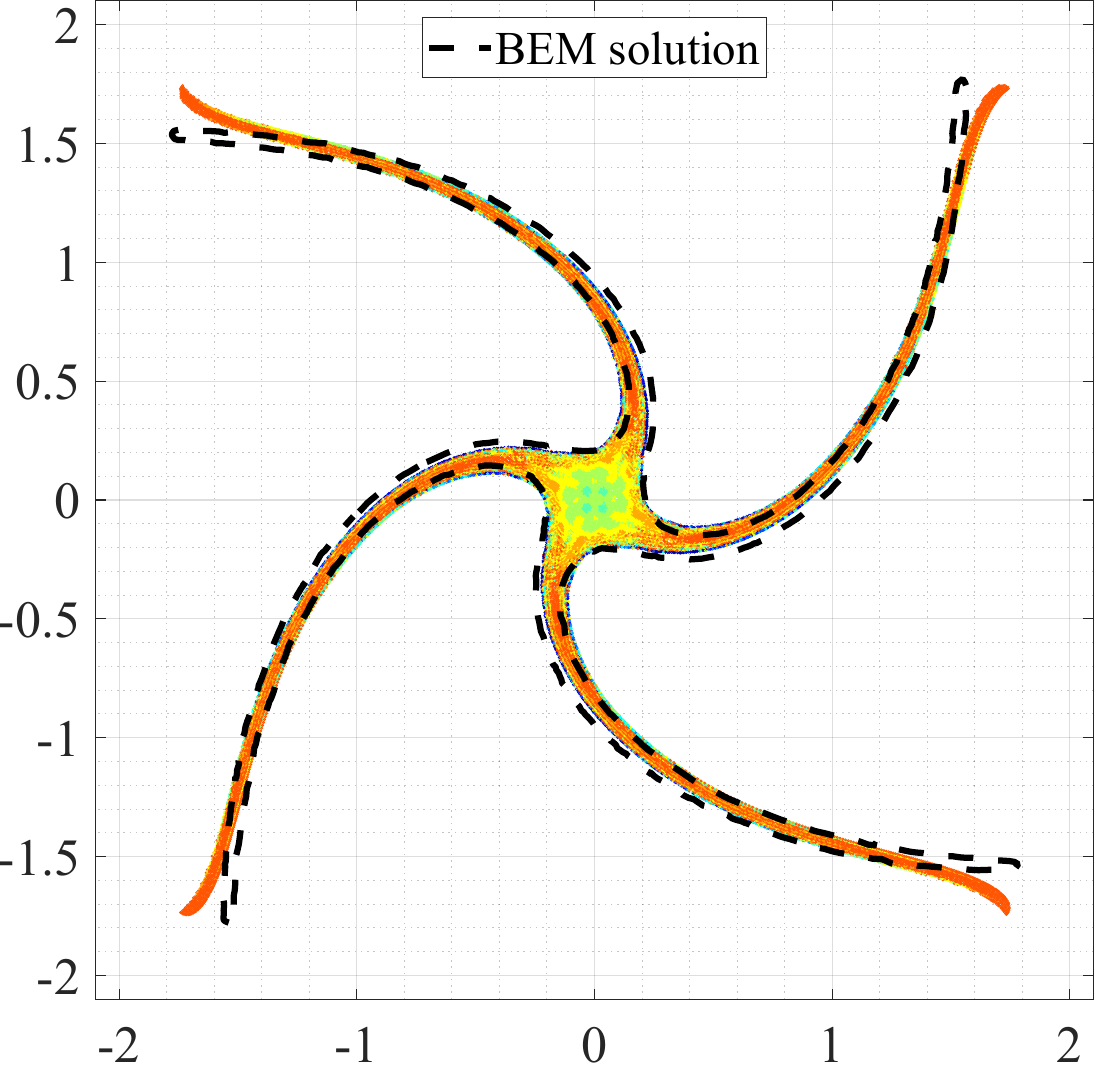}
\caption{$t\omega = 4$}
\end{subfigure}
\caption{Plot of the deformed shapes of the fluid patch with adaptive kernel and constant artificial viscosity parameters $(\gamma_1,\gamma_2)=(0.8,0.8)$. The dashed lines represent the solution obtained by BEM (\citep{oger2016sph} and \citep{sun2017deltaplus}). Particles are colored with pressure contours.}
\label{figure 17}
\end{figure}
\begin{figure}[htp]
\begin{subfigure}[b]{.5\textwidth}
\includegraphics[width=\textwidth]{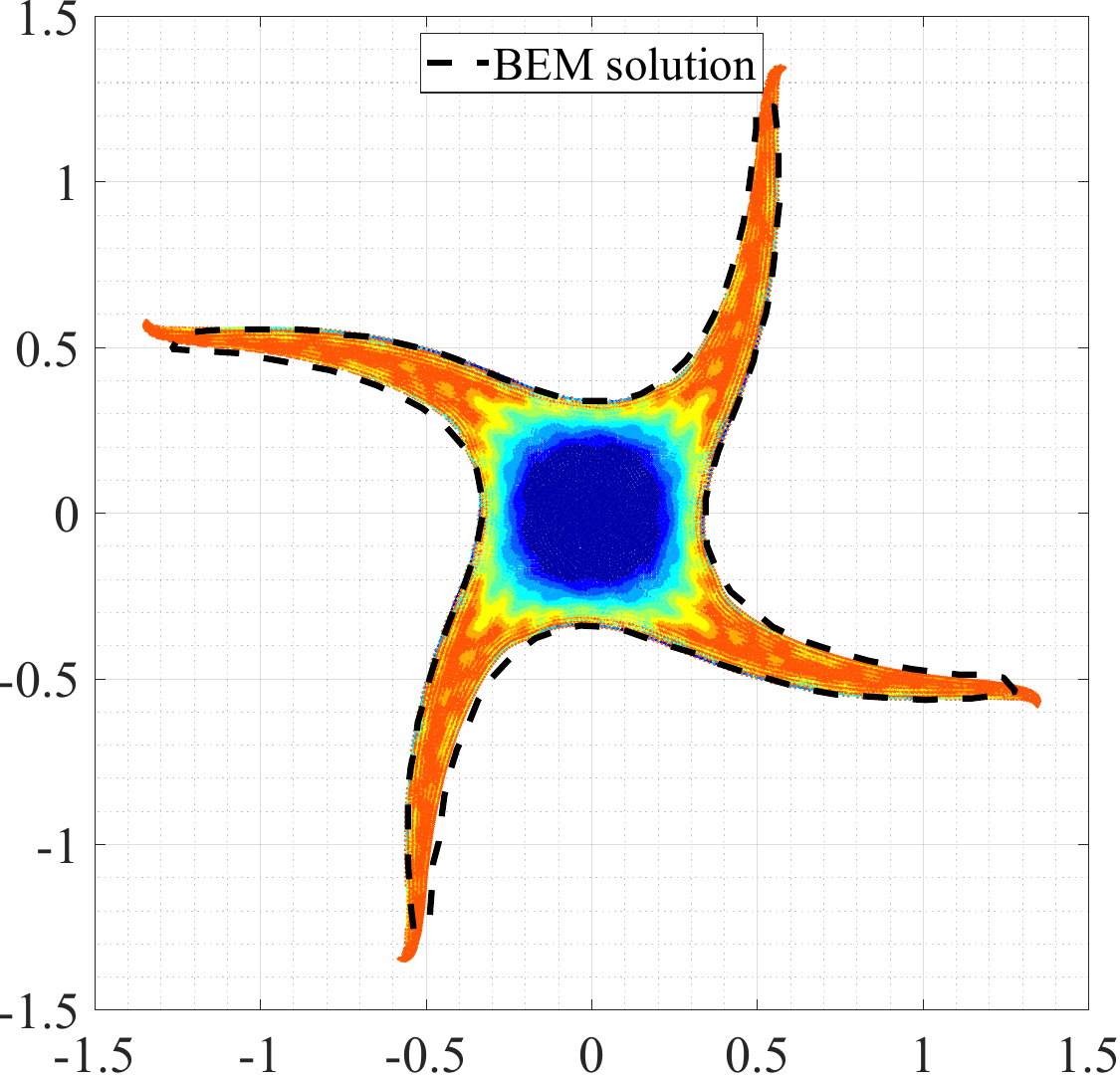}
\caption{$t\omega = 2$}
\end{subfigure}
\begin{subfigure}[b]{.5\textwidth}
\includegraphics[width=\textwidth]{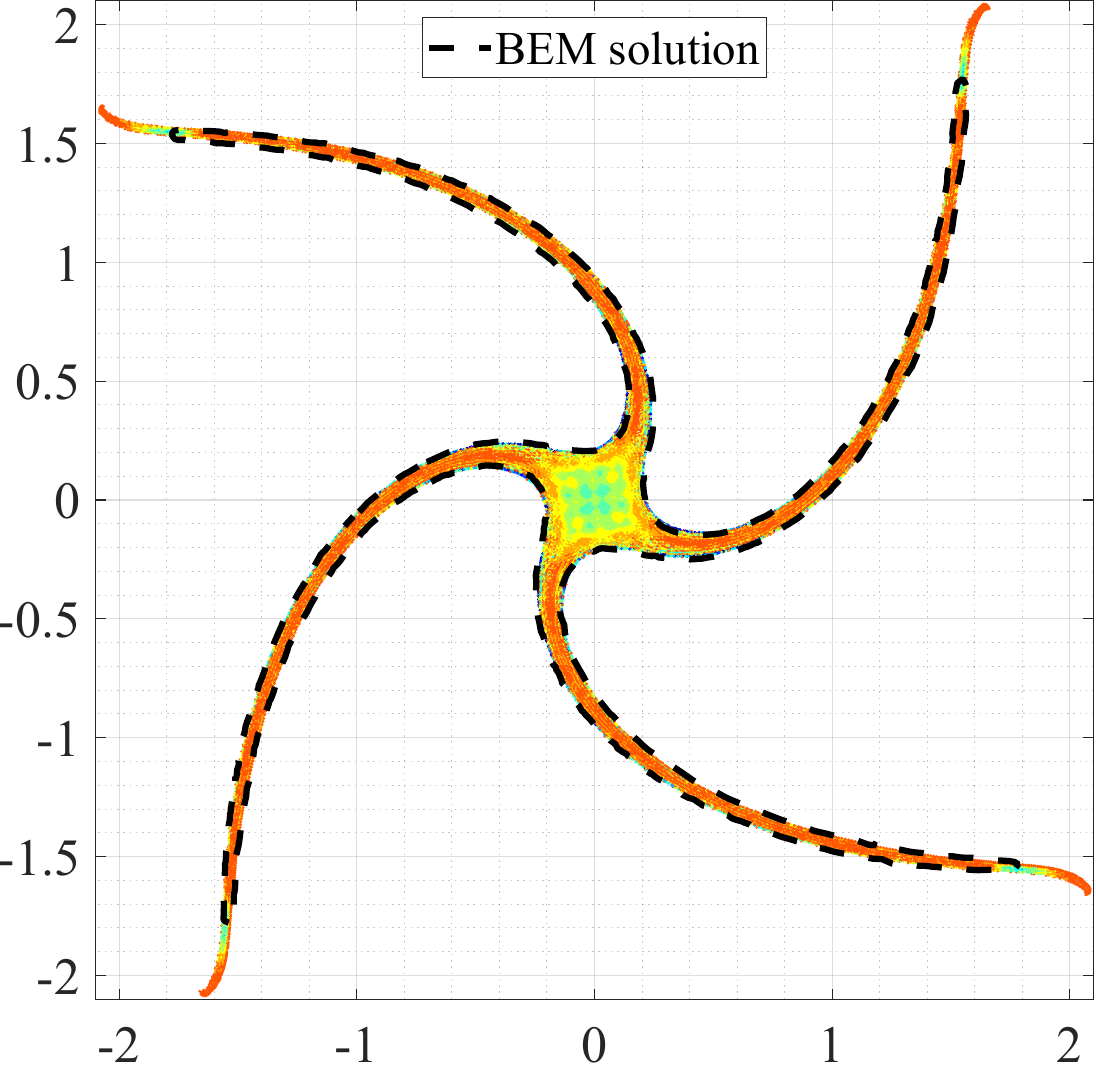}
\caption{$t\omega = 4$}
\end{subfigure}
\caption{Plot of the deformed shapes with adaptive kernel and with artificial viscosity varying spatially according to the pressure. Knot values have not been averaged for an interacting particle pair. The dashed lines represent the solution obtained by BEM (\citep{oger2016sph} and \citep{sun2017deltaplus}). Particles are colored with pressure contours.}
\label{figure 18a}
\end{figure}

\begin{figure}[htp]
\begin{subfigure}[b]{.5\textwidth}
\includegraphics[width=\textwidth]{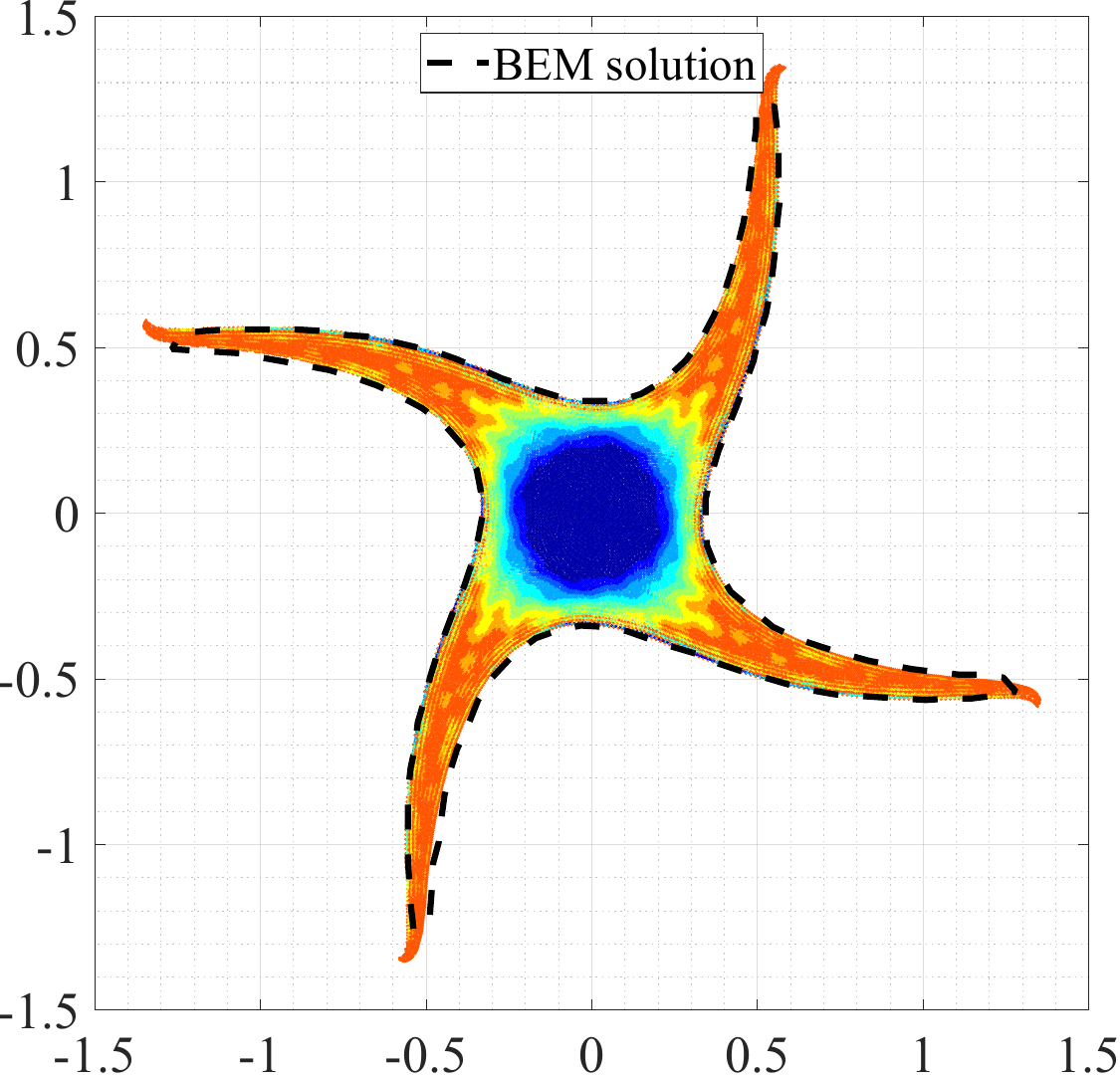}
\caption{$t\omega = 2$}
\end{subfigure}
\begin{subfigure}[b]{.5\textwidth}
\includegraphics[width=\textwidth]{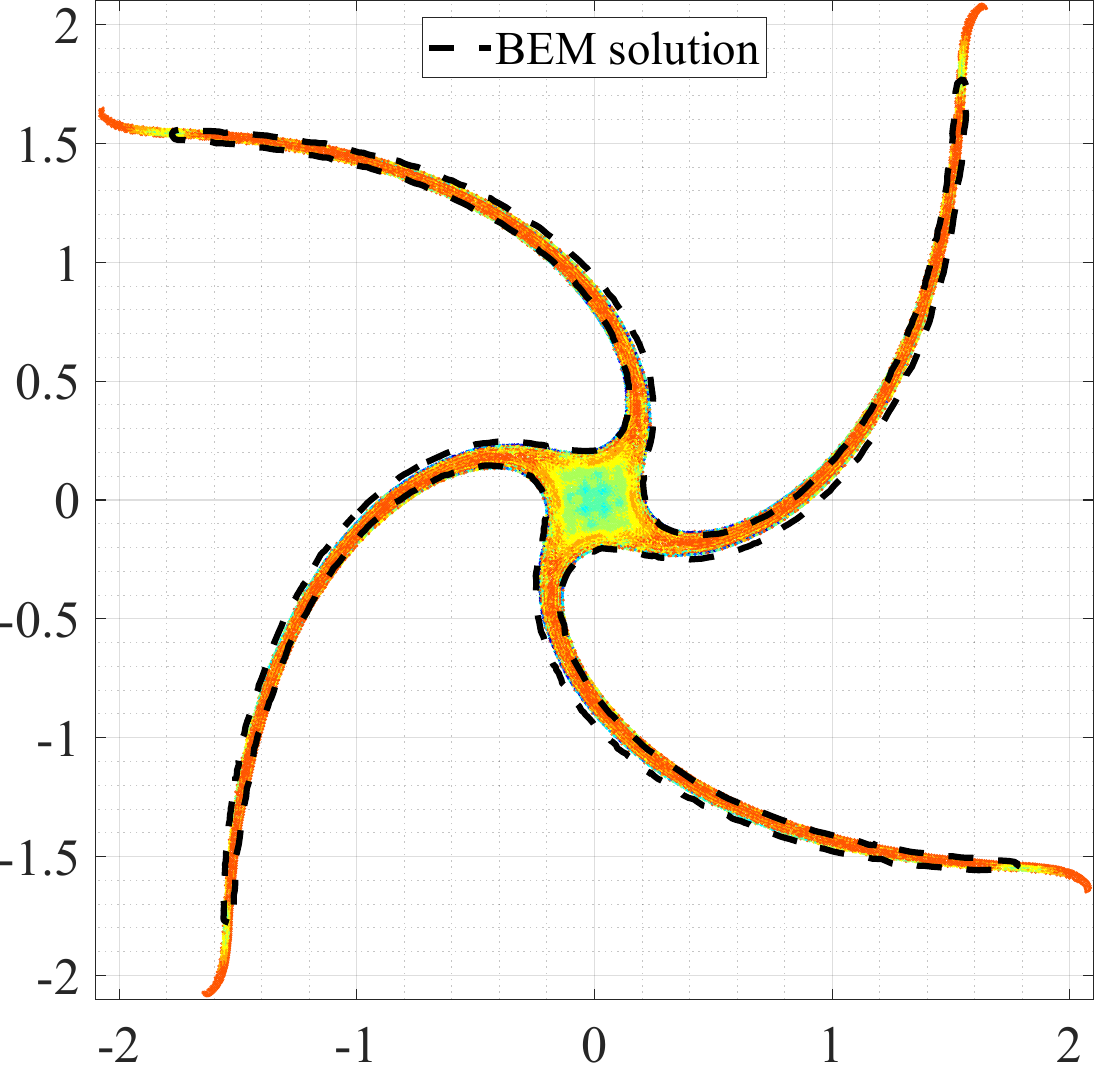}
\caption{$t\omega = 4$}
\end{subfigure}
\caption{Plot of the deformed shapes with adaptive kernel and with artificial viscosity varying spatially according to the pressure. Knot values are averaged for an interacting particle pair. The dashed lines represent the solution obtained by BEM (\citep{oger2016sph} and \citep{sun2017deltaplus}). Particles are colored with pressure contours.}
\label{figure 18b}
\end{figure}
Figure \ref{figure 19} shows the plot of the pressure at the centre of the patch, and it is compared with the BEM solution, and the SPH solution from \cite{le2013critical}. Similar to the result of \citep{le2013critical}, the pressure has an oscillation. The mean of the pressure distribution starts from a slightly higher value as compared to the BEM result. Nevertheless, the result obtained has a considerable agreement with the reference solutions. As the averaging of knot values has not been considered, it should lead to a violation of the conservation of energy. In Figure \ref{figure 20}, the total energy is plotted against time. The deviation in total energy is found to be less than $0.5\%$. As the final simulation, the averaging of the knot values between an interacting particle pair is considered. The deformed shapes are plotted in Figure \ref{figure 18b}, and a comparison with Figure \ref{figure 18a} shows that there is a negligible difference between the two approaches. Figure \ref{figure 20} shows that the energy is conserved in this approach.
\begin{figure}[htp]
\centering
\includegraphics[scale=0.4]{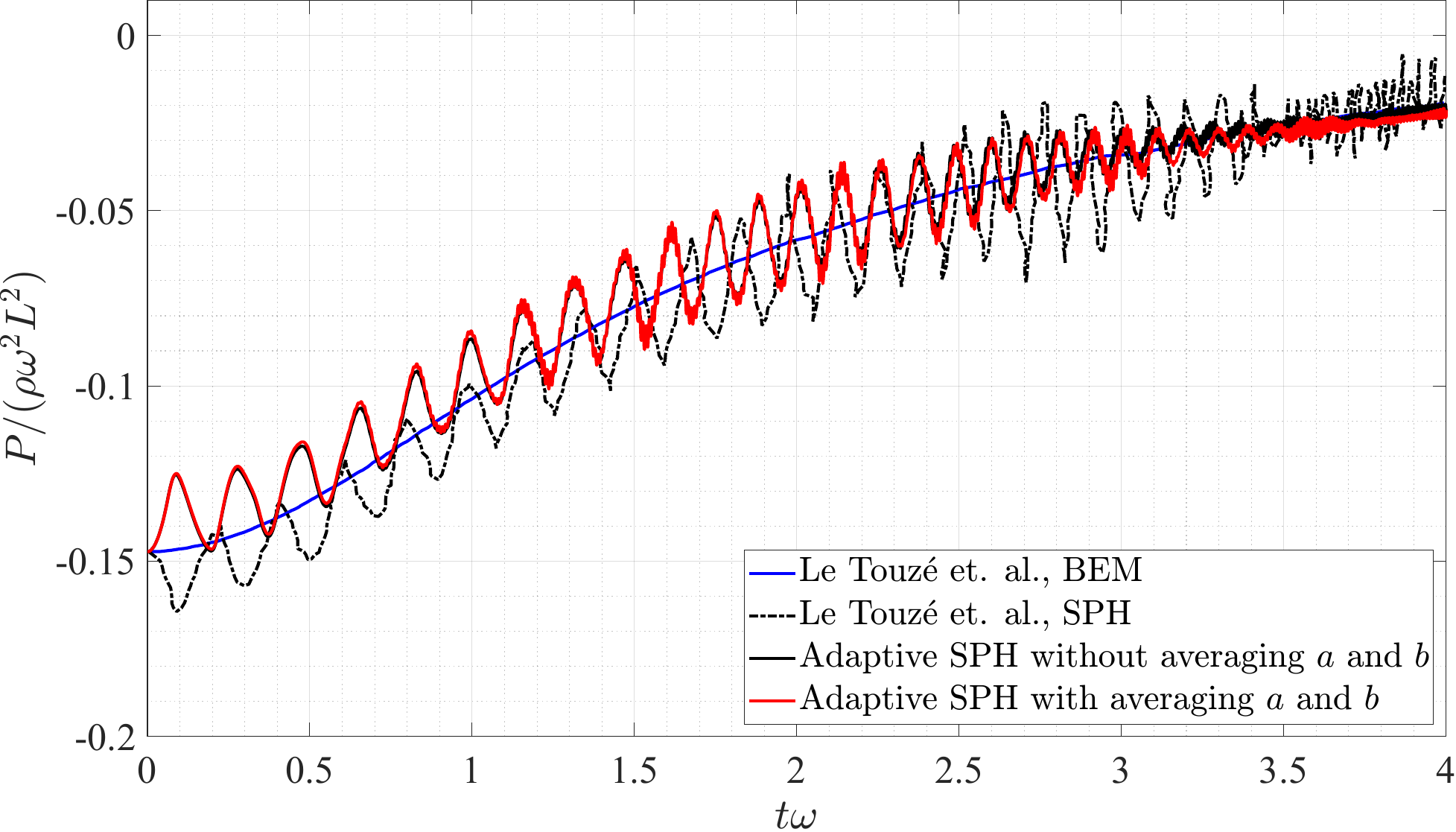}
\caption{Time history of the pressure at the center of the fluid patch. Comparisons have been made between the present SPH simulations with adaptive kernel (black and red solid lines), BEM solution (\citet{le2013critical}) (blue solid line), and the SPH simulation performed by \citet{le2013critical} (dashed dotted line).}
\label{figure 19}
\end{figure}

\begin{figure}[htp]
\centering
\includegraphics[scale=0.4]{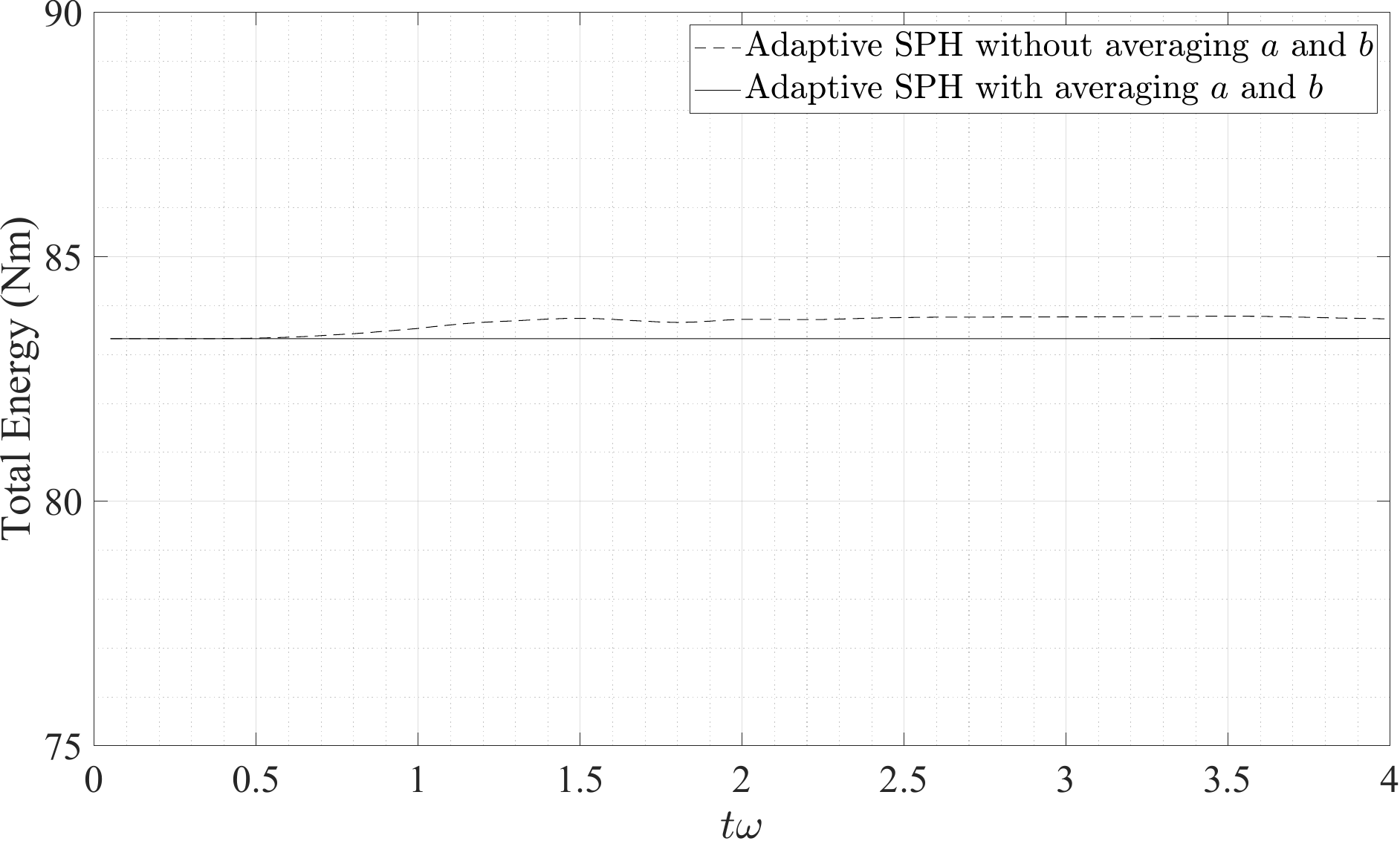}
\caption{Time history of the total energy of the system. Comparison is made between the two SPH simulations, with and without the averaging of knot values between an interacting particle pair.}
\label{figure 20}
\end{figure}
\section{Discussion and conclusions}
\label{s7}
The results can be summarised as follows:
\begin{enumerate}
  \item In this study, first, a 1D perturbation analysis is performed for an Oldroyd B material, both for the exact equations and for the SPH approximations to the exact equations. For the short wavelengths, the SPH dispersion relation shows Zero Energy Modes, which result in the clumping of particles in numerical simulations.
  \item A B-spline basis function is proposed as the SPH kernel, and the knot positions of the basis function can be moved to change the shape of the kernel. By changing the position of the knots, the extremum of the first derivative of the kernel function can be moved in order to satisfy Swegle's condition \citep{swegle1995smoothed}, and prevent \textit{tensile instability}.
  \item Though theoretically, the Zero Energy Modes can be present in the case of tension and compression both, the literature suggests preventing the instability in the direction of tensile stress (\citep{gray2001sph},\citep{monaghan2000sph}) or ensuring positive pressure in the entire domain (\citep{morris1997modeling},\citep{marrone2013accurate}), to eliminate the instability. Hence, in our study, a technique is adopted where the \emph{farthest immediate neighbour} is tracked, and the intermediate knot position of the kernel, $a$, is moved (\textit{a-adaptive}) such that the extremum of the $1$-st derivative of the SPH kernel is just beyond the \emph{farthest immediate neighbour}. If the strain in the direction of tension is large, then both knot positions $a$ and $b$ are shifted (\textit{ab-adaptive}), keeping the support size the same. This satisfies Swegle's condition in the direction of tension. An approach to estimate the \emph{farthest immediate neighbour} from the local strains is demonstrated.
  \item Via the perturbation analysis, it is shown how this proposed technique eliminates the short wavelength Zero Energy Modes but at the same time ensures the accuracy in the long wavelength range.
  \item Numerical simulations of a Newtonian drop and an Oldroyd B drop impacting a rigid surface show how this new algorithm prevents \textit{tensile instability} and gives good agreement with the literature results. In the rotation of a fluid patch problem, it is shown that the artificial viscosity is required in conjunction with the adaptive algorithm to fully eliminate the \textit{tensile instability}. It is also shown that the artificial viscosity alone cannot eliminate the instability. Because the artificial viscosity results in diffusion of the results, the parameters of the viscosity are varied spatially such that the values are maximum where the tensile pressure is the highest and the values have been reduced proportionally as the pressure reduces. A good comparison with the literature results is obtained.
   
\end{enumerate}

\section*{Acknowledgement}
The authors would like to acknowledge the Naval Research Board, DRDO, India for their support of this work.

\bibliographystyle{elsarticle-num-names}
\bibliography{Paper}

\end{document}